\documentclass[aps,prc,twocolumn,showpacs,preprintnumbers,
                          nofootinbib,float,superscriptaddress,longbibliography]{revtex4-1}
\usepackage{graphicx, fancybox}
\usepackage{amsmath,amssymb}
\usepackage[colorlinks=true, pdfstartview=FitV, linkcolor=red,
citecolor=blue, urlcolor=blue]{hyperref}
\usepackage{color}
\usepackage{soul}
\usepackage{url}

\usepackage[normalem]{ulem}

\begin{document}

\title{Collectivity and electromagnetic radiation in small  systems}

\author{Chun Shen}
\affiliation{Department of Physics, McGill University, 3600 University
Street, Montreal,
QC, H3A 2T8, Canada}

\author{Jean-Fran\c{c}ois Paquet}
\affiliation{Department of Physics and Astronomy, Stony Brook University, Stony Brook, New York 11794, USA}

\author{Gabriel S. Denicol}
\affiliation{Instituto de F\'{i}sica, Universidade Federal Fluminense, UFF, Niter\'{o}i, 24210-346, RJ, Brazil}

 \author{Sangyong Jeon}
 \affiliation{Department of Physics, McGill University, 3600 University
 Street, Montreal,
 QC, H3A 2T8, Canada}

 \author{Charles Gale}
 \affiliation{Department of Physics, McGill University, 3600 University
 Street, Montreal,
 QC, H3A 2T8, Canada}

\begin{abstract}

Collective behaviour has been observed in  hadronic measurements of high multiplicity proton+lead collisions at the Large Hadron Collider (LHC), as well as in (proton, deuteron, helium-3)+gold collisions at the Relativistic Heavy-Ion Collider (RHIC). To better understand the evolution dynamics and the properties of the matter created in these small systems,  a systematic study of the soft hadronic observables together with electromagnetic radiation from these collisions is performed, using a hydrodynamic framework. Quantitative agreement is found between theoretical calculations and existing experimental hadronic observables. The validity of the fluid dynamical description is estimated by calculating Knudsen and inverse Reynolds numbers.  Sizeable thermal  yields  are predicted  for low $p_T$  photons. Further predictions of higher order charged hadron anisotropic flow coefficients and  of thermal photon enhancement are proposed.

\end{abstract}

\maketitle
\date{\today}

\section{Introduction}

High energy nucleus-nucleus collision experiments conducted at the Relativistic Heavy-ion Collider (RHIC) and the Large Hadron Collider (LHC) probe QCD (Quantum Chromodynamics) under extreme conditions and create a novel state of matter: the quark-gluon plasma (QGP). This QGP has been found to be strongly-coupled and to exhibit striking collective behaviour. Relativistic hydrodynamics has been  a successful effective theory which provides a quantitative description of this collectivity. In fact, a program of quantitative comparison between theory and experiment now offers the genuine prospect of being able to even extract  transport properties of the QGP, as well as to set strong constraints on properties of the initial state \cite{Gale:2013da,*Heinz:2013th}. 

More specifically, some recent measurements have shown evidence of collectivity in the high multiplicity events of small collision systems such as p+Pb, (p, d, $^3$He)+Au collisions, at RHIC and LHC energies \cite{Chatrchyan:2013eya, Chatrchyan:2013nka,Aad:2013fja,Adare:2014keg, Adare:2015ctn, Loizides:2016tew}. 
These measurements suggest the creation of hot and strongly-coupled QGP droplets within a reaction zone of merely a few fm in size, suggesting that the mechanism responsible for the rapid approach to local equilibrium of plasma produced in heavy ion collisions may also operate in such  collisions.
Because the hydrodynamic description relies on well-separated distance/time scales between microscopic and macroscopic physics, the dynamics of these small collision systems appears to challenge the very validity of the fluid dynamical approach.
The origin of these collective features is therefore actively investigated, and one would like to elucidate whether they are inherited from properties of the initial state \cite{Dumitru:2010iy,Dusling:2012cg,Dusling:2013qoz,Dumitru:2014yza}, or appear during the collective expansion \cite{Nagle:2013lja,Bozek:2011if,Bozek:2012gr,Schenke:2014zha,Werner:2013ipa,Kozlov:2014fqa}. 

In order to unveil the  dynamics in these small systems, one must investigate multiple aspects of experimental observables within a consistent framework. A complete slate of measurements could include hadronic anisotropic flow \cite{Bozek:2013uha,Kozlov:2014fqa}, the mass ordering of identified particle $v_2$ \cite{Adare:2014keg, ABELEV:2013wsa}, particle interferometry \cite{Bozek:2013df,Shapoval:2013jca}, as well as penetrating probes such as QCD jets \cite{Shen:2016egw} and direct real and virtual photons \cite{Shen:2015qba}. 
This last observable, radiated throughout the dynamical evolution of the hot expanding medium and suffering from negligible final state interactions, is a particularly valuable probe since the local properties of the fluid at the photon's production point can be directly carried to the detectors.
Current experiments are able to isolate low-energy direct photons $p_T \lesssim 1$ GeV. Those photons are thus penetrating {\it and} soft: they enjoy a unique status among all the observables measured in hadronic  collisions.  

In this paper, we build on  previous work \cite{Shen:2015qba} by performing detailed comparisons with various hadronic measurements of high multiplicity p+Pb collisions at 5.02 TeV. A microscopic transport stage is employed to describe the out-of-equilibrium dynamics of the collision systems in the dilute hadronic phase. Its effect on hadronic observables in small systems is quantified. Systematic studies of hadronic and direct photon observables in (p, d, $^3$He)+Au collisions at the top RHIC energy are also presented within the same theoretical framework.

The rest of the paper is organized as follows. A description of the components of the hydrodynamical modelling, together with the choice of parameters, appear in Sec. II. A variety of collective signals involving soft hadronic observables are compared with existing experimental measurements in Sec. III. In Sec. IV, direct photon production from the small collision systems is studied in detail. The electromagnetic tomography and the viscous corrections to photon observables are discussed. Final remarks and conclusions   appear in Sec. V.

\section{Hydrodynamic modelling}

In this work, all the collision systems are numerically simulated using an hydrodynamics + hadronic cascade framework \cite{Shen:2014vra}. Fluctuating initial conditions in the transverse plane are generated using the Monte-Carlo-Glauber (MC-Glauber) model. 
The nucleon spatial configurations inside the heavy nuclei are sampled considering realistic repulsive 2-body nucleon-nucleon correlations \cite{Alvioli:2009ab}. For the light nuclei, the spatial topography of the deuteron's two-nucleon system is obtained from sampling the Hulthen wavefunction \cite{Adare:2013nff}, and  fluctuating $^3$He configurations come from results of Green's function Monte Carlo calculations using the AV18+UIX model interaction \cite{Carlson:1997qn}. 
Multiplicity fluctuations in every binary collision are randomly-sampled from a $\Gamma$-distribution at the wounded nucleon positions. The shape and scale parameters in the $\Gamma$-distribution are chosen to reproduce the multiplicity distribution measured in pp collisions at the same collision energy \cite{Shen:2014vra}. The inclusion of such multiplicity fluctuation was shown to be essential to reproduce the measured high multiplicity tail of the charged hadron yield distributions in small collision systems \cite{Alver:2008aq,Bozek:2013uha,Kozlov:2014fqa,Shen:2014vra}. 
In order to simulate efficiently high multiplicity events, centrality selection is determined by sorting minimum bias events according to their initial total entropy at  mid-rapidity, $dS/d\eta_s\vert_{\eta_s = 0}$, where $\eta_s = \frac{1}{2} \log \frac{t+z}{t-z}$ is the space-time rapidity. Triggering on initial total entropy of the system was shown to be a good approximation to determining centrality according to final charged hadron multiplicity, as done in  experiments \cite{Shen:2015qta}.

Starting from $\tau_0 = 0.6$ fm/$c$, an individual fluctuating entropy density profile is then evolved using a (2+1)D viscous hydrodynamics, {\tt VISH2+1}, with a lattice-QCD based equation of state (EoS), s95p-v1.2 \cite{Huovinen:2009yb}. 
The effect of the longitudinal dynamics of the system is also investigated using (3+1)D viscous hydrodynamics \cite{Schenke:2010nt}. 
Second order non-linear terms are included in the evolution of the shear stress tensor \cite{Denicol:2012cn},
\begin{eqnarray}
\tau_\pi \dot{\pi}^{\langle \mu\nu \rangle} + \pi^{\mu\nu} = && 2\eta \sigma^{\mu\nu} - \delta_{\pi\pi} \pi^{\mu\nu} \theta + \phi_7 \pi_\alpha^{\langle \mu} \pi^{\nu \rangle \alpha}  \notag \\  
&& - \tau_{\pi \pi} \pi_\alpha^{\langle \mu} \sigma^{\nu \rangle \alpha}.
\end{eqnarray}
The transport coefficients, $\tau_\pi$, $\delta_{\pi\pi}$, $\phi_7$, and $\tau_{\pi\pi}$ are fixed using formulae derived from the Boltzmann equation near the conformal limit \cite{Denicol:2014vaa}. 
An effective specific shear viscosity $\eta/s = 0.08$ is chosen for the collisions at the top RHIC energy and a slightly larger value $\eta/s = 0.10$ is used for p+Pb collisions at 5.02 TeV. With these choices of $\eta/s$, the simulated results can reproduce the measured charged hadron anisotropic flow coefficients, $v_{2,3}$, in central collisions fairly well, as will be shown in Figs.~\ref{fig1} and \ref{fig2}. 
The hydrodynamic description is switched to a microscopic hadronic cascade, UrQMD v3.4 \cite{Bass:1998ca,Bleicher:1999xi},   at $T_\mathrm{sw} = 155$\,MeV as the collision system becomes more dilute and out-of-equilibrium. 

The equation of state (EoS) s95p-v1.2, fitted to lattice calculations at high temperature is matched in the confinement region to a hadron resonance gas which comprises the same hadronic content implemented in UrQMD v3.4 \footnote{There are 305 hadronic species in total, up to 2.25 GeV in their mass, included in the hadron resonance gas phase in s95p-v1.2 and UrQMD v3.4. Please see {\url{http://urqmd.org/itypes.html}} for details.}.
In this way, the total energy of system is ensured to remain the same during the conversion from fluid cells into hadrons.  
The value of $T_\mathrm{sw}$ is fixed to reproduce the correct $p$ to $\pi$ ratio measured in $0-20$\% d+Au collisions \cite{Adare:2013esx}.

At the end of hadronic scatterings and resonance decays, the particle spectra and flow observables of stable particles are analyzed.

The anisotropic flow coefficients of particle of interests are evaluated using the scalar-product method \cite{Luzum:2012da},
\begin{equation}
v_n\{\mathrm{SP}\} = \frac{\langle v_n(p_T) v^\mathrm{ref}_n \cos[n(\Psi_n(p_T) - \Psi_n^\mathrm{ref})]\rangle}{\sqrt{\langle (v^\mathrm{ref}_n)^2 \rangle }}.
\label{eq1}
\end{equation}
For hadronic flow, we choose $v_n$ of charged hadrons integrated from $p_T = 0.3$ to 3 GeV as the reference flow vector for p+Pb collisions at 5.02 TeV to compare with the CMS data \cite{Chatrchyan:2013nka}. The $v^\mathrm{ref}_n$ is set to $v^\mathrm{ch}_n$ integrated from $p_T = 0.2$ to 2 GeV for (p, d, $^3$He)+Au collisions at the top RHIC energy, in line with  PHENIX measurements. 

To accumulate enough statistics for the hadronic and direct photon observables, we evolve 300 fluctuating events through hydrodynamics for every centrality  class presented in this paper. Then 2000 hadronic cascade runs are simulated for each hydrodynamic event. Finally, these 2000 ``oversampled'' events from the same hydrodynamic event are combined  in the hadronic flow analysis.

\section{Collectivity in small systems}

In this section, the collective behavior of hadronic observables is explored, for small collision systems at RHIC and LHC energies.  

\subsection{Prelude: The effect of broken longitudinal boost-invariance}
\begin{figure*}[ht!]
  \centering
  \centering
  \begin{tabular}{cc}
  \includegraphics[width=0.45\linewidth]{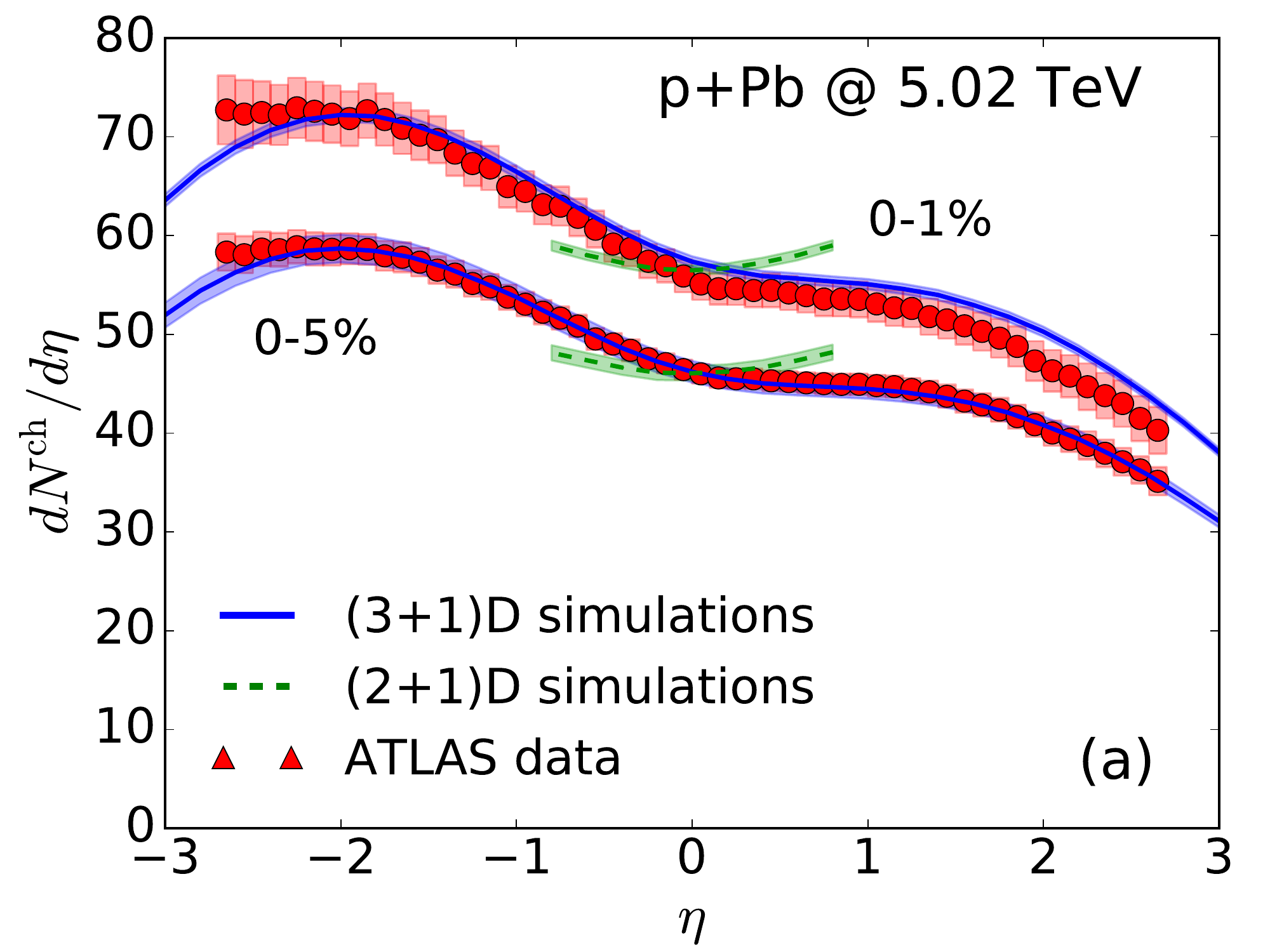} &
  \includegraphics[width=0.45\linewidth]{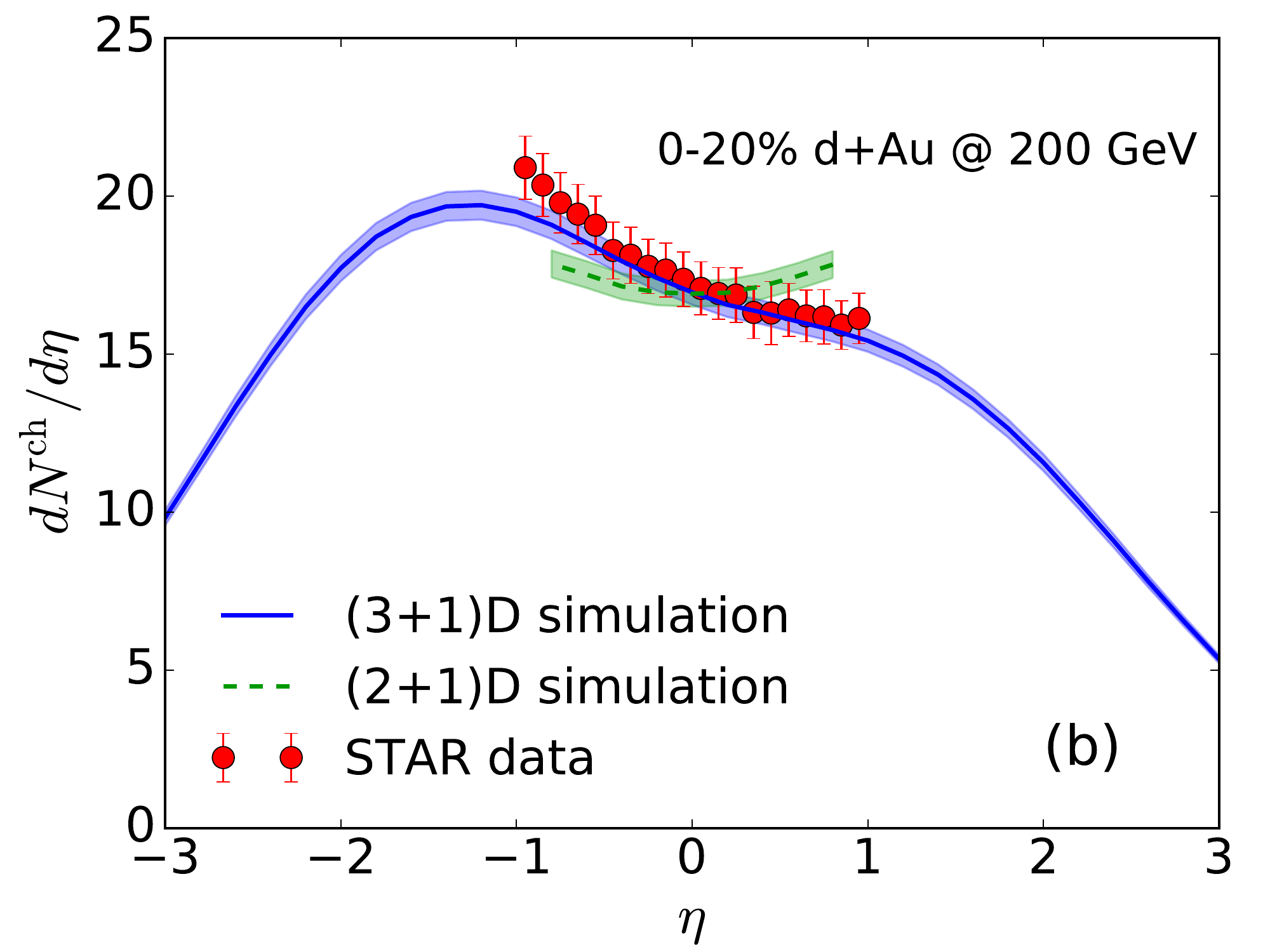}
  \end{tabular}
  \caption{(Color online) Pseudorapidity dependence of charged hadron multiplicity for central p+Pb collisions at 5.02 TeV (panel (a)) and d+Au collisions at 200 GeV (panel (b)) for $(2+1)$D and $(3+1)$D hydrodynamics, compared with ATLAS~\cite{Aad:2015zza} and STAR~\cite{Adams:2004dv} measurements. The shaded bands represent statistical uncertainty.}
  \label{figdNdeta}
\end{figure*}
%

\begin{figure*}[ht!]
  \centering
  \centering
  \begin{tabular}{cc}
  \includegraphics[width=0.45\linewidth]{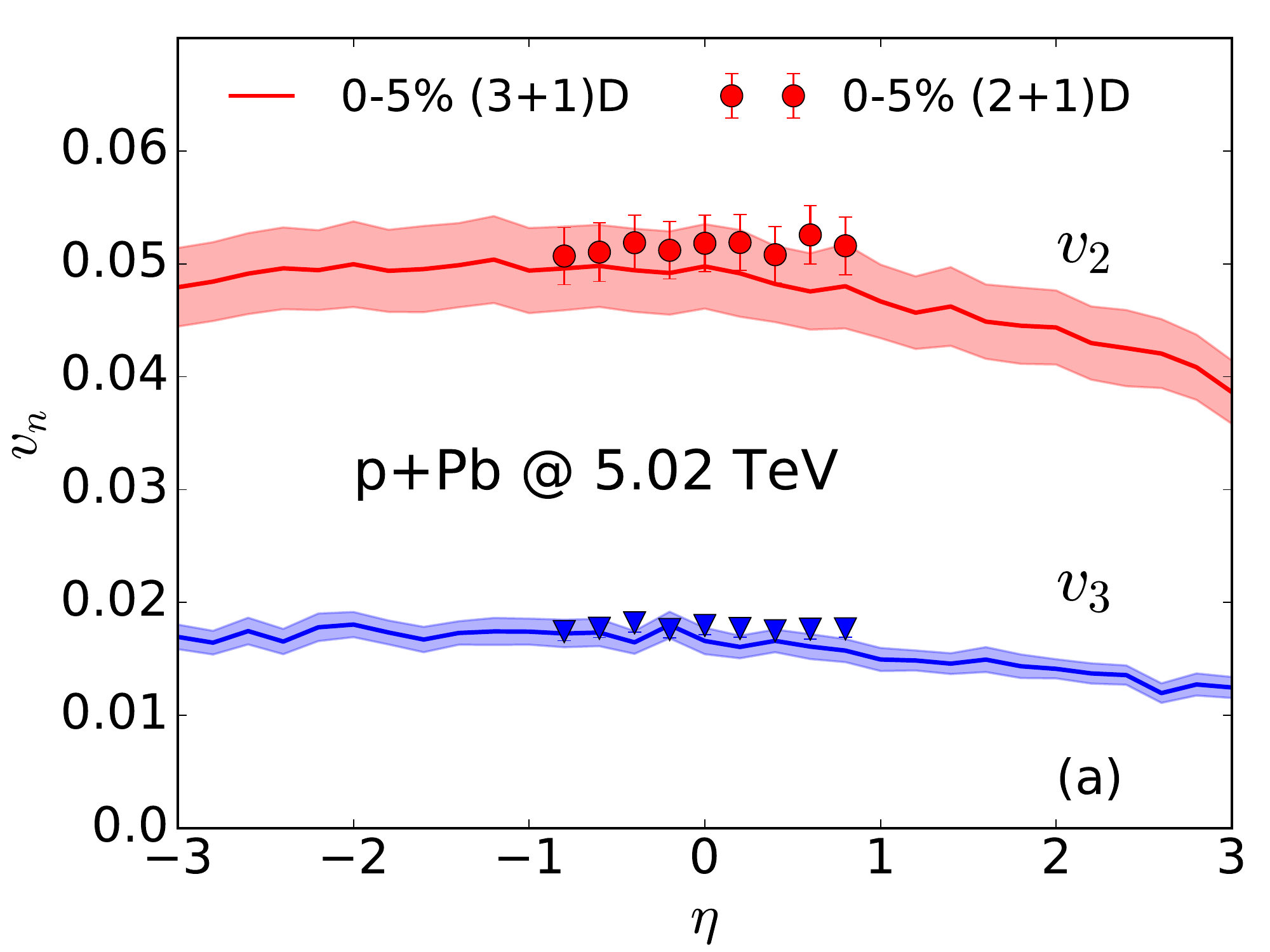} 
  \includegraphics[width=0.45\linewidth]{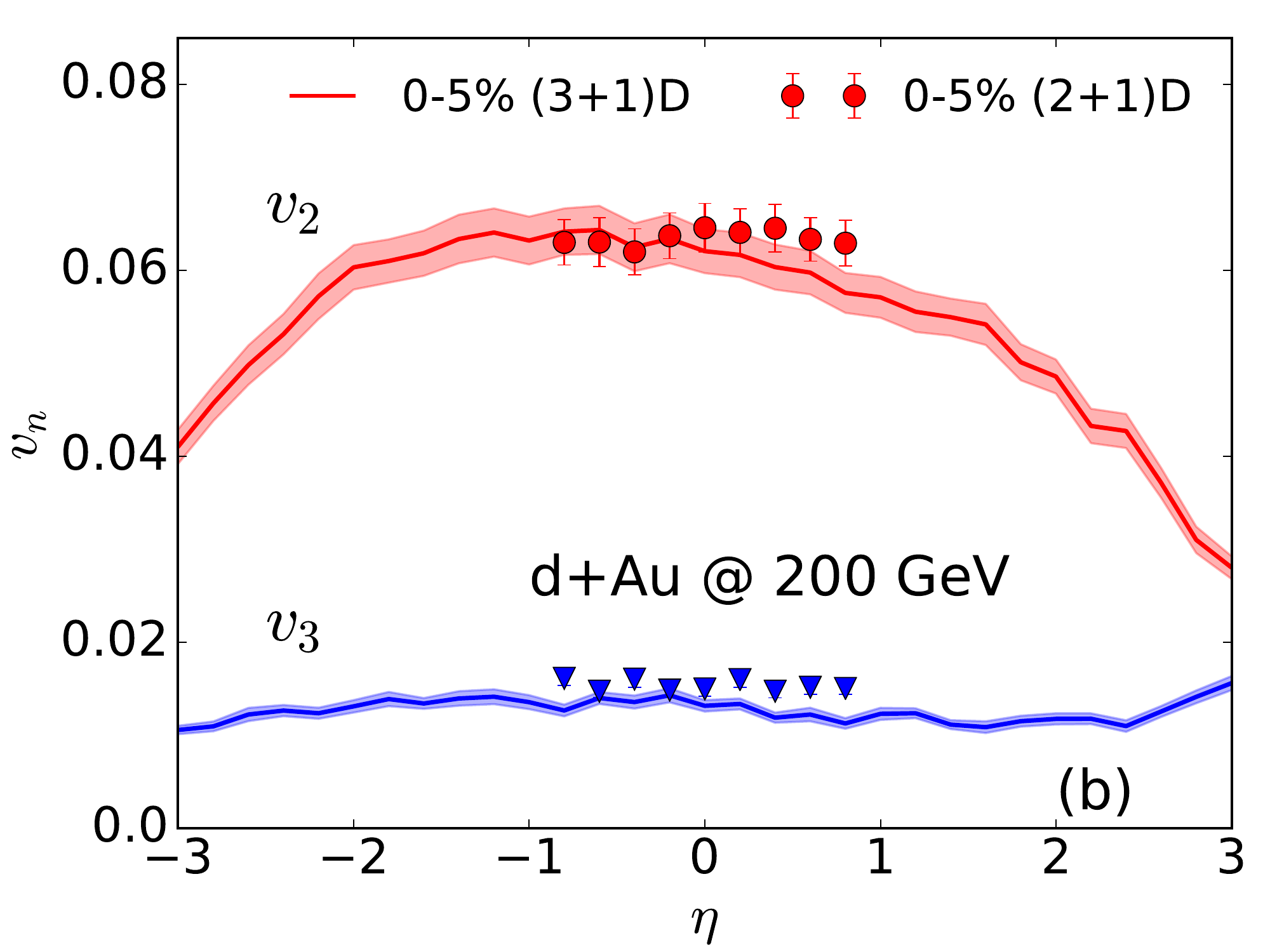}
  \end{tabular}
  \caption{(Color online) Charged hadron anisotropic flow coefficients $v_{2, 3}\{2\}$ as a function of particle's pseudorapidity in central p+Pb collisions at 5.02 TeV (panel (c)) and d+Au collisions at 200 GeV (panel (d)) for $(2+1)$D (points) and $(3+1)$D (lines) hydrodynamics. The shaded bands represent statistical uncertainty. }
  \label{figvnEta}
\end{figure*}
%
Light-heavy nuclei collisions are asymmetric: boost-invariance in the longitudinal direction is explicitly broken. This study starts by quantifying the effects of  longitudinal boost-non-invariance in small systems on hadronic and photon flow observables, at mid-rapidity. 

To model these collisions in three dimensions, we extend the MC-Glauber initial energy density profiles in the longitudinal direction with the following envelope functions \cite{Kozlov:2014fqa}: 
\begin{eqnarray}
e({\bf x_\perp}, \eta) =&&  f_L(\eta)\left[\sum_{i=1}^{N_\mathrm{part}^\mathrm{left}} \exp\left(-\frac{({\bf x_\perp - x}_i)^2}{2\sigma^2} \right) \right] \notag \\
&+& f_R(\eta) \left[\sum_{i=1}^{N_\mathrm{part}^\mathrm{right}} \exp\left(-\frac{({\bf x_\perp - x}_i)^2}{2\sigma^2} \right) \right],
\end{eqnarray}
where the envelope function $f_{L,R}(\eta)$ is given by
\begin{eqnarray}
&&f_{L,R}(\eta) = \left(1 \pm \frac{\eta}{\eta_\mathrm{max}} \right) \notag \\
&&\quad\times \left[\theta(\vert\eta\vert - \eta_0)\exp\left(-\frac{(\eta - \eta_0)^2}{2\sigma_\eta^2} \right) + \theta(\eta_0 - \vert \eta \vert) \right]
\end{eqnarray}
Here, $\eta_{\mathrm{max}}=y_\mathrm{beam}$, the beam rapidity. The parameters, $\eta_0$ and $\sigma_\eta$, are fixed such that the measured pseudo-rapidity dependence of charged hadron multiplicity is reproduced. 

The full (3+1)D hydrodynamic equations are solved with {\tt MUSIC} \cite{Schenke:2010nt}.
The transport parameters in the hydrodynamic simulation and transition to the hadronic cascade are the same as those in the (2+1)D simulations.

\begin{figure*}[ht!]
  \centering
  \begin{tabular}{cc}
  \includegraphics[width=0.45\linewidth]{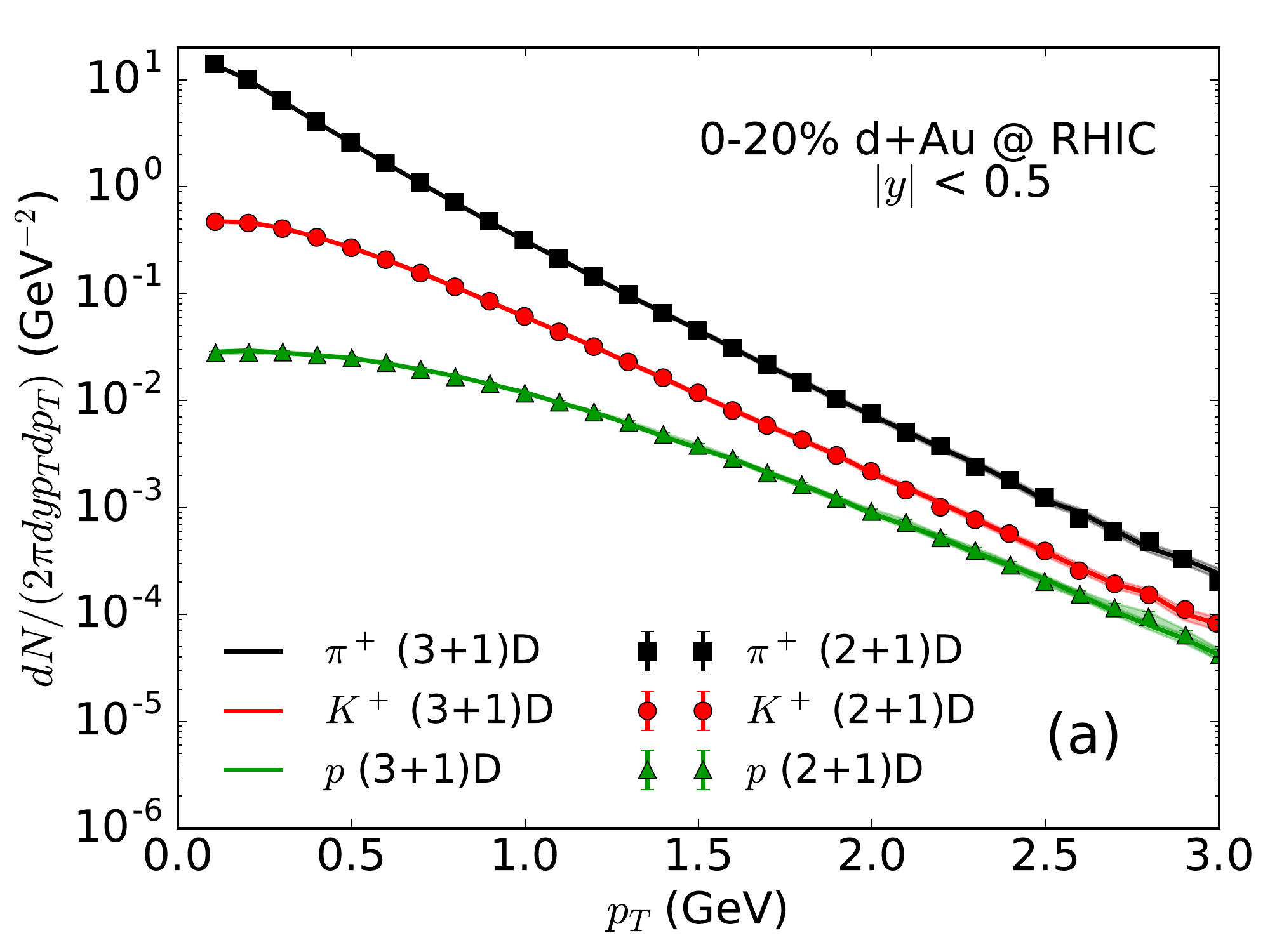} &
  \includegraphics[width=0.45\linewidth]{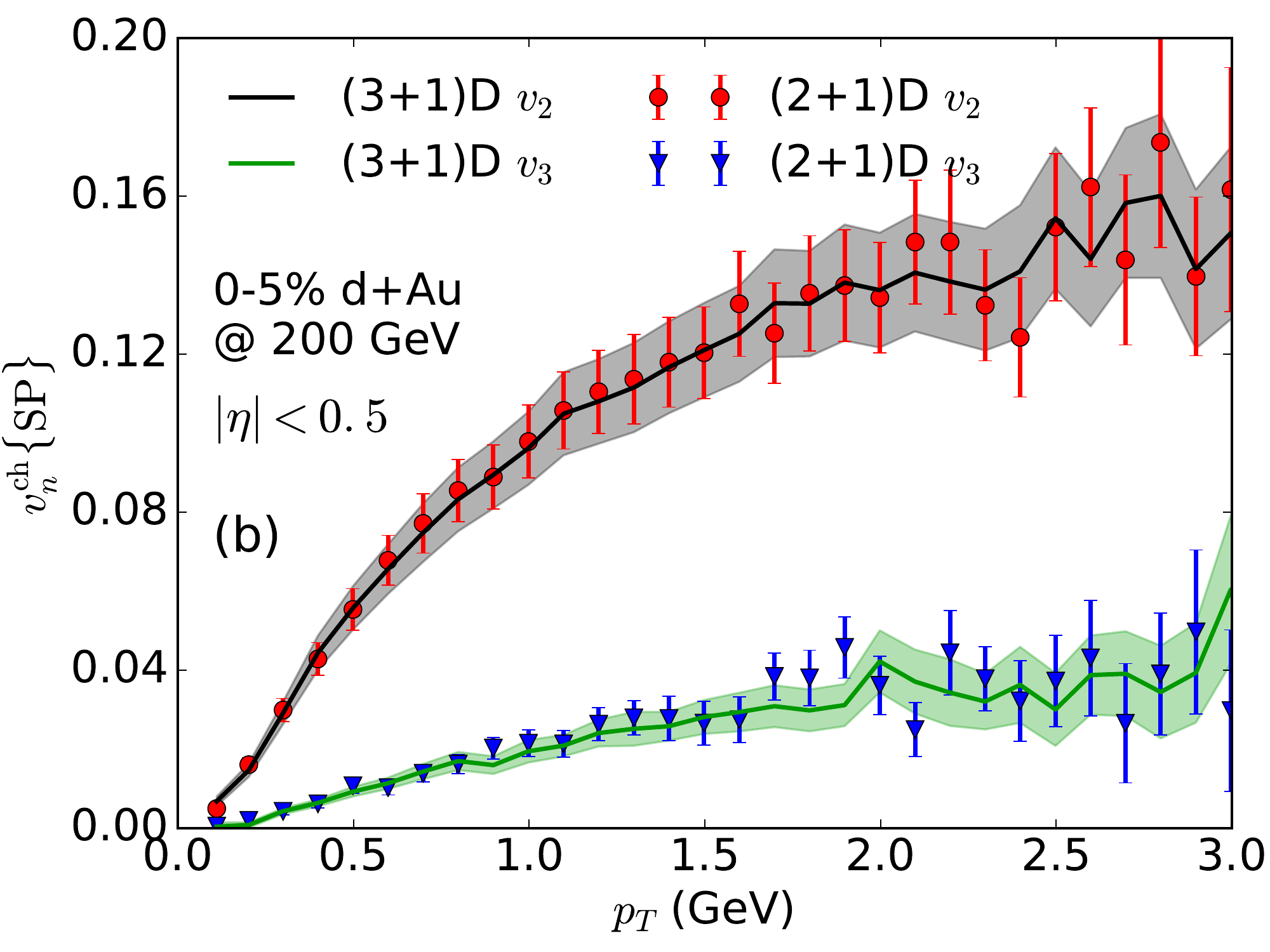}
  \end{tabular}
  \caption{(Color online) Comparisons of mid-rapidity identified particle spectra and charged hadron $p_T$-differential $v_{2,3}\{\mathrm{SP}\}(p_T)$ between (2+1)D simulations assuming longitudinal boost-invariance and full (3+1)D calculations for central d+Au collisions at 200 GeV.  The shaded bands represent statistical uncertainty.}
  \label{fig0.2}
\end{figure*}
%
\begin{figure*}[ht!]
  \centering
  \centering
  \begin{tabular}{cc}
  \includegraphics[width=0.45\linewidth]{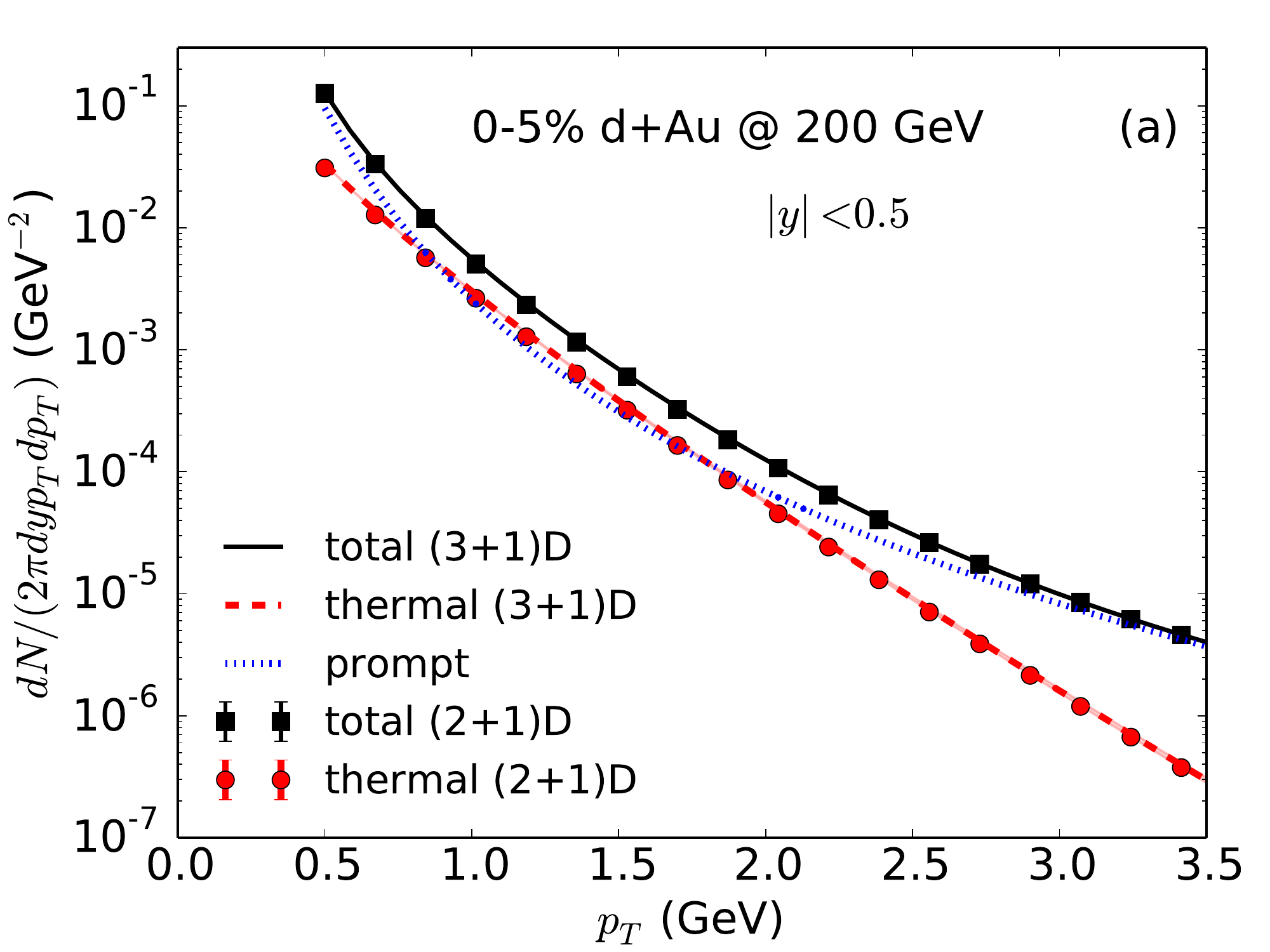} &
  \includegraphics[width=0.45\linewidth]{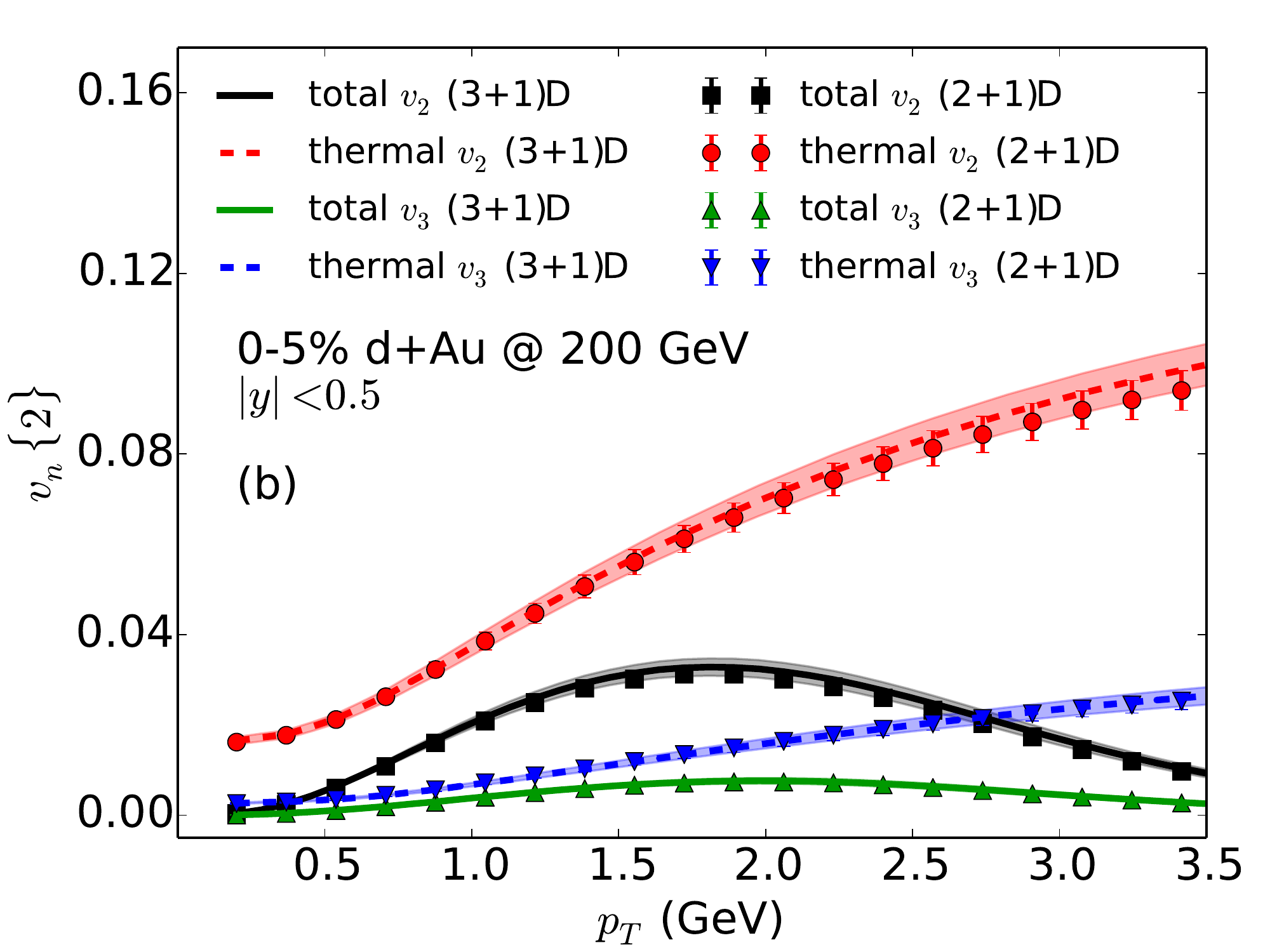}
  \end{tabular}
  \caption{(Color online) Comparisons of direct photon spectra and anisotropic flow coefficients at $y=0$ between (2+1)D simulations assuming longitudinal boost-invariance and full (3+1)D calculations for d+Au collisions at 200 GeV. The shaded bands represent statistical uncertainty.}
  \label{fig0.3}
\end{figure*}
%

Fig.~\ref{figdNdeta} shows the pseudo-rapidity dependence of charged hadrons multiplicity and their anisotropic flow coefficients in central p+Pb collisions at 5.02 TeV and central d+Au collisions at 200 GeV. The $\eta$-dependence in charged hadron yields is quite strong in these highly asymmetric collision systems. The results from boost-invariant simulations are indicated as the green dashed lines, within a pseudo-rapidity $\vert \eta \vert < 1$ interval. Although $dN^\mathrm{ch}/d\eta$ is a symmetric function in $\eta$ for the (2+1)D simulations, the integrated $dN^\mathrm{ch}/d\eta \vert_{\vert \eta \vert< 0.5}$ still agrees well with the full (3+1)D simulations. 

Fig.~\ref{figvnEta} compares the pseudo-rapidity dependence of charged hadron $v_n$  of (2+1)D and (3+1)D simulations, in both central p+Pb collisions and central d+Au collisions. The momentum anisotropy is relatively flat in the mid-rapidity region, $\vert \eta \vert < 1$ and the (2+1)D results (indicated by the circle and triangle markers) agree quite well with the full (3+1)D simulations.

Fig.~\ref{fig0.2} features the $p_T$-differential particle spectra and charged hadron $v_n(p_T)$, as obtained with (2+1)D and (3+1)D simulations at  mid-rapidity. The hadronic observables computed with (2+1)D simulations serve as a good approximation to the (3+1)D results at mid-rapidity, even in these highly asymmetric  systems. 

We postpone a discussion of photon observables until Section \ref{photon_section}, but it suffices to say here that thermal photons are  sensitive to the entire evolution history of the medium than hadronic observables. In Figs.~\ref{fig0.3}, the thermal photon spectra and the anisotropic flow coefficients are shown, for (2+1)D and (3+1)D simulations, for photons at mid-rapidity. Calculations show that boost-invariance is still a very good approximation for thermal photon production at mid-rapidity in d+Au collisions at  RHIC. We verified that similar results were found in p+Pb collisions at LHC energy.

\subsection{Hadronic flow observables}

As shown in the previous subsection, boost-invariance is still a good approximation for mid-rapidity observables in  asymmetric  systems. We shall thus henceforth conduct our systematic study using boost-invariant conditions. All  hadronic flow observables are computed for pseudo-rapidity $\vert \eta \vert < 0.5$.

\begin{figure*}[ht!]
  \centering
  \centering
  \begin{tabular}{cc}
  \includegraphics[width=0.4\linewidth]{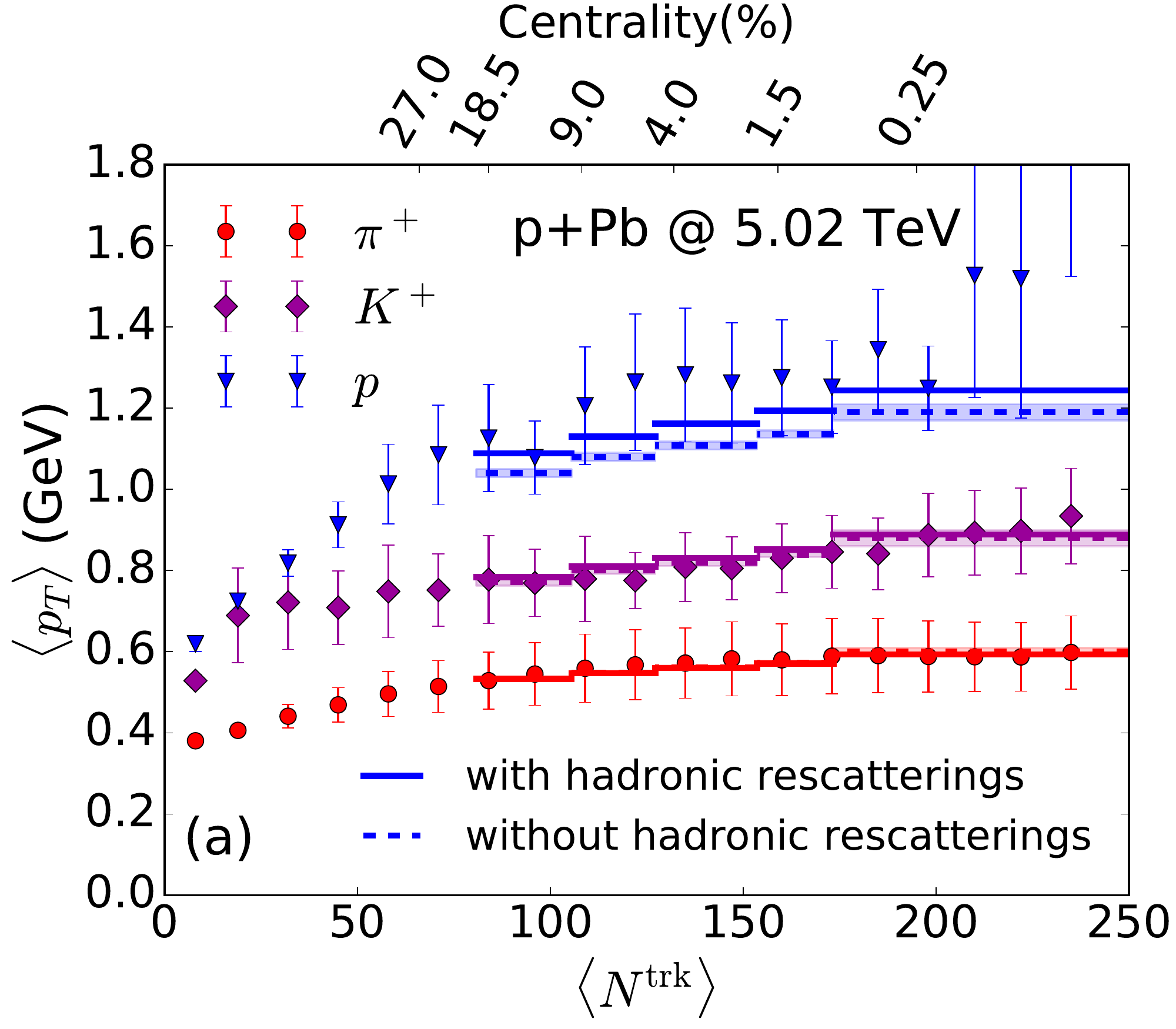} & 
  \includegraphics[width=0.4\linewidth]{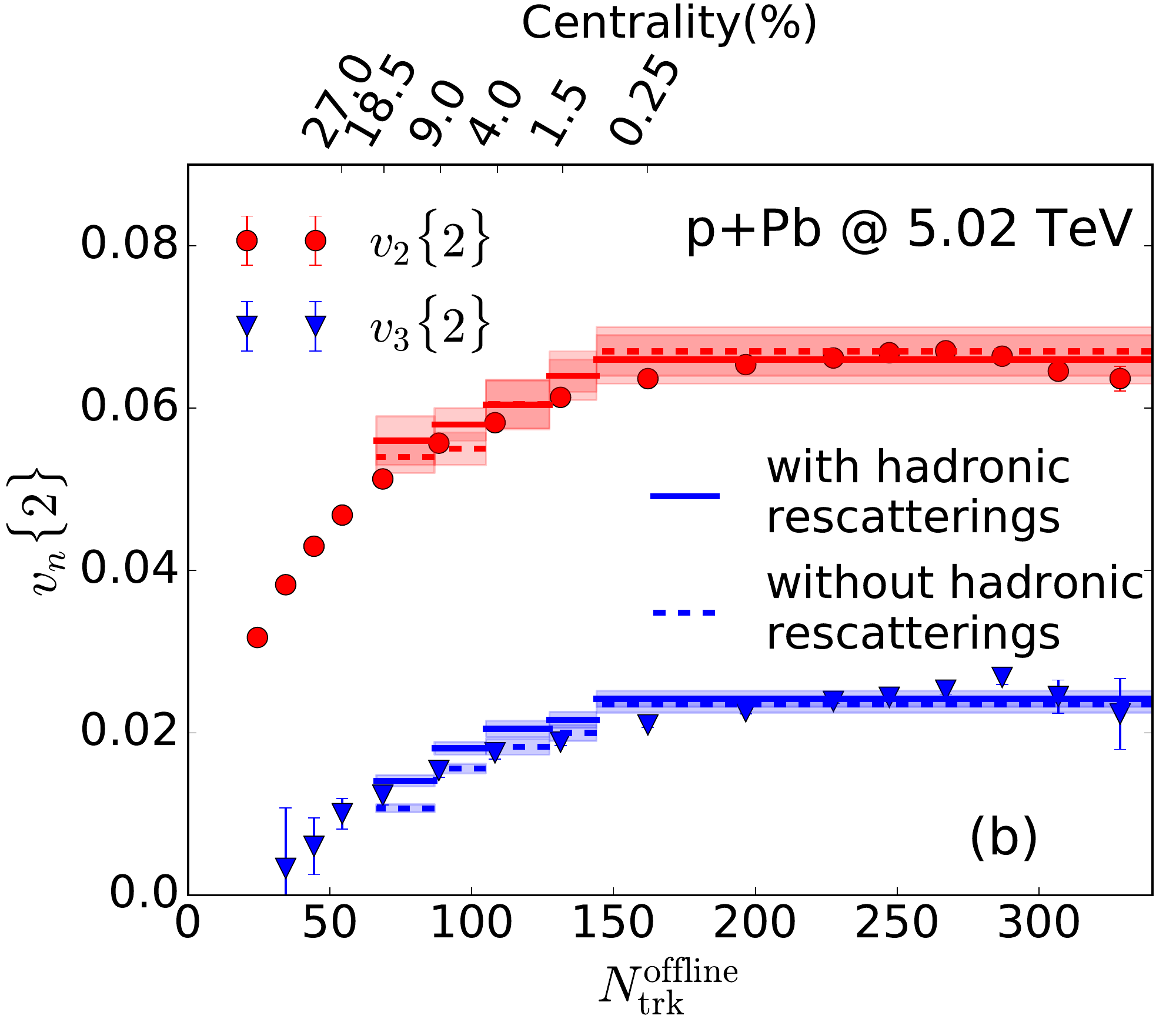}
  \end{tabular}
  \caption{(Color online) Identified particle averaged transverse momentum (a), and charged hadron anisotropic flow coefficients (b), compared to the CMS measurements \cite{Chatrchyan:2013eya,Chatrchyan:2013nka}. Charged hadron $v_n\{2\}$ is integrated from $p_T =0.3$ to 3.0 GeV. The mapping between centrality class boundaries and the number of particle tracks, $\langle N^\mathrm{trk} \rangle$ and $N_\mathrm{trk}^\mathrm{offline}$ is taken from Table 1 in Ref.~\cite{Chatrchyan:2013nka}. The shaded bands represent statistical uncertainty.}
  \label{fig1}
\end{figure*}
\begin{table*}[ht!]
  \centering
  \begin{tabular}{c|c|c|c|c|c|c}
  \hline \hline
  Collision system &  $\frac{dN^\mathrm{ch}}{d\eta} \big\vert_{\vert \eta \vert < 0.5}$  & $\langle p_T \rangle({\pi^+})$ (GeV) & $\langle p_T \rangle({K^+})$ (GeV) & $\langle p_T \rangle({p})$ (GeV) & $v_2^\mathrm{ch}\{2\}$ & $v_3^\mathrm{ch}\{2\}$  \\  \hline \hline
  0-5\% p+Au @ 200 GeV  & 11.8(1) & 0.52(1) & 0.72(2) & 0.98(3) & 0.037(1) & 0.0091(3) \\ \hline
  0-5\% d+Au @ 200 GeV  & 17.7(1) & 0.50(1) & 0.70(1) & 0.95(2) & 0.054(1) & 0.0114(4) \\ \hline
  0-5\% $^3$He+Au @ 200 GeV  & 22.9(1) & 0.49(1) & 0.69(1) & 0.93(2)& 0.059(1) & 0.0116(4) \\ 
  \hline \hline
  \end{tabular}
  \caption{Integrated observables of hadronic particle production and their anisotropic flow coefficients in (p, d, $^3$He)+Au collisions at 200 GeV. The statistical uncertainty on the last digit is indicated in parentheses.}
  \label{table1}
\end{table*}
%

We start our comparisons with integrated hadronic observables. Because it is difficult to estimate the number of particle tracks detected in the CMS experiments with our theoretical model, we use the conversion table (Table 1) in Ref.~\cite{Chatrchyan:2013nka} to map our calculations in different centrality bins to the experimental measurements. 
As shown in Fig.~\ref{fig1}, quantitative agreement is achieved between our hydrodynamic simulations, the measured identified particle averaged transverse momentum ($\langle p_T \rangle$) and  charged hadron anisotropic flow coefficients, $v_{2,3}\{2\}$ in top 20\% p+Pb collisions at 5.02 TeV. The proton $\langle p_T \rangle$ appears slightly underestimated but nevertheless falls within experimental uncertainties. The centrality dependence of these global observables is well reproduced by our approach. 

The effect of the hadronic afterburner is also highlighted in the same figures, with the solid lines including the effect of hadronic rescattering with UrQMD, and the dashed line including only hadronic decays, but no rescattering. The proton $\langle p_T \rangle$ is found to have a small but visible increase with the additional hadronic scattering, but the actual increase  is much smaller than in the case of Au+Au and Pb+Pb collisions \cite{Ryu:2015vwa}. The transport phase helps the small system to further develop a few percent of anisotropic flow. This effect is more pronounced in peripheral  bins, where the fireball lifetime is too short to convert all the system's spatial eccentricity into momentum anisotropy during the hydrodynamic stage. 

Predictions of the integrated hadronic observables in $0-5$\% (p, d, $^3$He)+Au collisions at 200 GeV are summarized in Table~\ref{table1} for future comparison. With respect to p+Pb collisions at 5.02 TeV, the identified particle mean $\langle p_T \rangle$ are about 20\% smaller at the top RHIC energy. This is because the system lifetime is about 20\% shorter compared to the collisions at LHC energy. This limits the development of radial flow during the hydrodynamic evolution. Moreover, the pressure gradients are also smaller at 200 GeV, which translates into a smaller expansion rate. These two factors yield  anisotropic flow coefficients about 50\% smaller in p+Au collisions at RHIC than in  p+Pb collisions at 5.02 TeV. The elliptic flow coefficients in d+Au and $^3$He+Au collisions, on the other hand, are comparable with the $v_2\{2\}$ in p+Pb collisions because of larger initial eccentricities in these systems. 

\begin{figure*}[ht!]
  \centering
  \includegraphics[width=0.95\linewidth]{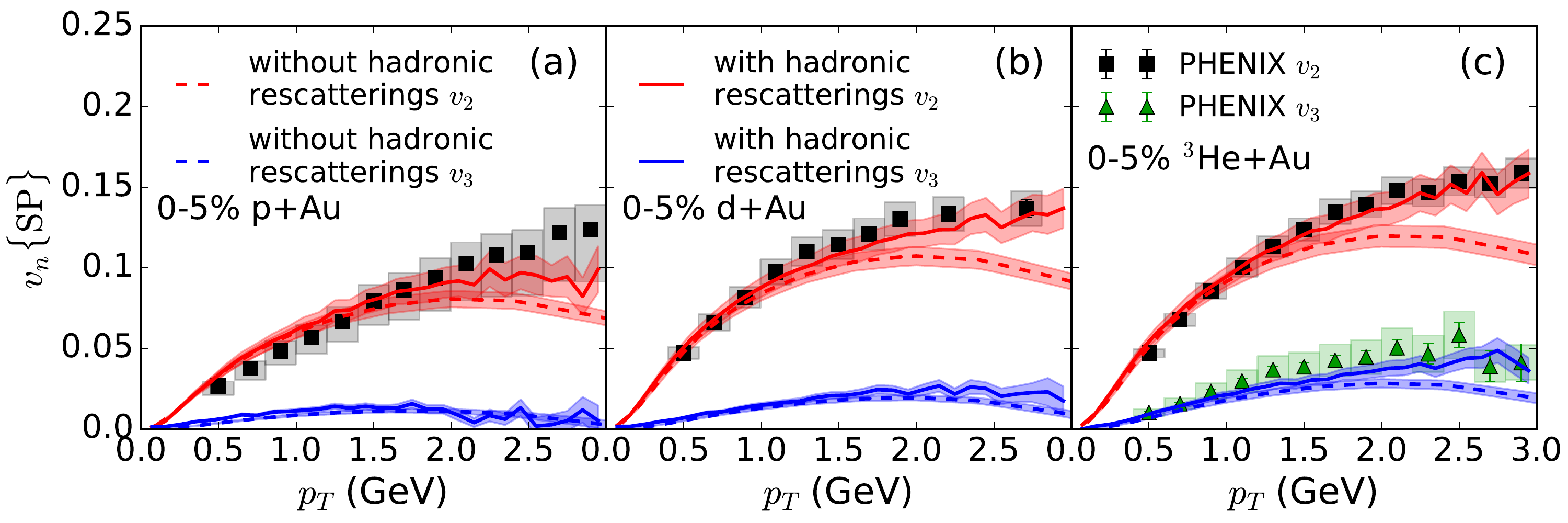}
  \caption{(Color online) Charged hadron $p_T$-differential anisotropic flow coefficients, $v_{2,3}\{\mathrm{SP}\}$, compared with PHENIX measurements for 0-5\% (p, d, $^3$He)+Au collisions at 200 GeV \cite{Adare:2014keg,Adare:2015ctn}. The  p + Au  elliptic flow preliminary data were shown at Quark Matter 2015 \cite{Nagle2015}. The shaded bands represent statistical uncertainty.}
  \label{fig2}
\end{figure*}
%
\begin{figure}[ht!]
  \centering
  \includegraphics[width=0.9\linewidth]{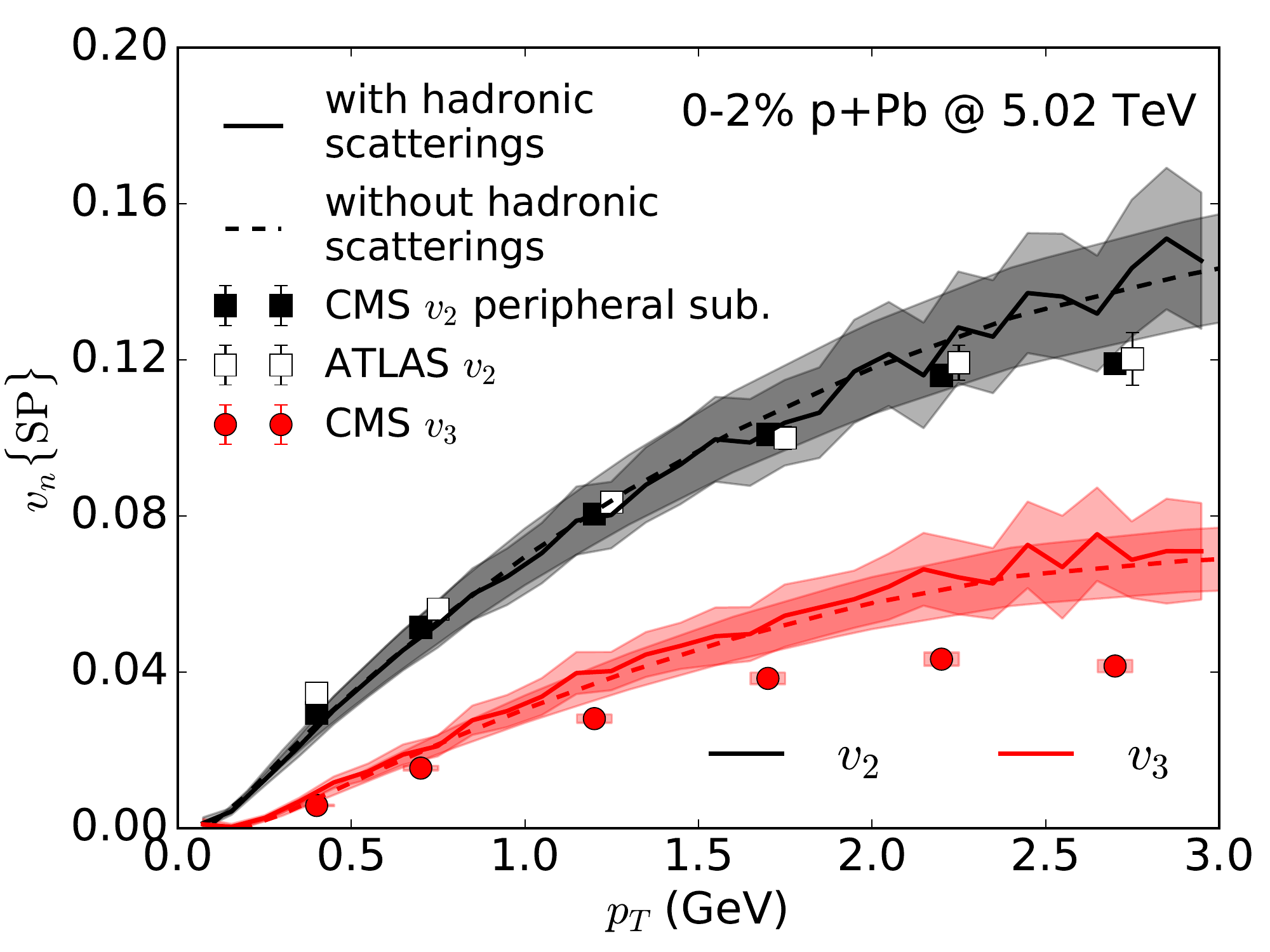}
  \caption{Charged hadron anisotropic flow $v_{2,3}\{\mathrm{SP}\}$ compared with the CMS \cite{Chatrchyan:2013nka} and the ATLAS measurements \cite{Aad:2013fja} in 0-2\% p+Pb collisions at 5.02 TeV. The shaded bands represent statistical uncertainty.}
  \label{fig3}
\end{figure}
%

Now, we take a closer look at $p_T$-differential observables. The charged hadron anisotropic flow coefficients, $v_{2,3}\{\mathrm{SP}\}(p_T)$, are compared with experimental measurements in Fig.~\ref{fig2} and Fig.~\ref{fig3} for small collision systems at RHIC and LHC energies, respectively. 
At the top RHIC energy, our hybrid approach with $\eta/s = 0.08$ for $T > 155$ MeV successfully provides a consistent description of the PHENIX anisotropic flow measurements in $0-5$\% p+Au, d+Au, and $^3$He+Au collisions. 
Hadronic rescattering from the transport phase is found to increases the high $p_T$ charged hadron $v_n\{\mathrm{SP}\}$ and improves the agreement with experimental data. Prediction of $p_T$-differential triangular flow $v_3\{\mathrm{SP}\}(p_T)$ in p+Au and d+Au collisions are shown for future comparison. 
In Fig.~\ref{fig3}, a same level of agreement for charged hadron $v_{2,3}\{\mathrm{SP}\}(p_T)$ is achieved in the top 2\% p+Pb collisions at 5.02 TeV with an effective $\eta/s = 0.10$ for $T > 155$ MeV. 
The effect of the hadronic cascade lessens as the collision energy is increased. This is because the larger pressure gradients at higher collision energy drive the system to develop hydrodynamic radial flow faster. Most of the spatial eccentricity has already been converted to momentum anisotropy before switching to hadronic transport. 

\begin{figure}[ht!]
  \centering
  \includegraphics[width=0.9\linewidth]{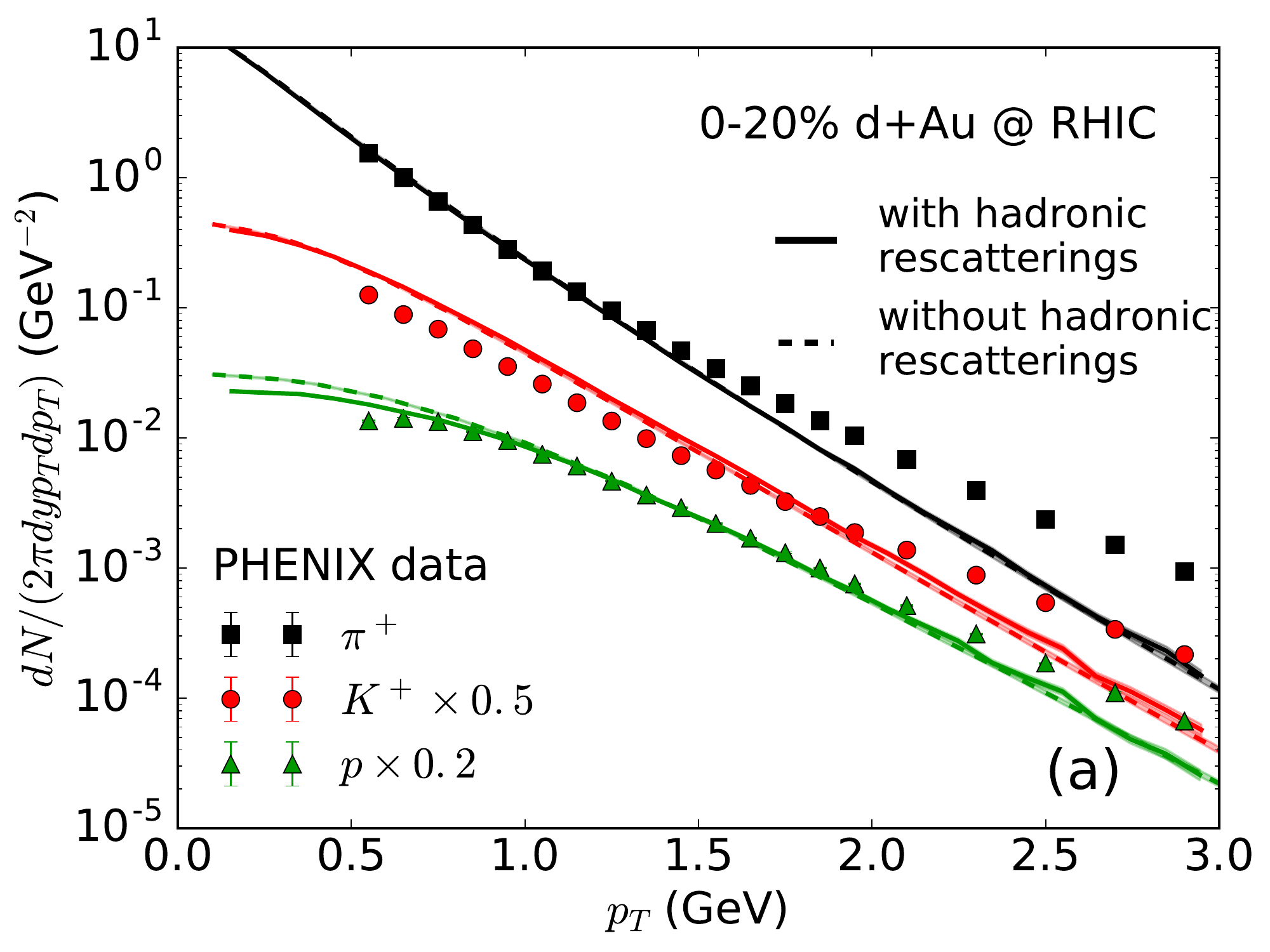} \\
  \includegraphics[width=0.9\linewidth]{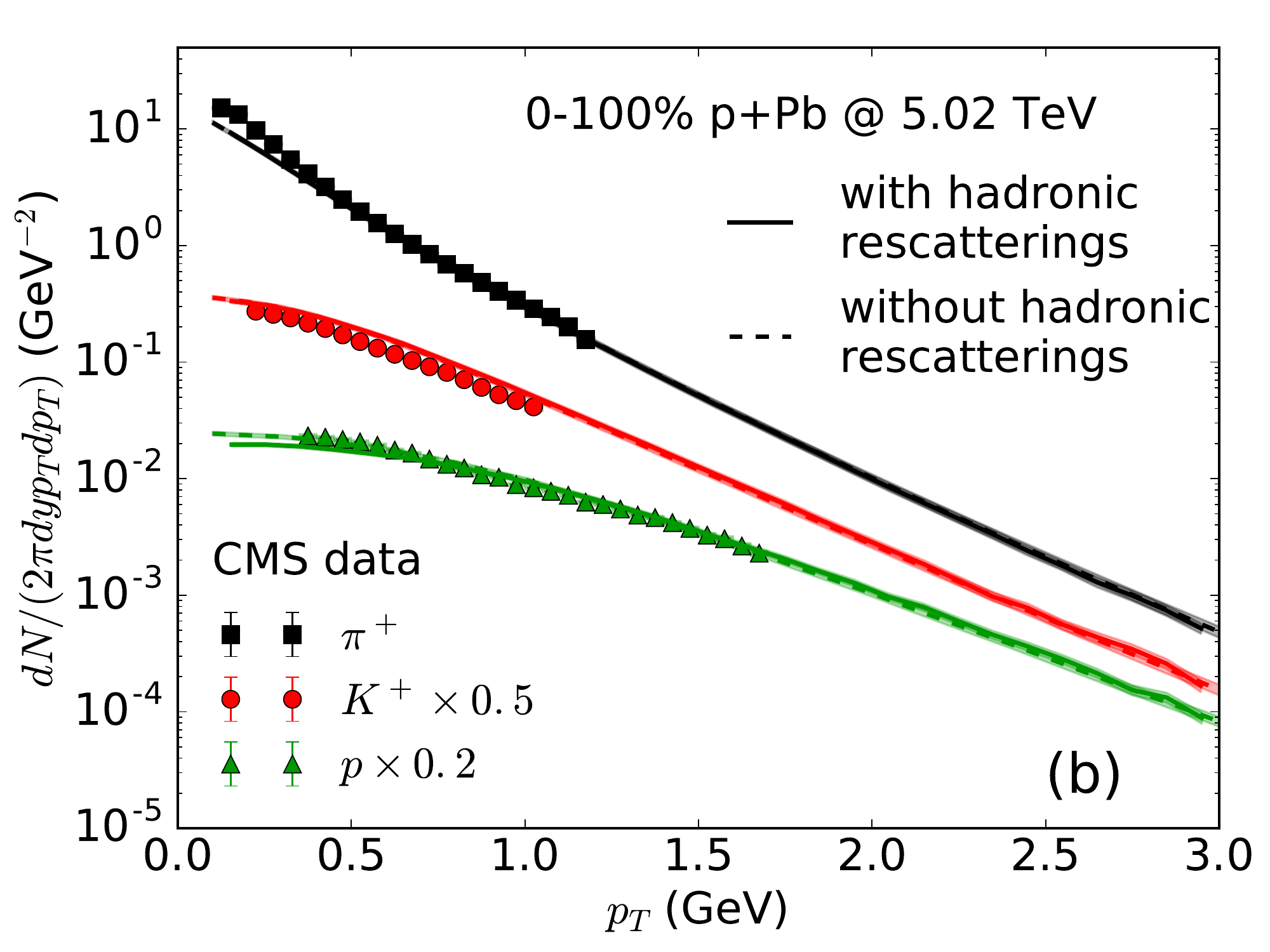}
  \caption{Identified particle spectra compared with experimental measurements in (a) 0-20\% d+Au collisions at 200 GeV \cite{Adare:2013esx} and (b) minimum bias p+Pb collisions at 5.02 TeV \cite{Chatrchyan:2013nka}. The shaded bands represent statistical uncertainty.}
  \label{fig4}
\end{figure}
%

In Fig.~\ref{fig4}, identified particle spectra are compared with experimental measurements for 0-20\% d+Au collisions at 200 GeV and minimum bias p+Pb collisions at 5.02 TeV. In 0-20\% d+Au collisions, our hybrid calculations provide a good description of the soft hadron spectra up to 1.5 GeV. 
Agreement with measurements at higher $p_T$ may require the contribution from recombination with jet shower partons, which is not considered here.

We note that the effect of hadronic rescattering, also shown in Fig.~\ref{fig4}, is very small for pions and kaons.
The proton spectrum at $p_T < 0.5$ GeV is reduced due to baryon anti-baryon annihilation in the transport phase. The fact that the high $p_T$ region of proton spectra remains the same suggests the hadronic rescatterings play a minor role in the dilute gas phase. Hence, the observed increase of the proton mean $\langle p_T \rangle$ observed earlier (Fig.~\ref{fig1}) owes  mainly to baryon anti-baryon annihilation.

\begin{figure}[ht!]
  \centering
  \includegraphics[width=1.0\linewidth]{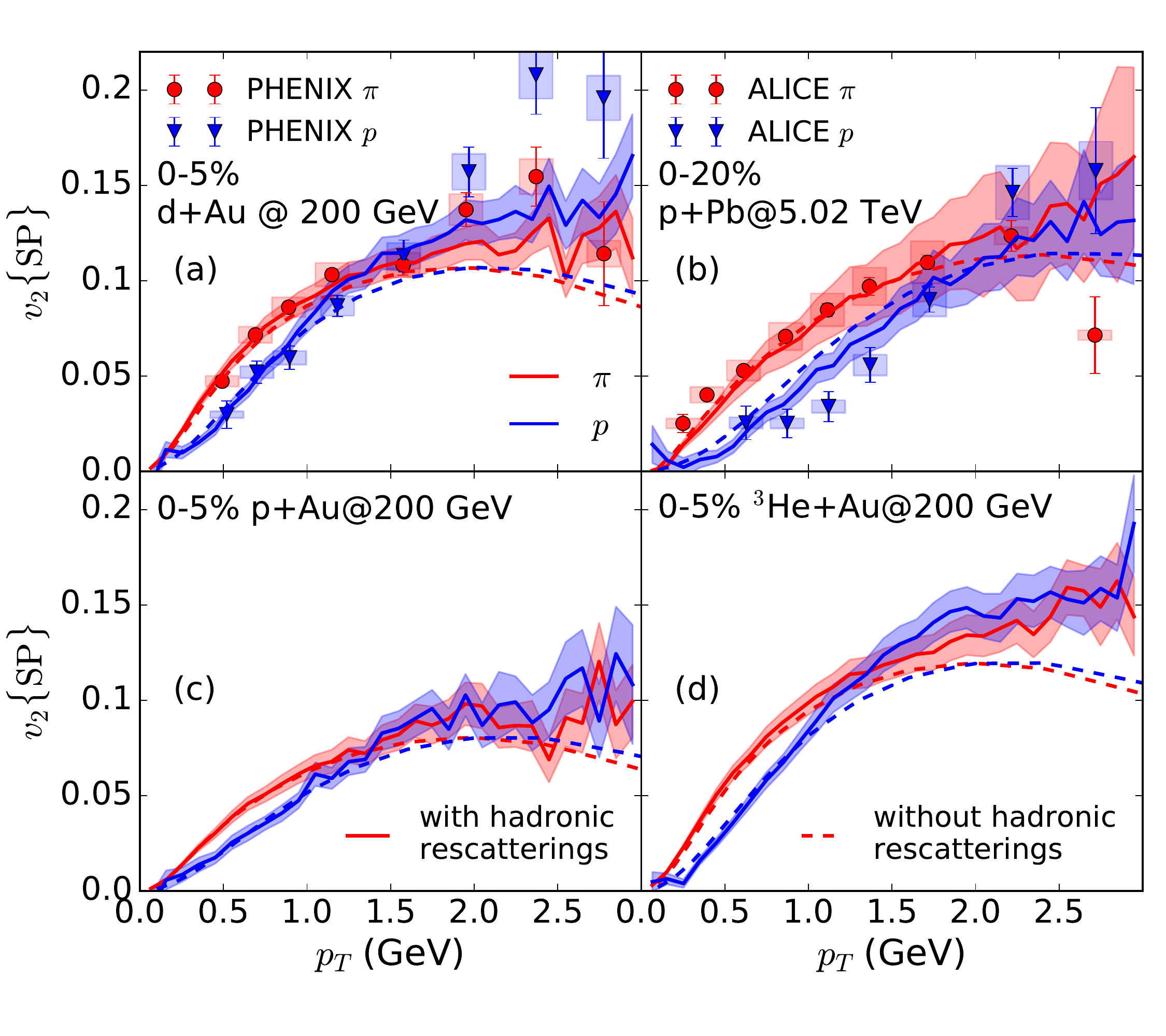}
  \caption{Identified pion and proton elliptic flow coefficients compared with the PHENIX \cite{Adare:2014keg} and ALICE measurements \cite{ABELEV:2013wsa} in 0-5\% d+Au collisions at 200 GeV (a) and 0-20\% p+Pb collisions at 5.02 TeV (b), respectively. Predictions of pion and proton $v_2\{\mathrm{SP}\}$ in top 5\% p+Au and $^3$He+Au collisions are shown in panels (c) and (d). The legends apply to all four panels. The shaded bands represent statistical uncertainty.}
  \label{fig5}
\end{figure}
%
The mass-ordering of hadronic flow coefficients has long been considered a hallmark of fluid-dynamical behavior \cite{Huovinen:2001cy}. Alternative interpretations have however recently appeared for asymmetric \cite{Zhou:2015iba,Schenke:2016lrs} and even symmetric \cite{Li:2016ubw} heavy-ion collisions. In the current work, mass ordering in the identified particle elliptic flow is investigated in Fig.~\ref{fig5}. The hydrodynamic model quantitatively produced the mass splitting between pion and proton $v_2\{\mathrm{SP}\}(p_T)$ measured in 0-5\% d+Au collisions at 200 GeV and 0-20\% p+Pb collisions at 5.02 TeV. Within hydrodynamic framework, the larger difference between pion and proton elliptic flow in p+Pb collisions at 5.02 TeV can be understood as the consequence of a stronger radial flow blue shifts the proton $v_2$ to high $p_T$ regions at higher collision energy. 
The effect of hadronic rescattering is found to be small, implying that that most of the mass splitting is developed in the hydrodynamic phase.
This suggests that a strongly-coupled QGP core in the small collision systems can be at the origin of the mass ordering in measured identified particle $v_2$. 
Predictions of pion and proton elliptic flow coefficients in p+Au and $^3$He+Au collisions are shown in Figs.~\ref{fig5}c and \ref{fig5}d for future comparison. The amount of mass splitting is found to be similar for the three systems studied, at the top RHIC energy. 

\section{Photon radiation}
\label{photon_section}

A hot and rapidly expanding QGP droplet radiates thermal photons. As shown in the previous section, the hydrodynamic medium has been well calibrated to reproduce various aspects of hadronic observables in the small collision systems.  In this section, the significance of thermal photon enhancement in the final measurable direct photon signal is addressed. 
More specifically, only thermal photon radiation from fluid cells whose temperature is higher than the switching temperature, $T_\mathrm{sw} = 155$\,MeV is considered. 
Since the number of hadrons and average number of collisions per hadrons is small in p+A collisions, photon emission in the late stage of the medium is not expected to be large.
An estimation of the contributions from temperature below $T_\mathrm{sw}$ will be presented in the next section.  

The leading-order QGP photon emission rate \cite{Arnold:2001ms} is used for temperature larger than 180 MeV,  and  hadronic photon production rates below. In the hot hadronic phase, photon production through meson-meson scattering \cite{Turbide:2003si}, from (the imaginary part of) many-body $\rho$-spectral function, from $\pi-\pi$ bremsstrahlung \cite{Rapp:1999ej,Rapp:1999qu,Liu:2007zzw,Heffernan:2014mla}, and from $\pi-\rho-\omega$ reaction channels \cite{Holt:2015cda} are considered. 
Shear viscous corrections to the 2 to 2 scattering processes in the QGP phase \cite{Shen:2014nfa} and to the meson-meson reactions in the hadronic phase \cite{Dion:2011pp, Shen:2014thesis} are included. Using the fact that the shear stress tensor - $\pi^{\mu\nu}$ - is symmetric, traceless, and orthogonal to the flow velocity, the photon emission rates can be written as \cite{Shen:2014nfa},
\begin{eqnarray}
&& \!\!\!\!\!\!\!\! E_q\frac{d \Gamma}{d^3 k}(E_q, T, \pi^{\mu\nu}) \notag \\ 
&& = \Gamma_0 (E_q, T) + \delta \Gamma (E_q, T, \pi^{\mu\nu}) \notag \\
&& = {\Gamma}_0(E_q, T) + \frac{\pi^{\mu\nu} \hat{q}_\mu \hat{q}_\nu}{2(e+P)} \chi\left(\frac{E_	q}{T}\right)  {\Gamma}_\pi (E_q, T),
\label{eq3}
\end{eqnarray}
where ${\Gamma}_0(E_q, T)$ and $\delta \Gamma (E_q, T)$ denote the equilibrium rate and first order shear viscous correction, respectively.
Finally, decay photons from short-lived resonances that can not be subtracted in the experimental cocktail background are included when computing the direct photon observables \cite{Rapp:1999qu,vanHees:2014ida, Shen:2014thesis,Paquet:2015lta}. A detailed list and a discussion of the non-cocktail decay channels that produce photons can be found in Chapter 21 of Ref.~\cite{Shen:2014thesis}\footnote{Compared to the list in Chapter 21 of Ref.~\cite{Shen:2014thesis}, we exclude photons from the decay channel $\rho_0 \rightarrow \pi^+ + \pi^- + \gamma$ in the short-lived resonance contribution. Photons from this channel are added in the decay photon cocktail.}.

    \subsection{Prompt photons}

\label{sec:prompt}

\begin{table}[tb]
  \centering
  \begin{tabular}{c|c|c|c}
  \hline \hline
  Collision system & Centrality & $N_\mathrm{coll}$ & $N_\mathrm{part}$  \\  \hline \hline
  p+Pb @ 5.02 TeV & 0-1\%     &  15.92(4) & 16.92(4)  \\ \hline
  			      & 0-5\%     &  14.47(2) & 15.47(2)  \\ \hline
                               & 0-20\%   &  12.51(1) &  13.51(1)  \\ \hline
                               & 0-100\% & 6.50(1)  &  7.50(1) \\ \hline \hline
  p+Au @ 200 GeV & 0-5\%   &  10.28(1) & 11.28(1)  \\ \hline
    			      & 0-20\%  &  8.62(1) & 9.62(1)  \\ \hline
    			      &0-100\% &  4.66(1) & 5.66(1)  \\ \hline \hline
  d+Au @ 200 GeV & 0-5\%   &  18.48(2) & 18.19(2)  \\ \hline
    			      & 0-20\%  &  15.75(2) & 15.40(1) \\ \hline
    			      & 0-100\% &  7.90(2) & 8.31(1) \\ \hline \hline
  $^3$He+Au @ 200 GeV & 0-5\%   &  26.48(2) & 25.35(2) \\ \hline
       				        & 0-20\%  &  22.67(1) & 21.59(1) \\ \hline
    				        & 0-100\% &  10.59(1) & 10.88(1) \\  \hline \hline
  \end{tabular}
	\caption{The averaged number of binary collisions $N_\mathrm{coll}$ and participant nucleons $N_\mathrm{part}$ in p+Pb collisions at 5.02 TeV and (p, d, $^3$He)+Au collisions at 200 GeV. Statistical uncertainties of the last digits are in parentheses.}
  \label{table2}
\end{table}

The photons produced by the very first nucleon-nucleon collisions are the prompt photons. These are evaluated with perturbative QCD at next-to-leading order in $\alpha_s$~\cite{Aurenche:1987fs,Aversa:1988vb,incnlo}, as in previous work~\cite{Paquet:2015lta,Paquet:2015Thesis}. The isospin effect is included to account for the different proton-to-neutron ratio of each colliding ion. Cold nuclear effects are taken into account with the nCTEQ15 nuclear parton distribution functions~\cite{Kovarik:2015cma}. 
The perturbative calculation of photon production is scaled up from the nucleon-nucleon result, by the number of binary collisions $N_\mathrm{coll}$ which is summarized in Table~\ref{table2} for the different  systems investigated in this work.

We note that the nCTEQ15 nuclear parton distribution functions were constrained using nuclear deep-inelastic scattering ($e A \to e + X$) and nuclear Drell-Yan ($p A \to l^+ l^- + X$). The default parametrization of nCTEQ15 was also constrained with pion production in d+Au collisions, although a separate parametrization that did not use these pion measurements was also made available. It is this second version of nCTEQ15 --- referred to by the authors as nCTEQ15-np --- that is used in this work. The rationale for this choice is that constraining nuclear parton distribution functions with hadronic measurements from small systems makes the explicit assumption that no QGP is formed, i.e. that no significant energy loss or thermal hadron production occurs in such collisions. Since the opposite assumption is made in the present work, it would not be consistent to use nuclear distribution functions partly constrained by hadronic measurements. This is also the reason why the widely used EPS09 nuclear parton distribution functions~\cite{Eskola:2009uj} are not used in this work\footnote{Since hadronic measurements represent only a small fraction of the measurements used to constrain nuclear parton distribution functions, it may appear that the use of hadronic data can only result in a small contamination of the extracted functions. It is important to note, however, that measurements are not necessarily given the same weight when constraining distribution functions. For example, hadronic measurements are given a large weight in EPS09 to provide better constraints on the gluon distribution, increasing the influence of this measurement on the nuclear parton distribution functions.}.

\begin{figure*}[ht!]
  \centering
  \centering
  \begin{tabular}{cc}
  \includegraphics[width=0.4\linewidth]{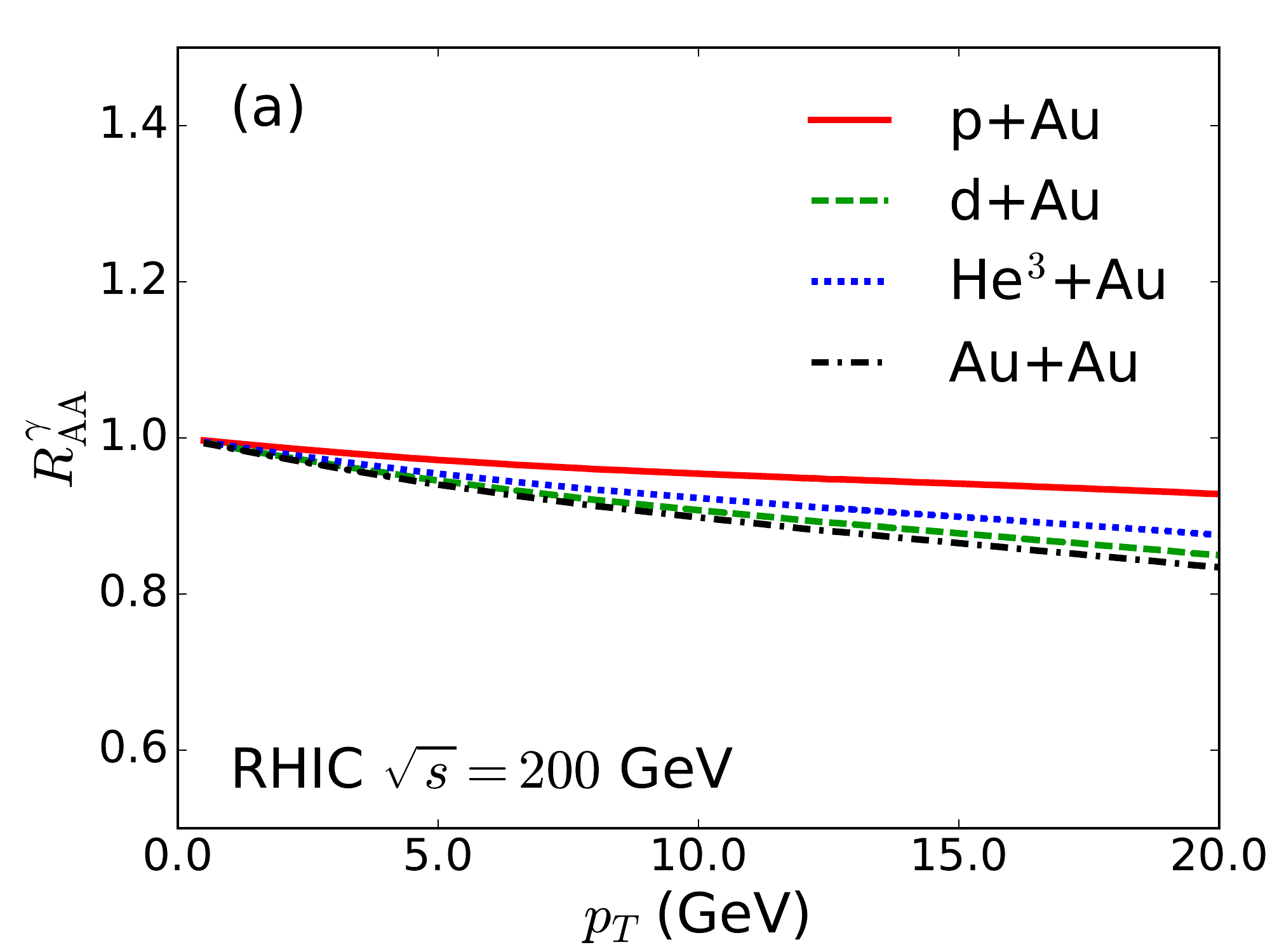} &
  \includegraphics[width=0.4\linewidth]{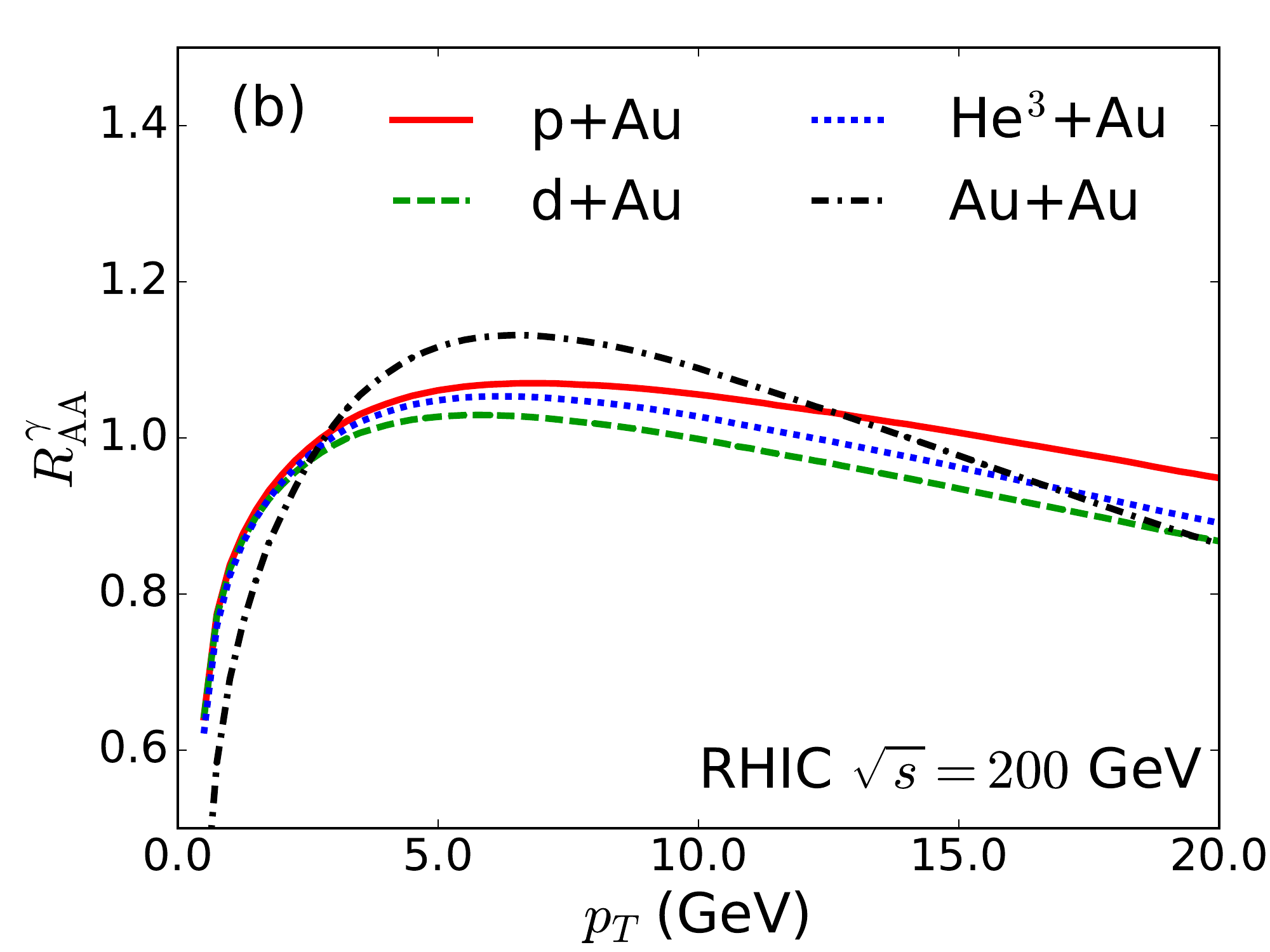}
  \end{tabular}
  \caption{Nuclear modification factor of prompt photons at RHIC $\sqrt{s_{NN}}=200$~GeV (a) with isospin corrections only and (b) with both isospin and nCTEQ15 nuclear parton distribution functions corrections.}
  \label{prompt_photon_RAA}
\end{figure*}
%

The isospin effect is only significant at high $p_T$, which is shown in Figure~\ref{prompt_photon_RAA}a by plotting $R_{\textrm{pAu}}$, $R_{\textrm{dAu}}$, $R_{\textrm{He$^3$Au}}$ and $R_{\textrm{AuAu}}$ at RHIC ($\sqrt{s_{NN}}=200$~GeV). The introduction of cold nuclear effects with nCTEQ15 differentiates $R_{\textrm{AuAu}}$ from that of the small systems (p+Au, d+Au, He$^3$+Au), as seen in Figure~\ref{prompt_photon_RAA}b. 

As investigated in Refs.~\cite{Arleo:2011gc} and \cite{Klasen:2013mga}, uncertainties on prompt photon production in nuclear collisions originate  from the factorization/renormalization/fragmentation scale dependence, photon fragmentation functions and nuclear parton distribution functions. The questionable  reliability of perturbative QCD at low $p_T$  further increases the uncertainty in this region of momentum. 
To constrain the scale dependence of the calculation, the factorization, renormalization and fragmentation scales are taken to be proportional to the photon transverse momentum, and the proportionality constant is fixed~\cite{Paquet:2015lta,Paquet:2015Thesis} using proton-proton measurements. 

The photon fragmentation function BFG-II~\cite{Bourhis:1997yu} is used. It appears to be slightly better than other fragmentation functions at describing low momentum photon measurements in proton-proton collisions~\cite{Klasen:2014xfa}. 

The nuclear distribution functions nCTEQ15 themselves have uncertainties~\cite{Kovarik:2015cma} reflecting the limited constraining power of available nuclear data. They can be used to provide an uncertainty band on prompt photon predictions. This uncertainty was studied in Ref.~\cite{Arleo:2011gc} for different nuclear parton distribution functions, EPS09. Given that the nuclear distribution function uncertainties of EPS09 and nCTEQ15 are of the same order, this previous work can be used as a guide for the size of the uncertainties due to the nuclear parton distribution functions. Thus, uncertainties from nCTEQ15 of order 10\% are expected for the nuclear modification factor of prompt photons. 

Combining all the above uncertainties in prompt photons, along with possible final state effects on prompt photon production (e.g. the effect of parton energy loss on fragmentation photon production), it is clear that work remains to be done to determine the precise contribution of this source of photons in small collision systems. Nevertheless, we believe that the calculation presented above provides a sufficiently good determination of prompt photons to establish if thermal photons can be observed above the prompt photon background.

\begin{figure*}[ht!]
  \centering
  \centering
  \begin{tabular}{cc}
  \includegraphics[width=0.45\linewidth]{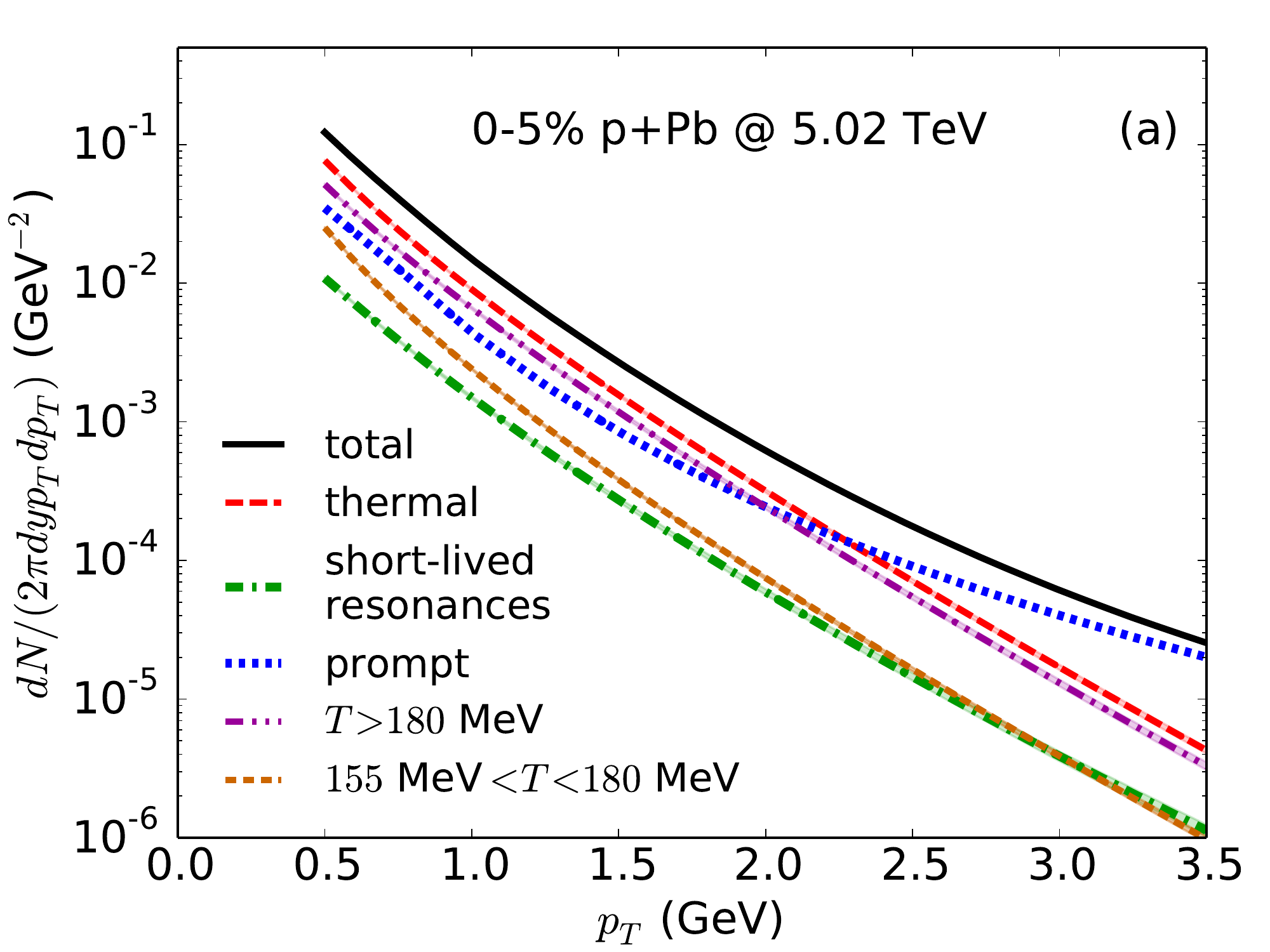} &
  \includegraphics[width=0.45\linewidth]{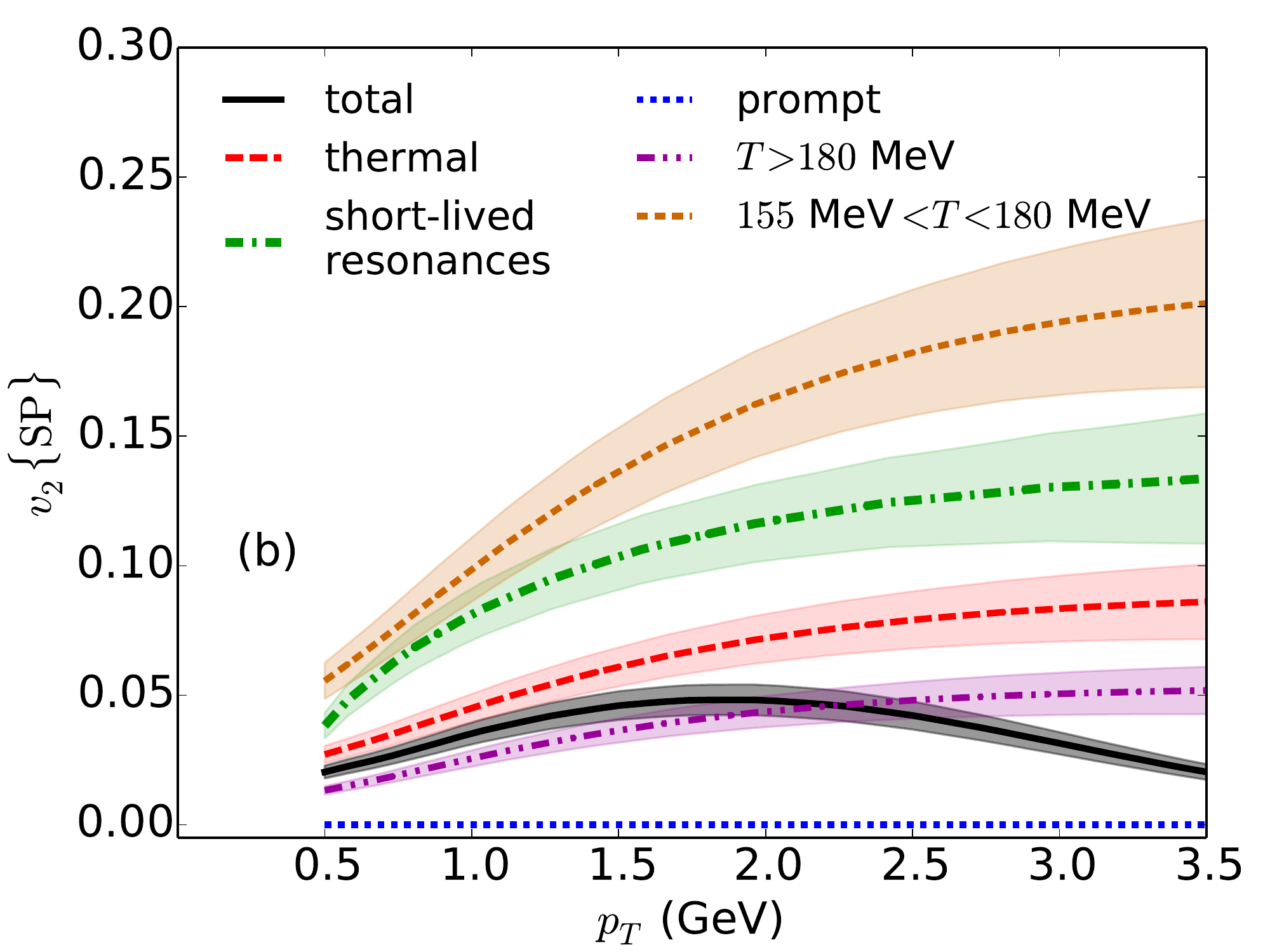}
  \end{tabular}
  \caption{Direct photon spectra (a) and $p_T$-differential elliptic flow coefficients (b) in 0-5\% p+Pb collisions at 5.02 TeV. Contributions from individual channels are shown. Thermal photons from the QGP phase are denoted as $T > 180$\,MeV, and those from the hot hadronic phase are represented by the curve labelled $155$ MeV$< T < 180$ MeV. The ``thermal;'' curve is the sum of those last two sources. The shaded bands represent statistical uncertainty. }
  \label{fig6}
\end{figure*}
%

\subsection{Direct photon spectra and $v_n$}

Fig.~\ref{fig6} presents the direct photon spectra and elliptic flow coefficients in 0-5\% most central p+Pb collisions at 5.02 TeV. Contributions from individual production channels are shown. 
We find that the thermal sources are significant in the total direct photon signal for $p_T < 2$ GeV. Thermal radiation 
represents about 1.6 times that of the prompt contribution. Decay photons coming from the short-lived resonances contribute only about 10\% to the direct photon yield. 
In the thermal photon signal, almost 80\% of the photons come from $T > 180$ MeV region:
 indeed radiation from the QCD phase diagram at temperatures above that of the crossover is observed. 
 Contribution of thermal photons from spacetime regions with $T<180$ MeV
is as small as decay photons coming from the short-lived resonances. The relative importance of photon originating from the low and high temperature region of the plasma is an important feature of the calculation that we discuss in more details in the next section.

In Fig.~\ref{fig6}b, the net $p_T$-differential elliptic flow coefficient of direct photons is shown together with the contributions from individual channels. Prompt photons are assumed to carry zero anisotropy. The direct photon elliptic flow in $0-5$\% p+Pb collisions reaches its maximum $\sim$ 0.05 at $p_T \sim 1.7$ GeV. The strength of this signal is comparable to that of the direct photon $v_2$ measured in $0-40$\% Pb+Pb collisions \cite{Lohner:2012ct}. The rise and fall in the direct photon elliptic flow reflects the competition between thermal and prompt sources at different $p_T$ \cite{Shen:2013cca}. Although the elliptic flow of hadronic photons and decay photons from short-lived resonances are large, their contribution to the total direct photon elliptic flow is limited, owing to their small yield as shown in Fig.~\ref{fig6}a. 

\begin{figure}[ht!]
  \centering
  \includegraphics[width=0.9\linewidth]{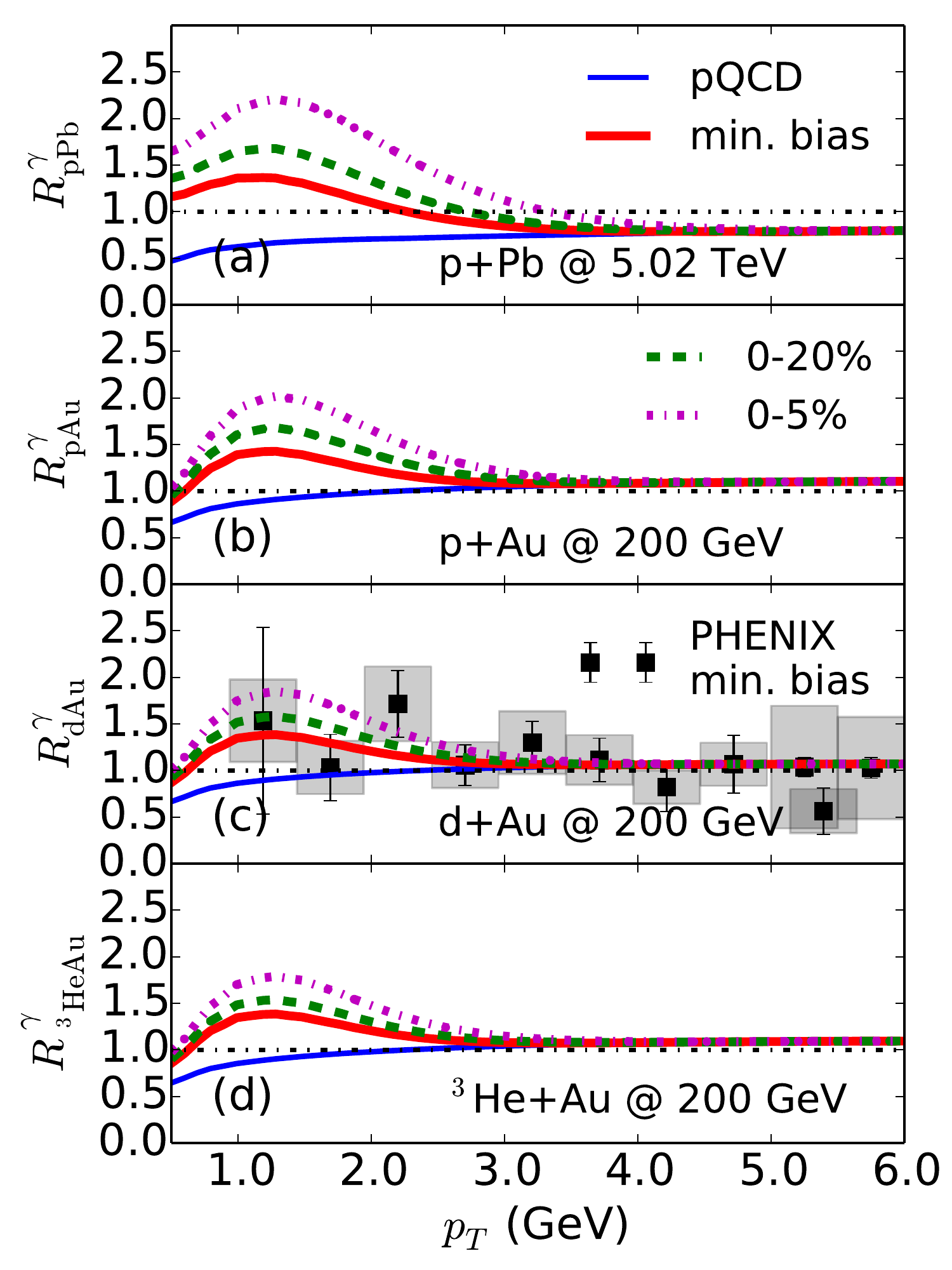}
  \caption{(Color online) {\it Panels (a-d):} Thermal photon enhancement in 0-5\%, 0-20\%, and minimum bias p+Pb  collisions at 5.02 TeV and (p, d, $^3$He  collisions at 200 GeV. In Panel (c), the PHENIX measurement is from Ref. \cite{Adare:2012vn}.
  }
  \label{fig7}
\end{figure}
%

We summarize the thermal photon enhancement in all four small collision systems as the nuclear modification factor $R_{pA}^\gamma$. From panels (a) to (d), one sees that collision systems at RHIC and the LHC exhibit qualitatively similar behaviour. We find sizeable enhancements in the direct photon signals from thermal radiation with respect to the prompt production for $p_T < 3$ GeV. The prompt photon $R_{pA}^\gamma$ is below 1 in the low $p_T$ region because of nuclear shadowing effects (see Section~\ref{sec:prompt}). 
The  thermal enhancement over this baseline is the largest in p+Pb collisions at 5.02 TeV. At the top RHIC energy, we find that direct photons $R_\mathrm{pAu}^\gamma > R_\mathrm{dAu}^\gamma > R_{^{3}\mathrm{HeAu}}^\gamma$. This is because the number of binary collisions from p+Au to $^{3}$He+Au collisions (see Table \ref{table2}) increases more rapidly than the thermal radiation; this makes for a larger relative weight of the prompt photon production. As a function of centrality, direct photon $R_{pA}^\gamma$ is larger in more central collisions. In $0-5$\% most central collisions, direct photon $R_{pA}^\gamma$ are predicted to reach up to 2 at $p_T \sim 1.2$ GeV for all four systems. Although this thermal enhancement is smaller than what it is in Au+Au or Pb+Pb collisions \cite{Adare:2008ab,Adam:2015lda,Shen:2016odt}, it still can serve as a precious signature of the hot medium in small collision systems. In Fig.~\ref{fig7}c, our result in minimum bias d+Au collisions is compared to the available PHENIX measurement \cite{Adare:2012vn}. The current experimental uncertainty does not permit a distinction between the  scenarios with, and without the thermal signal. 

\begin{figure}[ht!]
  \centering
  \includegraphics[width=0.9\linewidth]{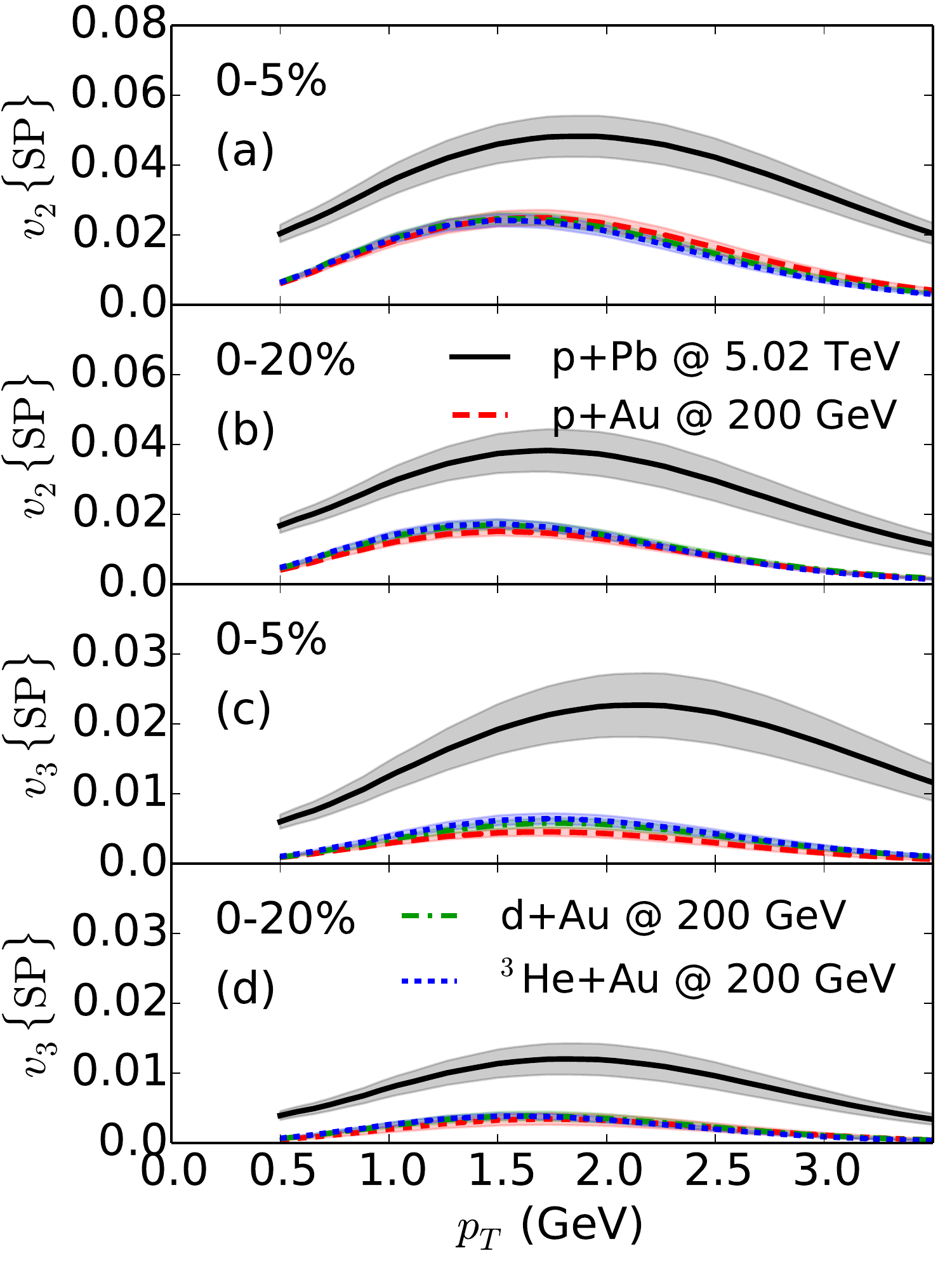}
  \caption{(Color online) {\it Panels (a-d):} Direct photon $v_{2,3}$ in 0-5\% and 0-20\% centralities of the small collision systems. The shaded bands represent statistical uncertainty.}
  \label{fig7b}
\end{figure}
%

The direct photon anisotropic flow coefficients $v_{2,3}\{\mathrm{SP}\}$ are shown in Figs.~\ref{fig7b} for all four systems investigated.
Direct photons from p+Pb collisions at 5.02 TeV carry the largest anisotropic flow, owing to the combined effects of higher temperatures achieved and larger pressure gradients. The direct photon anisotropic coefficients at top RHIC energy are similar, for p+Au, d+Au, and $^3$He+Au collisions despite the initial eccentricities $\varepsilon_n$ being quite different \cite{Nagle:2013lja}. This similarity in the momentum anisotropy of these three small systems was observed previously in hadrons (see Fig.~\ref{fig5}), and the considerable damping effect of non-flowing prompt photons makes it even harder to distinguish differences in the direct photon $v_n$ of these systems.

\subsection{Centrality, system size and center-of-mass energy dependence}

\begin{figure}[ht!]
  \centering
  \includegraphics[width=0.95\linewidth]{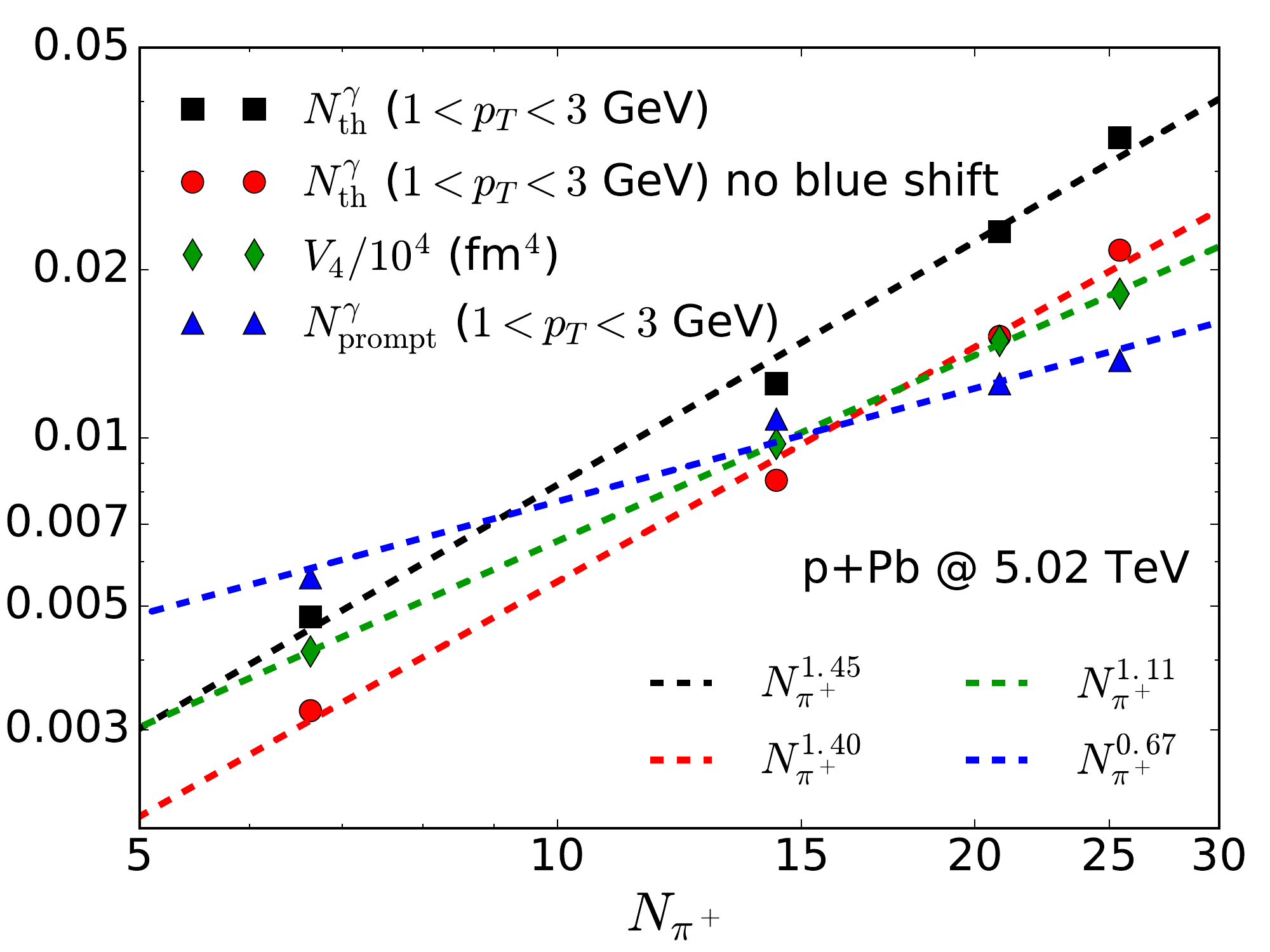}
  \caption{(Color online) The centrality dependence of $p_T$-integrated thermal and prompt photon yields compared with charged pion multiplicity $N_{\pi^+}$ in p+Pb collisions at 5.02 TeV in the mid-rapidity region, $\vert y \vert < 0.5$. The meaning of the ordinate is specified in the legend.}
  \label{fig7.1}
\end{figure}
%

In this section, we take a closer look at the centrality, system size and center-of-mass energy dependence of photon production in relativistic nuclear collisions. All three of these parameters can be mapped reasonably well to the pion multiplicity $N_{\pi^+}$: more central collision, larger system size and higher center-of-mass energy all translate into larger pion production. The pion multiplicity is thus used as proxy for these quantities. To facilitate comparisons of different systems, the photon spectra is integrated in transverse momentum to obtain the photon multiplicity, which we evaluate in this section with the cuts $1<p_T^\gamma<3$~GeV.

As observed in Fig.~\ref{fig7}, the thermal photon enhancement is more pronounced in central collisions than at minimum bias.
This can also be seen in Fig.~\ref{fig7.1}, which shows the centrality dependence of the thermal and prompt photon multiplicities in p+Pb collisions at 5.02 TeV: the thermal photon multiplicity increases approximately with $N_{\pi^+}^{1.45}$, while the prompt photon multiplicity goes much more slowly as $N^\gamma_\mathrm{prompt} \propto N_{\pi^+}^{0.67}$.
The increase of thermal photons can be understood as a combination of changes in the systems' space-time volume, average temperature and blueshift effect. First note that the thermal photon multiplicity shown in Fig.~\ref{fig7.1} is not independent of the blueshift, because of the limited integration range used in $p_T^\gamma$ ($1<p_T^\gamma<3$~GeV). This effect can be quantified by evaluating the photon multiplicity without including blueshift, shown in Fig.~\ref{fig7.1}. This results in a slightly smaller slope of $N_{\pi^+}^{1.40}$, implying that the contribution of the blueshift to the centrality dependence of the thermal photon multiplicity is small. 
The growth of the spacetime volume $V_4$ as a function of centrality is slower compared to $N^\gamma_\mathrm{th}$, with a scaling exponent 1.11 shown in Fig.~\ref{fig7.1}. 
The remaining difference between $N_{\pi^+}^{1.40}$ and $N_{\pi^+}^{1.11}$ can be attributed to the increase of systems' average temperature, which is thus the dominant factor in the centrality dependence of thermal photons in small systems (as verified by a direct calculation). It is found that thermal and prompt photon production in small systems at RHIC show a similar centrality dependence as found above for p+Pb collisions at 5.02 TeV.

\begin{figure}[ht!]
  \centering
  \includegraphics[width=0.95\linewidth]{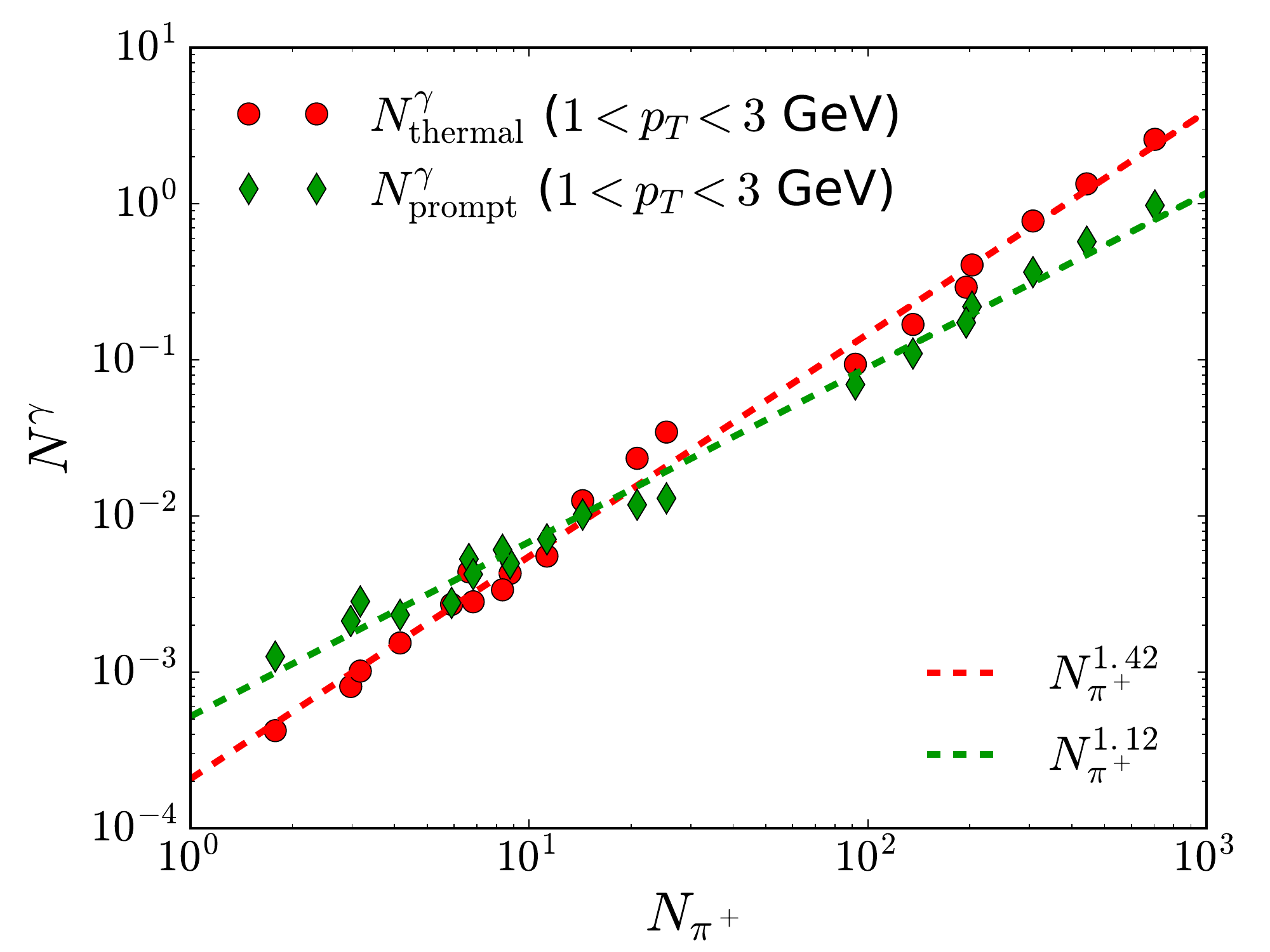}
  \caption{(Color online) Multiplicity of thermal and prompt photons ($1<p_T^\gamma<3$~GeV) as a function of the multiplicity of pions for all collision systems and centralities investigated in this work, along with the results from Au+Au collisions at RHIC and Pb+Pb collisions at the LHC from Ref.~\cite{Paquet:2015lta}}
  \label{mult_photons_pions}
\end{figure}
%

A slightly different picture emerges when investigating the overall centrality, system size and center-of-mass energy dependence of all systems studied in this work, along with calculation for Au+Au collisions at RHIC ($\sqrt{s_\mathrm{NN}}=200$~GeV) and Pb+Pb collisions at the LHC ($\sqrt{s_\mathrm{NN}}=2760$~GeV)~\cite{Paquet:2015lta}. Figure~\ref{mult_photons_pions} provides an overview for all these systems. Thermal photons are found to grow as $N_{\pi^+}^{1.42}$, which is very similar to the power found above for the centrality dependence at a fixed center-of-mass energy (see Fig.~\ref{fig7.1}). The prompt photon multiplicity is found to go as $N_{\pi^+}^{1.12}$, which is slower than thermal photons but significantly larger than the $N^\gamma_\mathrm{prompt} \propto N_{\pi^+}^{0.67}$ observed in Fig.~\ref{fig7.1} for the centrality dependence in p+Pb collisions. This difference in the scaling of prompt photons is not expected: at a fixed center-of-mass energy, the centrality and system size dependence of prompt photons originates mainly from the number of binary collisions that multiply the perturbative QCD calculation of prompt photons (see Section~\ref{sec:prompt}). On the other hand, changes in the center-of-mass energy of the collisions have a significant effect on both the number of binary collisions and the perturbative QCD calculation. The larger exponent found in Fig.~\ref{mult_photons_pions} reflects more closely the combination of these two effects. 

Note that the pion multiplicity is also a good proxy for the number of decay photons being produced in a collision, since $\pi^0\to\gamma \gamma$ is the dominant source of decay photons in heavy ion collisions. The observation in Figure~\ref{mult_photons_pions} that the thermal and prompt photon multiplicities grow faster than the pion multiplicity
summarizes the fact that the direct photon signal grows faster than the decay photon background as the collision energy and system size increases, making its measurement easier. Inversely, direct photons tend to be increasingly difficult to measure in small system. This subject is addressed in the next section.

\subsection{Inclusive and decay photons}

In order to estimate the difficulty in measuring the proposed direct photon observables, we investigate the ``photon decay cocktail'' in the small  systems, at top RHIC and LHC energies. For the cocktail content we include decays from $\pi^0$, $\eta$, $\omega$, $\eta^\prime$, $\phi$, $\Sigma_0$, and $\rho_0$: the same species included in the ALICE direct photon analysis in Pb+Pb collisions \cite{LohnerThesis, Adam:2015lda}. These particles contribute over 99\% of the decay photons in the calculation. The remaining less than 1\% of the decay photons from other short-lived resonances are included in the direct photon signal.  

\begin{figure*}[ht!]
  \centering
  \centering
  \begin{tabular}{cc}
  \includegraphics[width=0.35\linewidth]{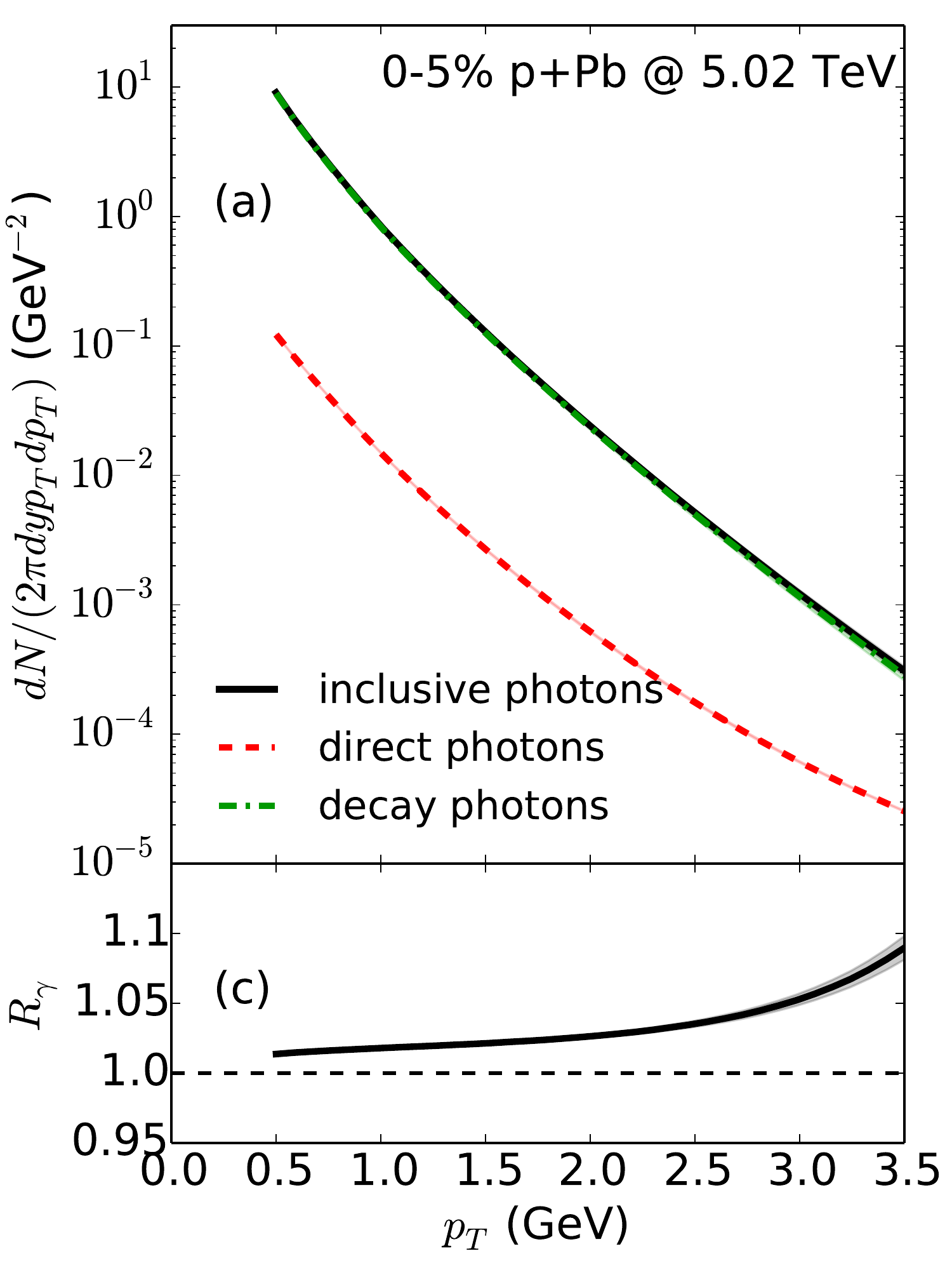} &
  \includegraphics[width=0.35\linewidth]{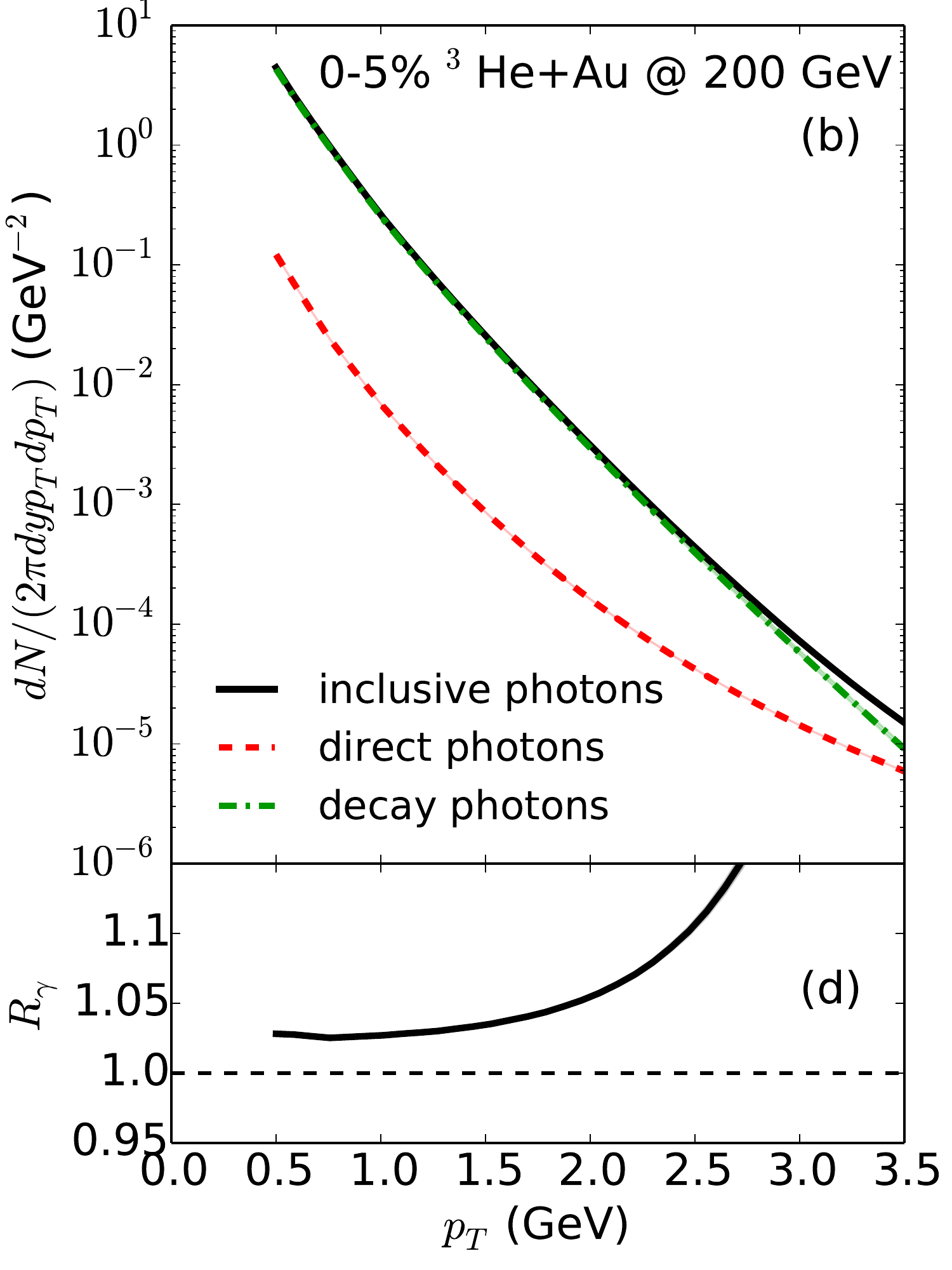}
  \end{tabular}
  \caption{Inclusive, decay, and direct photon spectra in 0-5\% p+Pb collisions at 5.02 TeV (a) and in 0-5\% $^3$He+Au collisions at 200 GeV (b). The corresponding ratio of inclusive over decay photons are shown in panels (c) and (d). }
  \label{fig8}
\end{figure*}

In Figs.~\ref{fig8}, we show the direct, decay, and inclusive (direct + decay) photon spectra in $0-5$\% p+Pb collisions at 5.02 TeV, and $^3$He+Au collisions at 200 GeV. Because the $\pi^0$ production in p+Pb collisions scales faster than number of binary collisions (see Figs.~\ref{fig7.1} and \ref{mult_photons_pions}), the signal to background (direct/decay photons) ratio is worse than the one in Au+Au or Pb+Pb collisions. This translates in a big gap between the direct and decay photon spectra at low $p_T$ in Figs.~\ref{fig8}(a) and (b). In panels (c) and (d), the ratio of inclusive to decay photons, $R_\gamma$, is  shown as a function of $p_T$. At 5.02 TeV, the value of $R_\gamma$ is about 1.02$ - $1.03 for $p_T < 2$ GeV and increases up to $\sim 1.05$ at $p_T = 3$ GeV. $R_\gamma$ is larger at the top RHIC energy. In 0$-$5\% $^3$He+Au collisions, $R_\gamma$ is about $4-5$\% higher than unity for $p_T < 2$ GeV. It increases rapidly for $p_T > 2$ GeV, where the prompt photon signal becomes more and more dominant. The larger values of $R_\gamma$ in $^3$He+Au collisions compared to those in p+Pb collisions can be understood as a consequence of the faster growth of $N_\mathrm{coll}$ as a function of centrality in the less asymmetric collision systems. The prompt photon yield increases relatively faster in $^3$He+Au collisions than in p+Pb collisions. The faster increase of $R_\gamma$ as a function of $p_T$ in $^3$He+Au collisions is because a weaker hydrodynamic radial flow is generated at top RHIC energy. The hadron spectra who produce decay photons are less blueshifted at lower collision energy. The direct photon signal for $p_T > 2$ GeV shines out more easily at RHIC than at the LHC. The $R_\gamma$ values in the small collision systems are about factor of $4\sim5$ smaller compared to those measured in Au+Au and Pb+Pb collisions \cite{Adare:2014fwh,Adam:2015lda}. This indicates that direct photon measurements in small collision systems are challenging, and that measurements at 200 GeV should be somewhat easier than at 5.02 TeV. 

\subsection{Thermal photons in pp collisions}

\begin{figure*}[ht!]
  \centering
  \centering
  \begin{tabular}{cc}
  \includegraphics[width=0.4\linewidth]{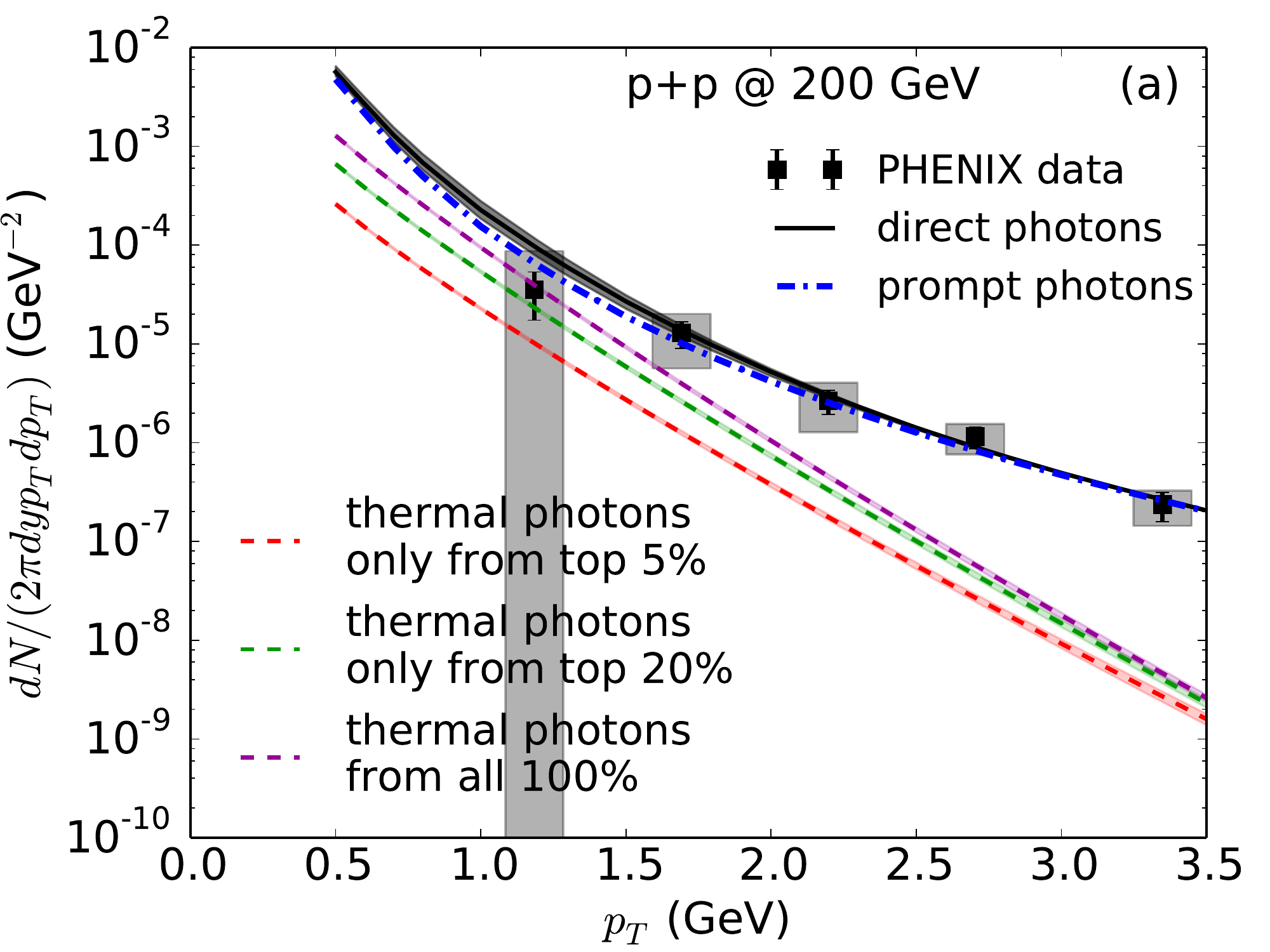} &
  \includegraphics[width=0.4\linewidth]{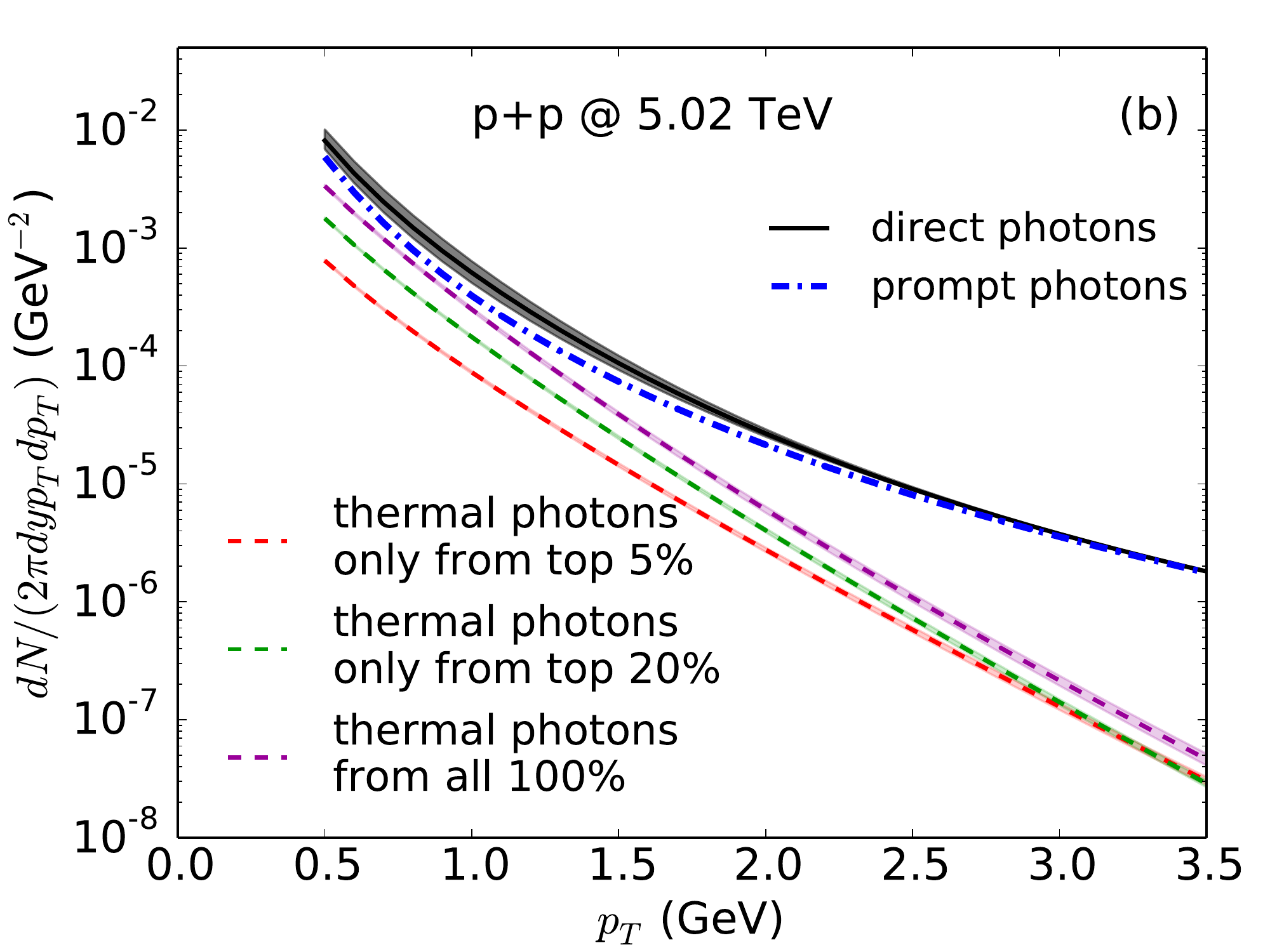}
  \end{tabular}
  \caption{(Color online) Direct photon spectra in minimum bias pp collisions at 200 GeV and 5.02 TeV. The shaded boxes around the data (left panel) \cite{Adare:2009qk} represent systematic uncertainties.  The band in the right panel illustrates the effect on considering the different centrality classes discussed in the text.}
  \label{fig8.1}
\end{figure*}
%

Collective flow signatures were also found in high multiplicity pp collisions at LHC energies \cite{Khachatryan:2016txc,Aad:2015gqa}. We estimate the amount of thermal photon radiation in pp collisions at 200 GeV and 5.02 TeV in Figs.~\ref{fig8.1}. Since the size of the medium created in the pp collisions is extremely small, systems may not achieve quasi-thermal equilibrium in every collision event. Hence, we explore the following three scenarios for thermal photon production: considering thermal radiation from only the top 5\% highest multiplicity events,  thermal radiation from only the top 20\% events, and thermal radiation from all events. 
For all these three cases, the thermal photon yields are smaller compared to the prompt component in minimum bias pp collisions at both 200 and 5.02 TeV. At 200 GeV, our direct photon spectrum agrees reasonably with the PHENIX measurements \cite{Adare:2009qk}. The relative size of thermal radiation to the prompt production increases with the collision energy from 35\% to 40\% at $p_T \sim 1$ GeV.

\section{Theoretical uncertainties}
\label{secV}

In this section we estimate some of the theoretical uncertainties inherent in some of calculations presented in this work.
%
\begin{figure*}[ht!]
  \centering
  \centering
  \begin{tabular}{cc}
  \includegraphics[width=0.4\linewidth]{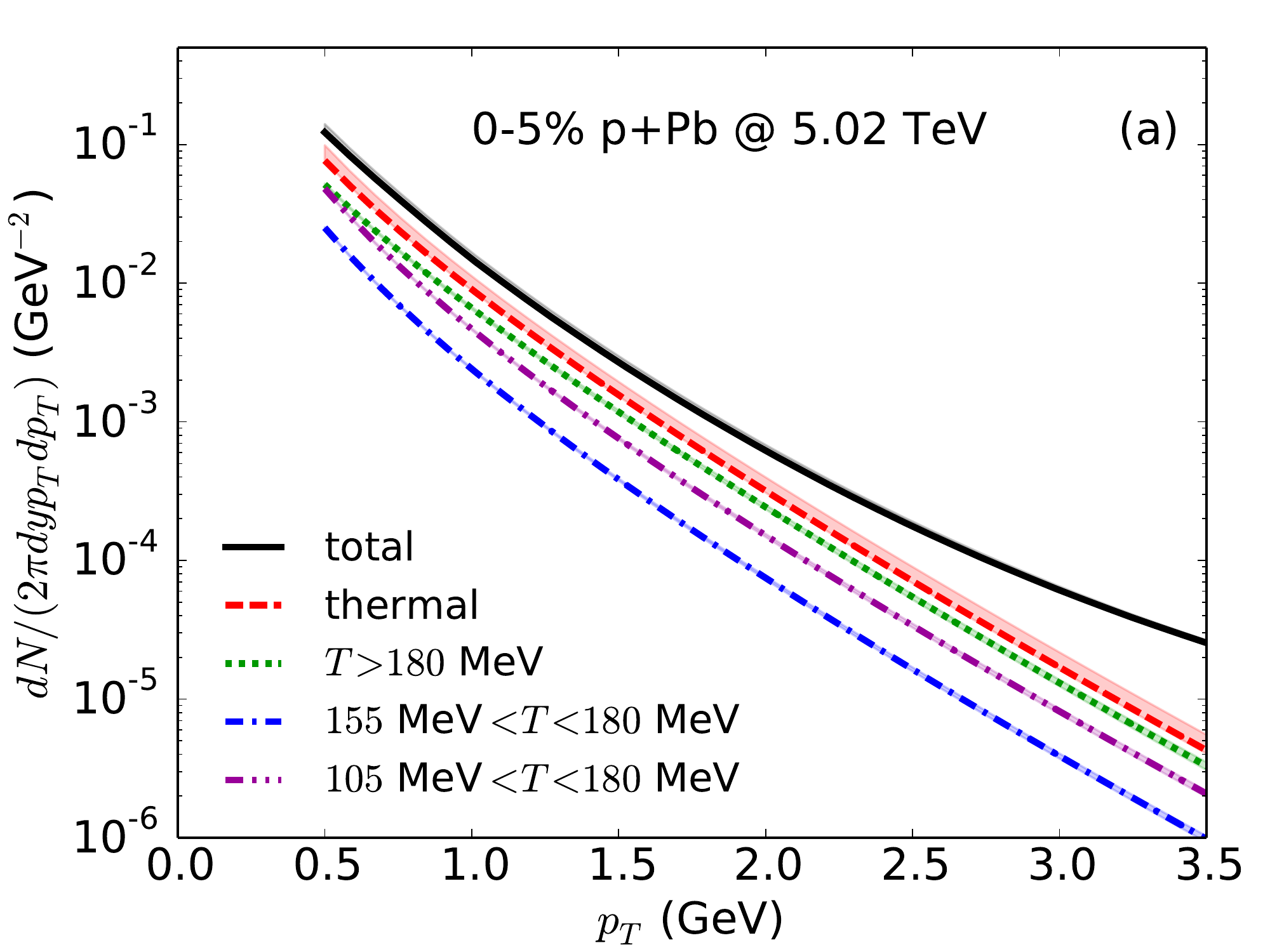} &
  \includegraphics[width=0.4\linewidth]{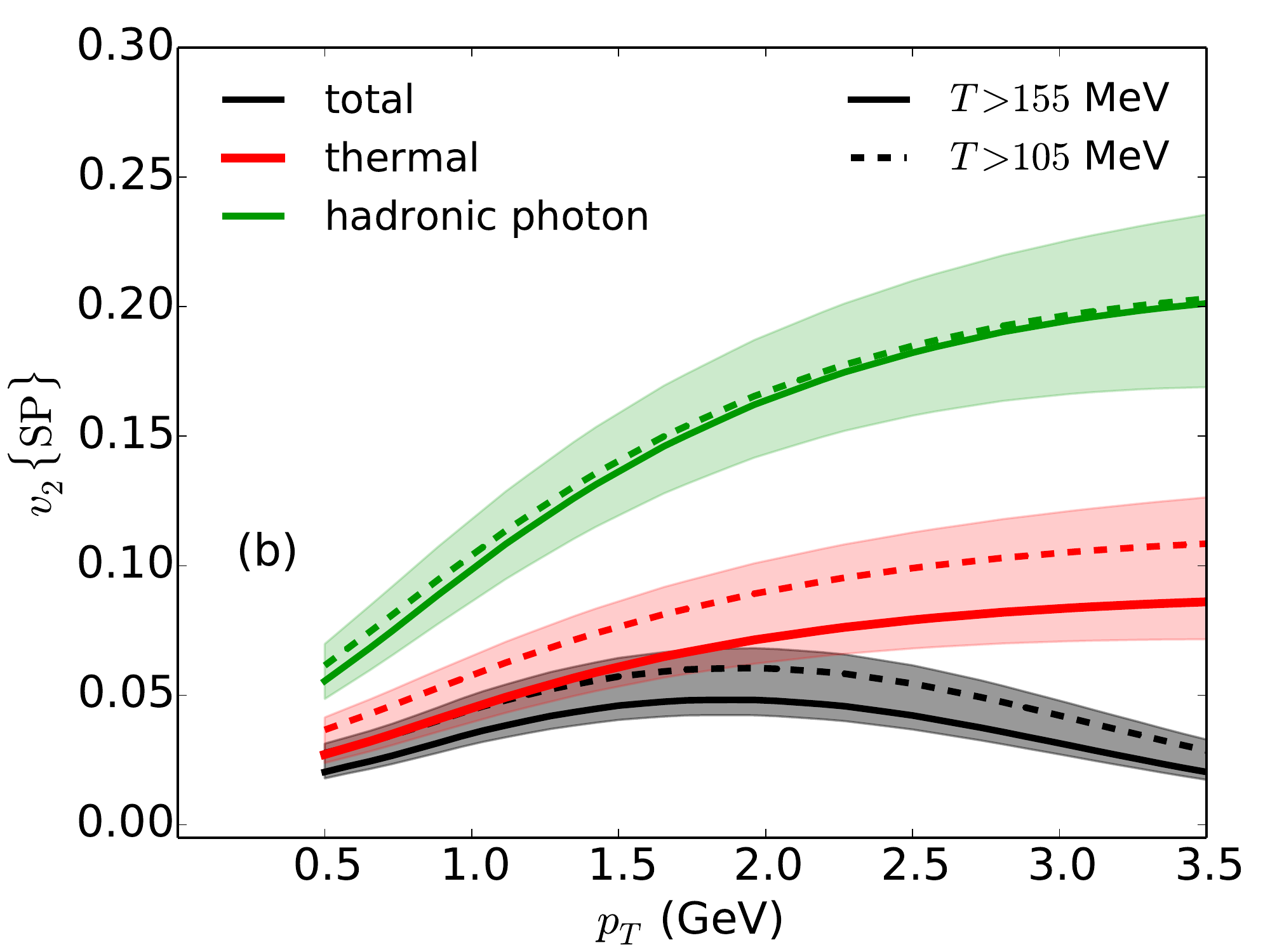}
  \end{tabular}
  \caption{(Color online) Direct photon spectra and elliptic flow results with hadronic photon emission extended to $T = 105$ MeV. The shaded band in the left panel shows the difference made by the inclusion of photon emission down to the lower temperature. The bands on the elliptic flow values are obtained from combining the statistical uncertainties  present for each of the two temperature regions described in the legend.}
  \label{fig9}
\end{figure*}
%

\subsection{Photon emission from the dilute hadronic phase}

The results presented up to this point included thermal photon radiation from fluid cells with temperatures higher than the switching temperature to the hadronic afterburner, $T_\mathrm{sw} = 155$ MeV. The system's dynamical evolution in the dilute hadronic phase is modelled by microscopic transport simulations. However, electromagnetic radiation is not currently included in the transport phase.  To estimate the additional thermal radiation from the dilute hadronic stage, it is possible to extend the hydrodynamic evolution below $T_\mathrm{sw}$ and evaluate thermal photon production from this hydrodynamic medium's profile. The hydrodynamic evolution is stopped at $T = 105$\,MeV,  as done in Ref. \cite{Paquet:2015lta}. 

Since inelastic scatterings that associated with species changing processes are expected to stop quickly, we use an equation of state that implements partial chemical freeze-out at 150 MeV in order to maintain correct hadronic chemical contents. Fugacity enhancements in hadronic photon emission rates are included in the calculations.  The choice of 150 MeV as partial chemical freeze-out temperature, and 105 MeV as the temperature as which the additional hydrodynamic evolution is stopped are obviously not unique, but are chosen as reasonable parameters to be used to estimate late stage photon production.

In Fig.~\ref{fig9}a, one sees that the hadronic photon emission, defined as photon emitted at temperature below 180 MeV, roughly doubles if thermal radiation is included  down to $T = 105$\,MeV. The final direct photon spectra increases by about 10\%. 
The $v_2$ of photons emitted below $T=180$ MeV remains almost unchanged by the addition of lower temperature photon emission, as shown in Fig.~\ref{fig9}b. This indicates that the momentum anisotropy of the system is fully built up and saturated around 155 MeV. In the end, the addition of these large-$v_2$ late stage photons increase the thermal and direct photon $v_2$ by about 30\% and 25\%, respectively. 

The bands in plotting the photon elliptic flow coefficients indicate our estimation of the theoretical uncertainties generated by the inclusion (or not) of the additional thermal contribution from the dilute hadronic phase. 

We note that there are reasons to believe that production of photons in the late stage of the medium should be significantly lower in small systems than in heavy ion collisions. In the hadronic transport simulations, we find the collision rate in central p+Pb collisions is only $\sim$0.2 collisions per particle below 155 MeV, in contrast to $\gtrsim$ 0.5 collisions per particle in semi-peripheral Pb+Pb collisions. This smaller collision rate is in part a consequence of the large expansion rate found at low temperature in small systems, which is around four times larger than what it is in heavy-ion collisions in the same temperature range. Considering the smaller number of collisions per hadrons, and the overall smaller number of hadrons present in the final state of small systems, it is thus less likely that a significant number of photons are produced at late time. In this sense, we expect photon production in small systems to be more biased towards the earlier and hotter regions of the plasma.

\begin{figure*}[ht!]
  \centering
  \centering
  \begin{tabular}{cc}
  \includegraphics[width=0.4\linewidth]{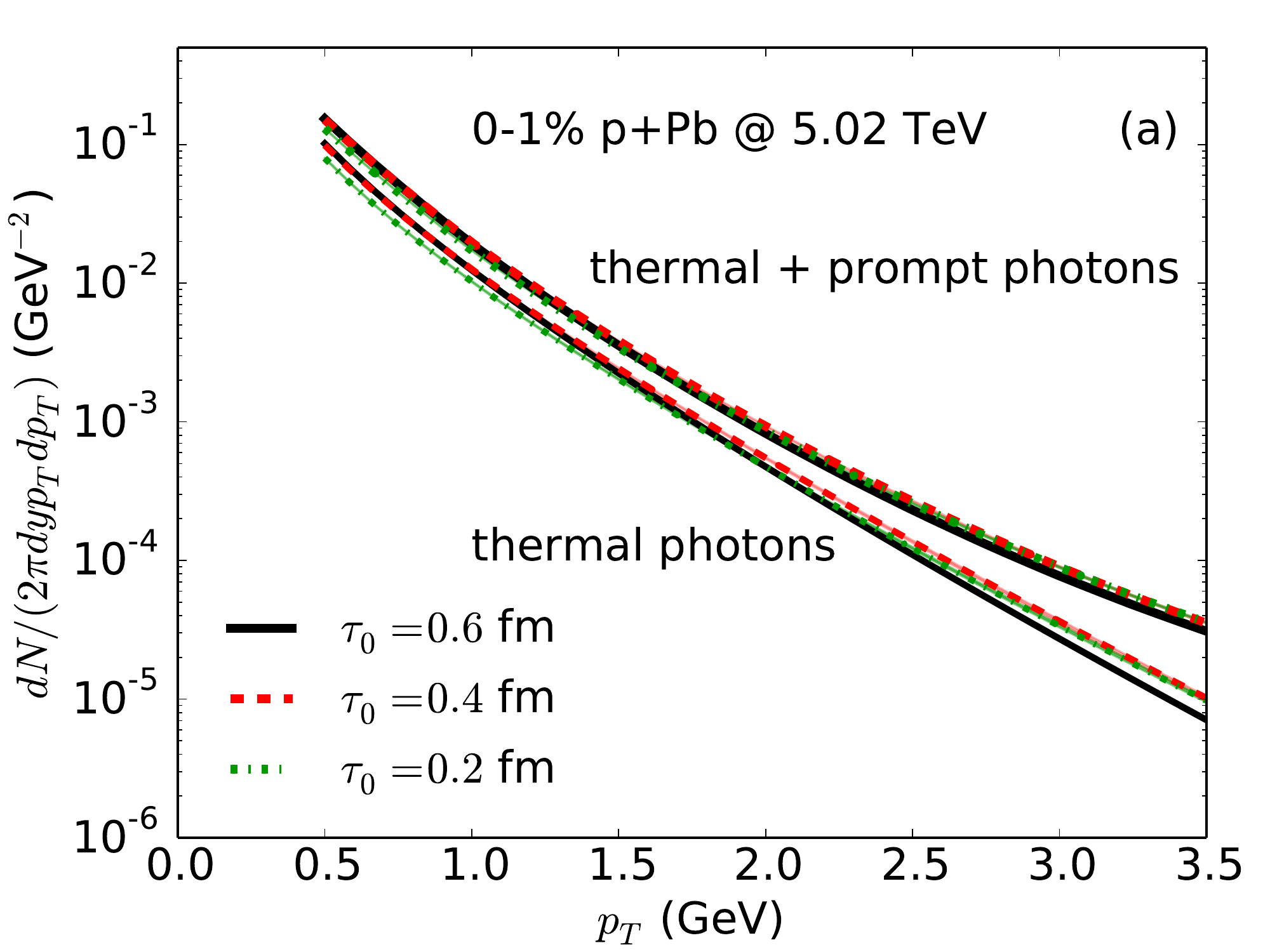} &
  \includegraphics[width=0.4\linewidth]{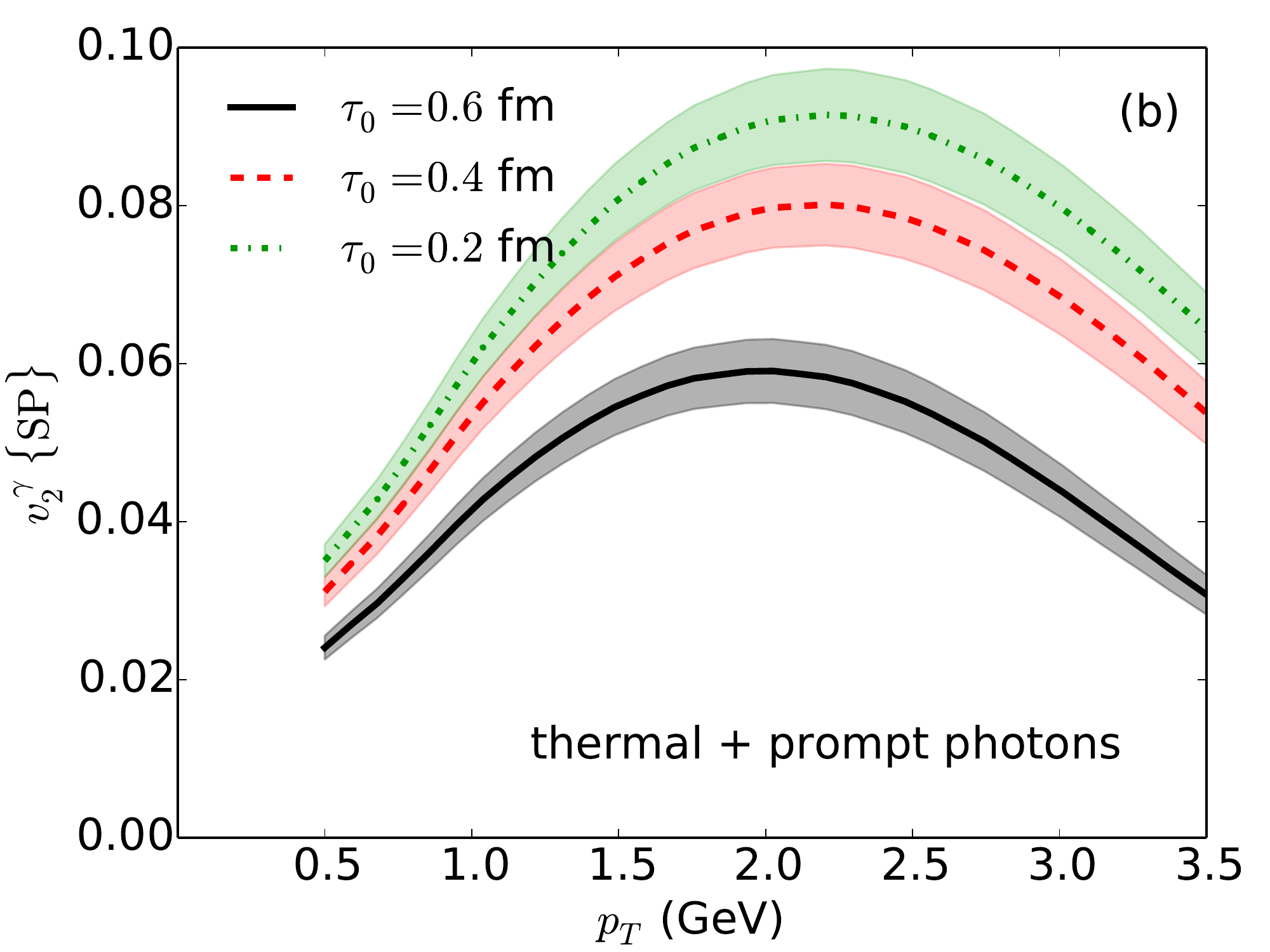}
  \end{tabular}
  \caption{(Color online)
Dependence of direct photon spectra (a) and elliptic flow coefficients (b) in 0-1\% p+Pb collisions on the thermalization time, $\tau_0$. Three values of $\tau_0$ , 0.2, 0.4, and 0.6 fm/$c$, were used. The charged hadron multiplicity was kept fixed for the different values of $\tau_0$ by adjusting the energy normalization of the initial conditions. The other parameters of the hydrodynamic model were not modified, meaning that only for $\tau_0$=0.6 fm/$c$ are hadronic observables globally well described (see Fig.~\ref{fig10b} and text for details). As such, the direct photon calculations at $\tau_0=$0.2, 0.4 fm/$c$ are meant to represent the sensitivity of the photon calculation to $\tau_0$, and not predictions for direct photons if such values of $\tau_0$ were used. The shaded bands represent statistical uncertainty.}
  \label{fig10a}
\end{figure*}
%

\begin{figure*}[ht!]
  \centering
  \centering
  \begin{tabular}{cc}
  \includegraphics[width=0.4\linewidth]{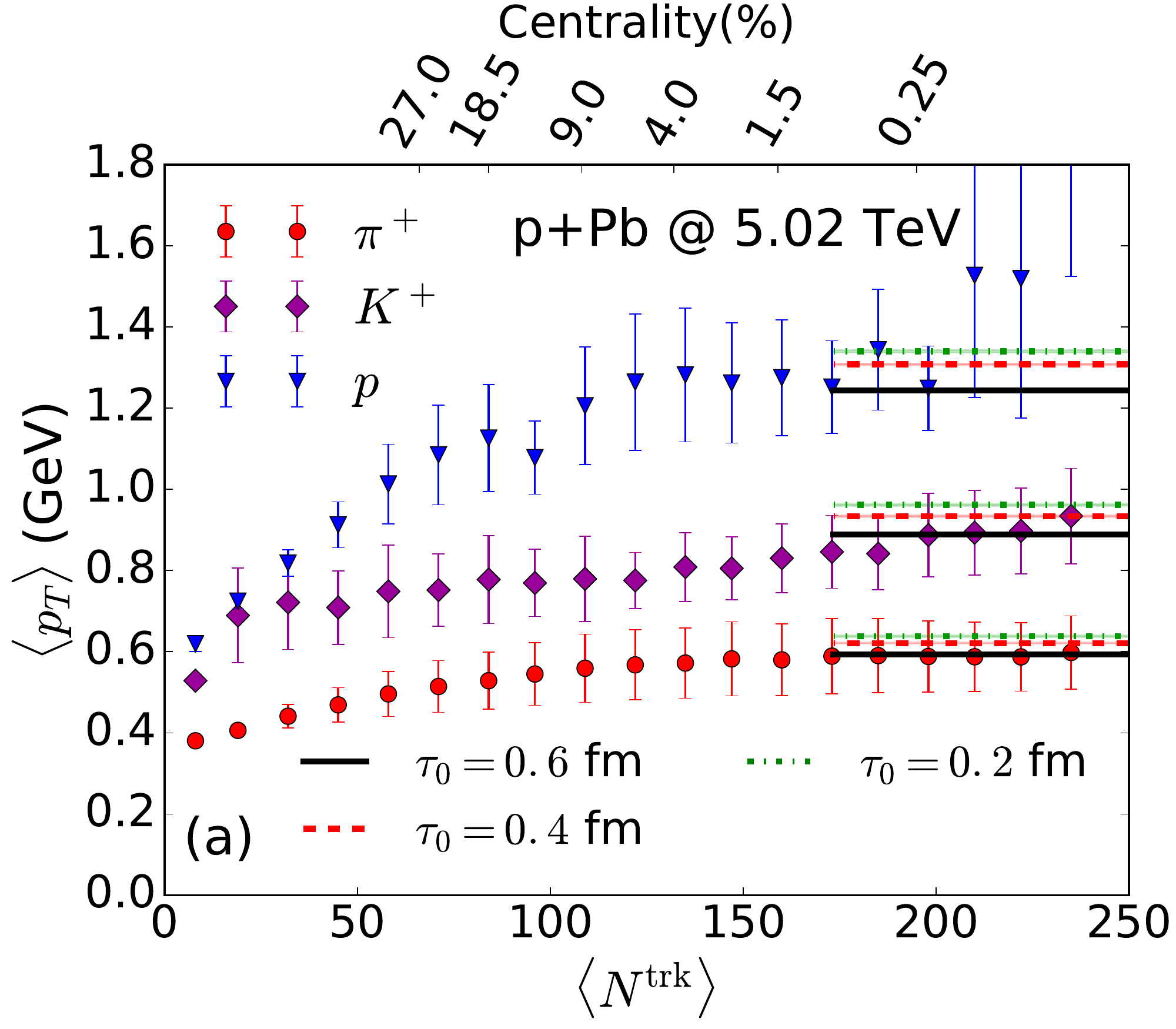} &
  \includegraphics[width=0.4\linewidth]{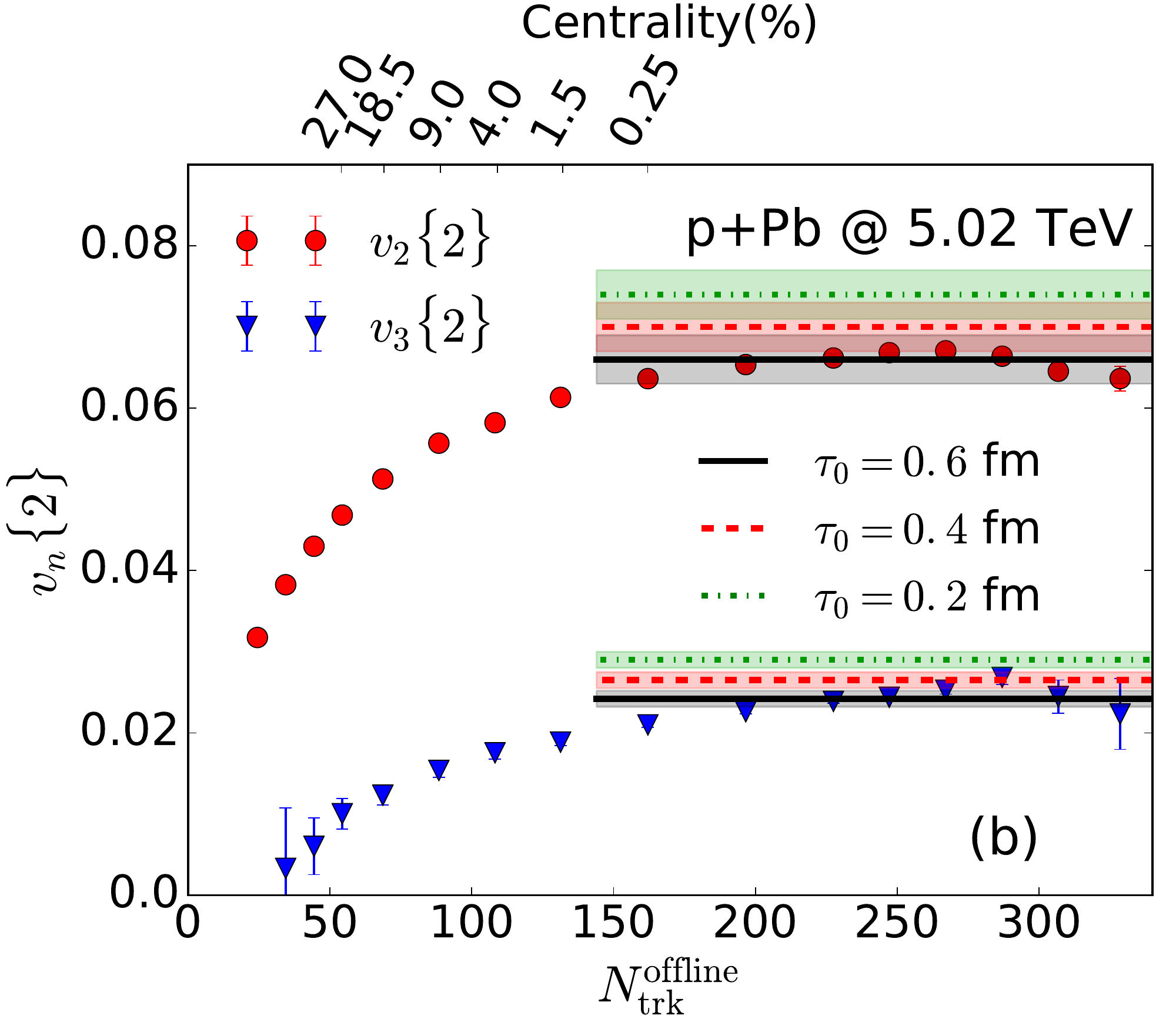}
  \end{tabular}
  \caption{(Color online) Identified particle mean $p_T$ (a) and charged hadron anisotropic coefficients, $v_{2,3}\{\mathrm{SP}\}$ (b) in 0-1\% p+Pb collisions with different thermalization time, $\tau_0 = 0.2$, 0.4, and 0.6 fm/$c$. The shaded bands represent statistical uncertainty. }
  \label{fig10b}
\end{figure*}
%
\subsection{Pre-equilibrium contributions}

Because the lifetime of the plasma produced in small systems is shorter ($\sim4$\,fm/$c$) than in heavy ion collisions ($\sim15-20$\,fm/$c$), the pre-equilibrium dynamics of the system may have a more sizeable influence on experimental observables. Non-trivial initial flow velocity profiles and early-stage electromagnetic probe production are two examples of the possible influence of pre-equilibrium dynamics on observables \cite{Vujanovic:2016anq}. A rigorous treatment of these effects would require a detailed model of pre-equilibrium dynamics, which still the subject of much active research and is currently an open question. Nevertheless, within the hydrodynamic framework used in this paper, it is possible to address the more modest question of the relative sensitivity of hadronic and photonic observables to the time  $\tau_0$ at which thermalization is assumed to occur, which was fixed to $\tau_0 = 0.6$ fm/$c$ for the results presented up to this point.

In Figs.~\ref{fig10a}, we investigate the effect of $\tau_0$ on
direct photon observables by choosing smaller values of $\tau_0$ for starting the hydrodynamics. Calculations with different $\tau_0$ are tuned such that the final charged hadron multiplicity remains the same. We find about 15\% more viscous entropy production if hydrodynamic evolutions are started at $\tau_0 = 0.2$ fm/$c$ compared to ones started at $\tau_0 = 0.6$ fm/$c$. 

The thermal photon spectrum is flatter with a smaller $\tau_0$ in Fig.~\ref{fig10a}. This is a consequence of the following two main factors. Firstly, more high $p_T$ photons are emitted from high temperature hot spots at the early time. At $\tau_0 = 0.2$ fm/$c$, the peak temperature of the medium can reach up to 650 MeV. Secondly, with a smaller $\tau_0$ the system's pressure gradients accelerate fluid cells earlier and develop more radial flow. It gives a stronger blueshift to the thermal photons emitted at the late stage. Both effects make the emitted thermal photon spectrum harder. 
Meanwhile, the large expansion rate shortens the fireball lifetime by $\sim 10\%$. The reduction of the space-time volume results a smaller thermal photon production with $\tau_0 = 0.2$ fm/$c$ compared to the collision events who started its transverse expansion at $\tau_0 = 0.6$ fm/$c$.
Finally, the large pressure gradients at the early time also help the anisotropic flow to develops faster during the hydrodynamic evolution. Direct photons $v_2$ in Fig.~\ref{fig10a} is found to be considerably larger with a smaller $\tau_0$. 

In Fig.~\ref{fig10b}, we verify the $\tau_0$ sensitivities on hadronic observables. Both identified particle mean $p_T$ and charged hadron $v_{2,3}$ increase as $\tau_0$ gets smaller. This is consistent with the direct photon observables in Fig.~\ref{fig10a} that hydrodynamic flow is developed earlier and larger with a smaller $\tau_0$. 

The effect of $\tau_0$ on photons and hadrons thus appear to be qualitatively similar, with photons being slightly more sensitive to the initial time than hadrons, as one can reasonably expect from probes that can be emitted at earlier times.

\begin{figure*}[ht!]
  \centering
  \centering
  \begin{tabular}{ccc}
  \includegraphics[width=0.32\linewidth]{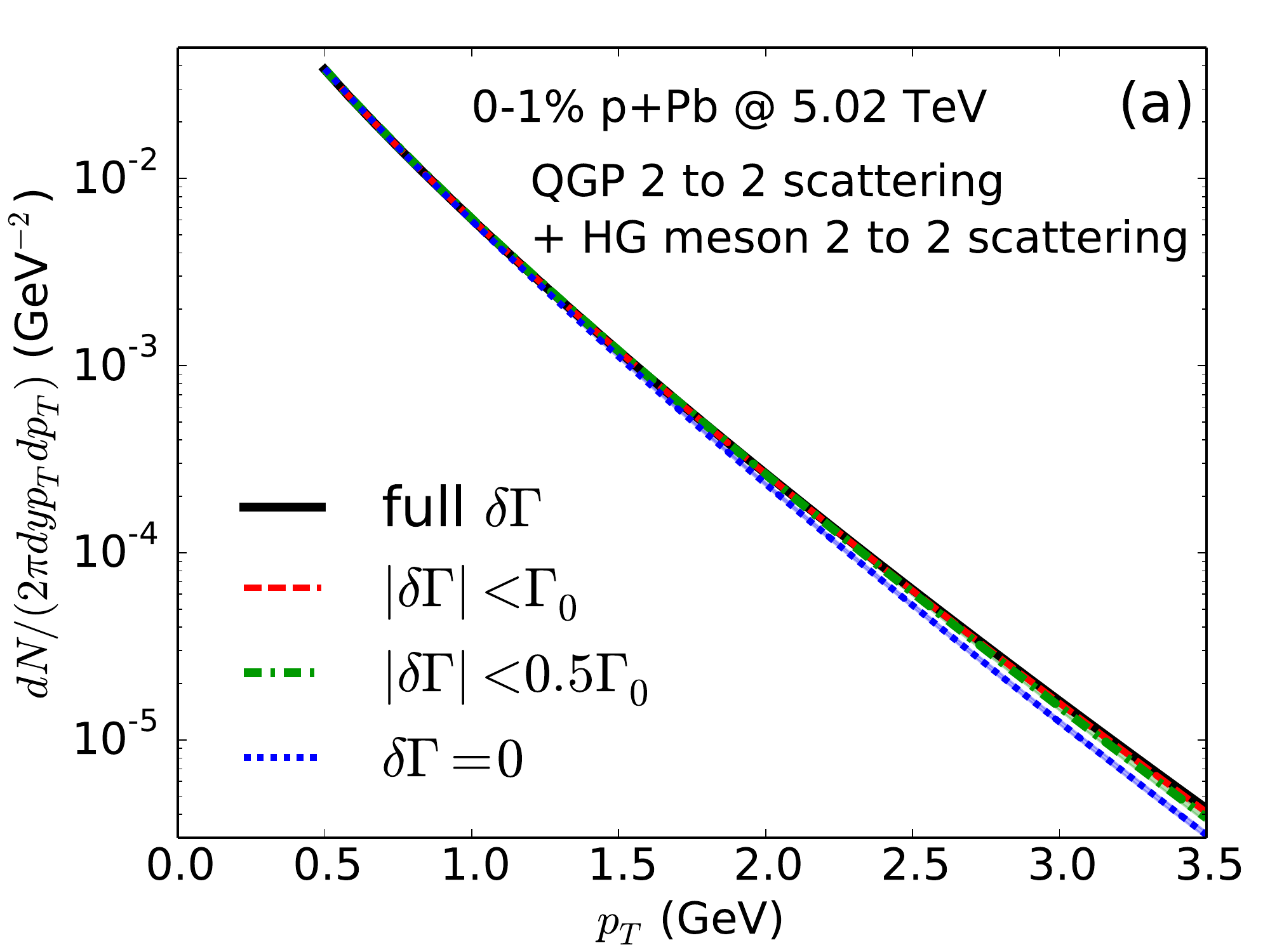} &
  \includegraphics[width=0.32\linewidth]{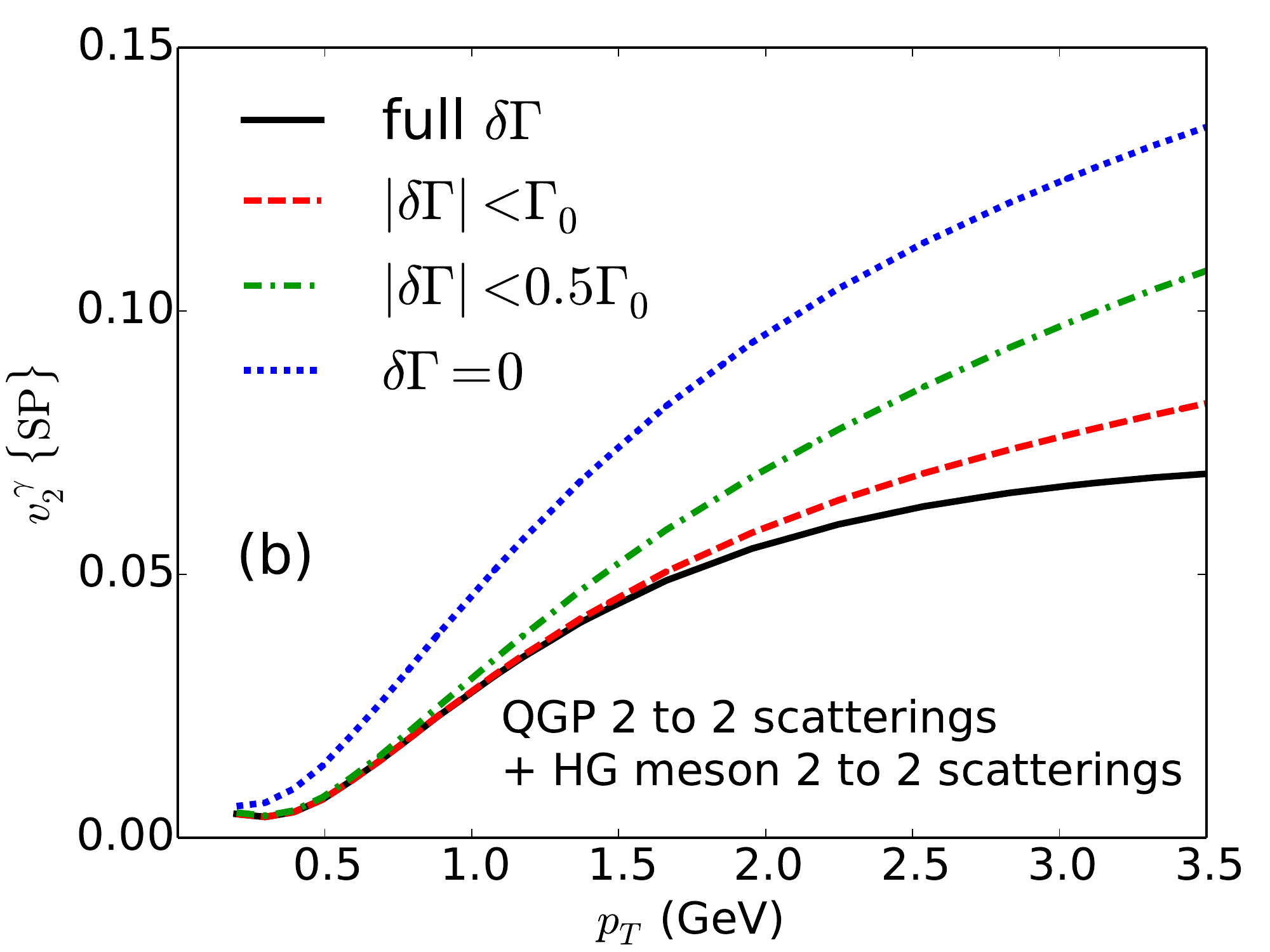} & 
  \includegraphics[width=0.32\linewidth]{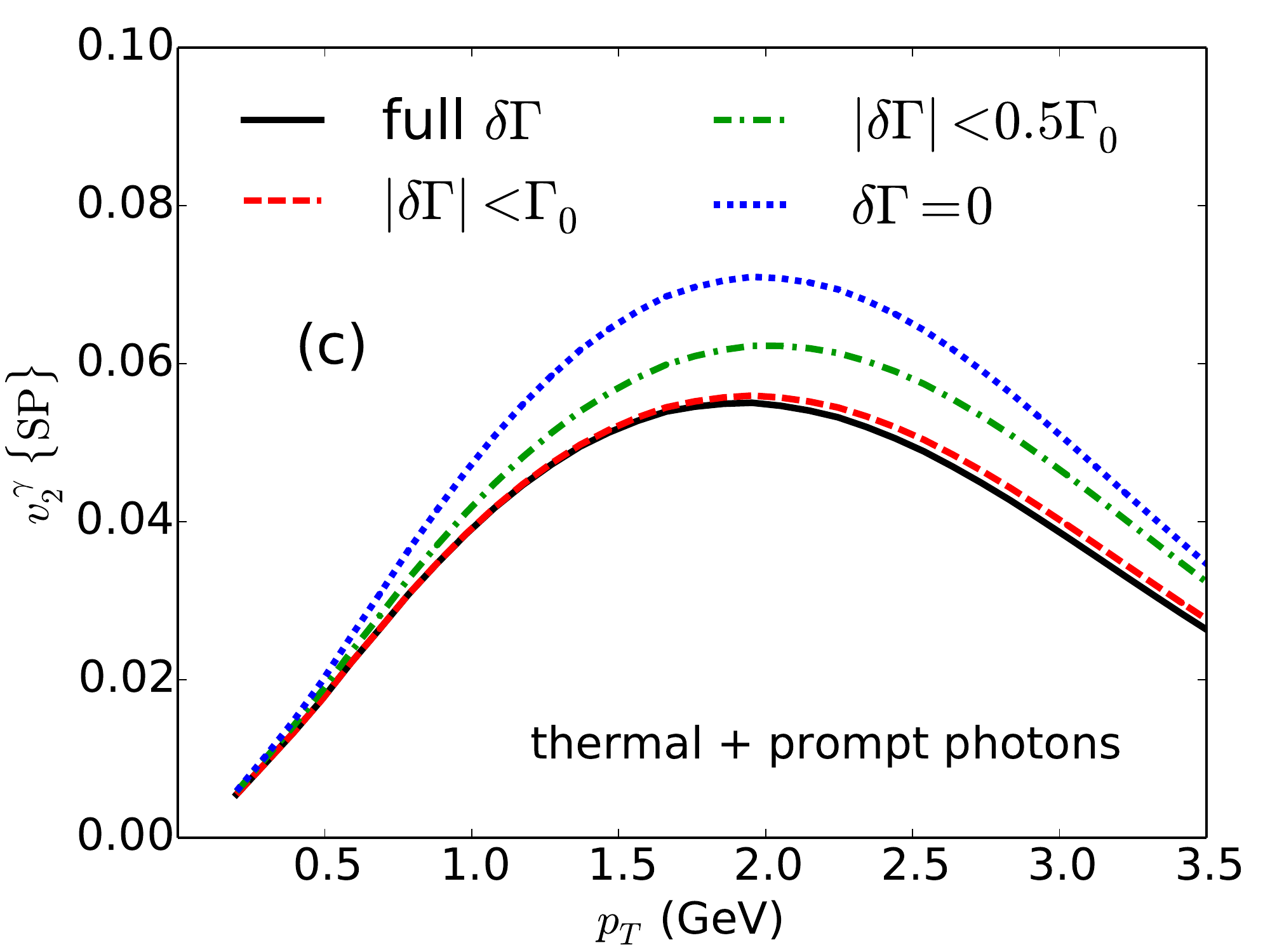}
  \end{tabular}
  \caption{(Color online) The effects of restricting the size of shear viscous corrections $\delta f$ to photon emission rate on (a) thermal photon spectra, (b) thermal photon elliptic flow, and (c) full direct photon $v_2$. }
  \label{fig11}
\end{figure*}
%

\subsection{Out-of-equilibrium corrections}

\label{section:viscousEffects}

Temperature and velocity gradients are larger in small systems than in heavy ion collisions. Consequently, the out-of-equilibrium corrections could also be expected to be larger.
 In the last part of this section, we investigate the shear viscous corrections to direct photon observables. 

We explore the sensitivity of direct photon spectra and $v_2$ to the shear viscous corrections $\delta \Gamma$ by constraining it to be smaller than a certain fraction of its corresponding equilibrium contribution. Based on Eq.~(\ref{eq3}), we compute the following  ratio for a photon with energy $E_q$ in the local rest frame of a fluid cell, 
\begin{equation}
r(E_q, T, \pi^{\mu\nu};a) = \frac{\vert \delta \Gamma (E_q, T, \pi^{\mu\nu}) \vert}{a \Gamma_0(E_q, T)},
\label{eq4}
\end{equation}
where the coefficient $a$ is a constant parameter, which determines the maximum allowed fraction of equilibrium contribution for the $\delta f$. Then the constrained photon emission rate in a fluid cell is evaluated as,
\begin{eqnarray}
&& \!\!\!\!\!\!\!\!\!\! E_q\frac{d \Gamma}{d^3 k}(E_q, T, \pi^{\mu\nu};a) = \notag \\
&& \Gamma_0 (E_q, T) + \frac{\delta \Gamma (E_q, T, \pi^{\mu\nu}) }{\mathrm{max}\{1, r(E_q, T, \pi^{\mu\nu}; a)\}}.
\label{eq5}
\end{eqnarray}

Because shear viscous correction is only available for those photons emitted from the 2 to 2 scatterings in the QGP phase and meson-meson reactions, we will focus on the thermal photon flow observables from these two channels in Figs.~\ref{fig11}(a) and (b).  We compared the thermal photon spectra and their elliptic flow coefficients with the fraction parameter $a = 0, 0.5, 1$, and $\infty$ in Eq. (\ref{eq5}). By choosing $a = 1$, we allow the maximum size of $\delta \Gamma$ to be equal to its equilibrium part. In this case, the thermal photon observables are very close to the no-constraint case ($a = \infty$) after integrated over all space-time volume. Some noticeable differences are present only for $p_T > 2.5$ GeV. If the constraint increases to $a = 0.5$ ($\vert \delta \Gamma \vert < 0.5 \Gamma_0$), thermal photon elliptic flow starts to show some sizeable variation. In Fig.~\ref{fig11}(c), we show the sensitivity of total direct photon $v_2$ to different choices of the parameter $a$. Among the cases $a = 0.5, 1.0$, and $\infty$, the largest variation of the direct photon $v_2$ reaches up to $\sim 15\%$ in $2 < p_T < 3$ GeV.

\section{Conclusion}

In this work, we have presented a systematic study of hadronic collective observables and of direct photon probes in small collisions systems at RHIC and LHC energies,  within a consistent dynamical approach.

It was found that hydrodynamic simulations can provide a good description of the hadronic flow observables for both inclusive charged hadrons and  identified particles. The effects of the hadronic transport description of the dilute hadronic phase on hadronic observables were quantified and found to be small in general. It was also found that, in the absence of longitudinal fluctuations, a boost-invariant assumption can provide a reasonable description of mid-rapidity photonic and hadronic observables.

A thermal photon enhancement  - over the case where no thermal sources are present - of the differential spectra  was predicted in  high multiplicity collision events, in small systems at both RHIC and LHC energies. In addition, we found that direct photons carry sizeable anisotropic flow $v_{2,3}\{\mathrm{SP}\}$. These proposed signals can serves as independent tests of the hydrodynamic description of small hot and dense systems. An analysis of decay cocktail and inclusive photons was presented to provide guidance to future experimental measurements.

Several aspects of theoretical uncertainties in describing the dynamics of small systems were explored in the last section of this work. Photon emissions from the dilute hadronic phase were found to contribute up to $\sim10\%$ in the total direct photon signal. 
The direct photon elliptic flow was shown to have a larger sensitivity than hadrons to the choice of thermalization time $\tau_0$.
The shear viscous corrections to the photon production calculations were examined and found to be under control.

Future work will include the study of more realistic sub-nucleon fluctuations in the initial state \cite{Mantysaari:2016ykx,Mantysaari:2016jaz}, of longitudinal fluctuations \cite{Khachatryan:2015oea,Aaboud:2016jnr}, as well as the interplay between soft-hard components in the intermediate $p_T$ region, and the inclusion of bulk viscosity \cite{Ryu:2015vwa}. 

It is worth re-emphasizing that  measurements of low $p_T$ photons represent precious information which completes and complements what is learned with hadronic observables. These photons are both penetrating {\it and} soft: they are a unique characterization tool for systems of all sizes and shapes.

\begin{acknowledgements}
This work was supported in part by the Natural Sciences and Engineering Research Council of Canada. JFP was supported by the U.S. D.O.E. Office of Science, under Award No. DE-FG02-88ER40388.  
Computations were made in part on the supercomputer Guillimin from McGill University, managed by Calcul Qu\'ebec and Compute Canada. The operation of this supercomputer is funded by the Canada Foundation for Innovation (CFI), NanoQu\'ebec, RMGA and the Fonds de recherche du Qu\'ebec - Nature et technologies (FRQ-NT). 
\end{acknowledgements}

\appendix
\section{Space-time structure of medium evolution and photon emissions in small systems}

In this appendix, the space-time evolution of the fluid-dynamical description of small systems is scrutinized. 

\begin{figure*}[ht!]
  \centering
  \centering
  \begin{tabular}{cc}
  \includegraphics[width=0.42\linewidth]{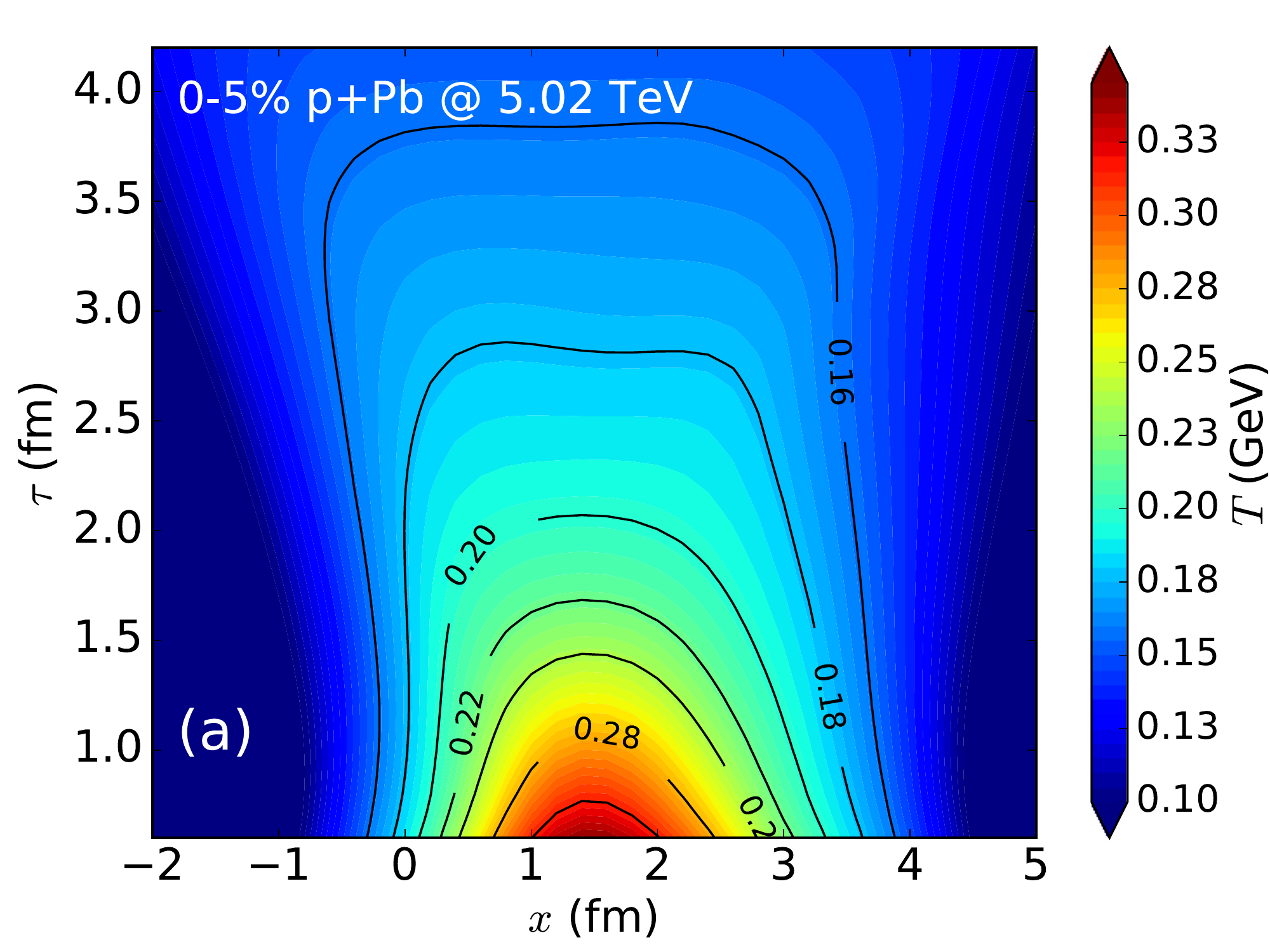} &
  \includegraphics[width=0.42\linewidth]{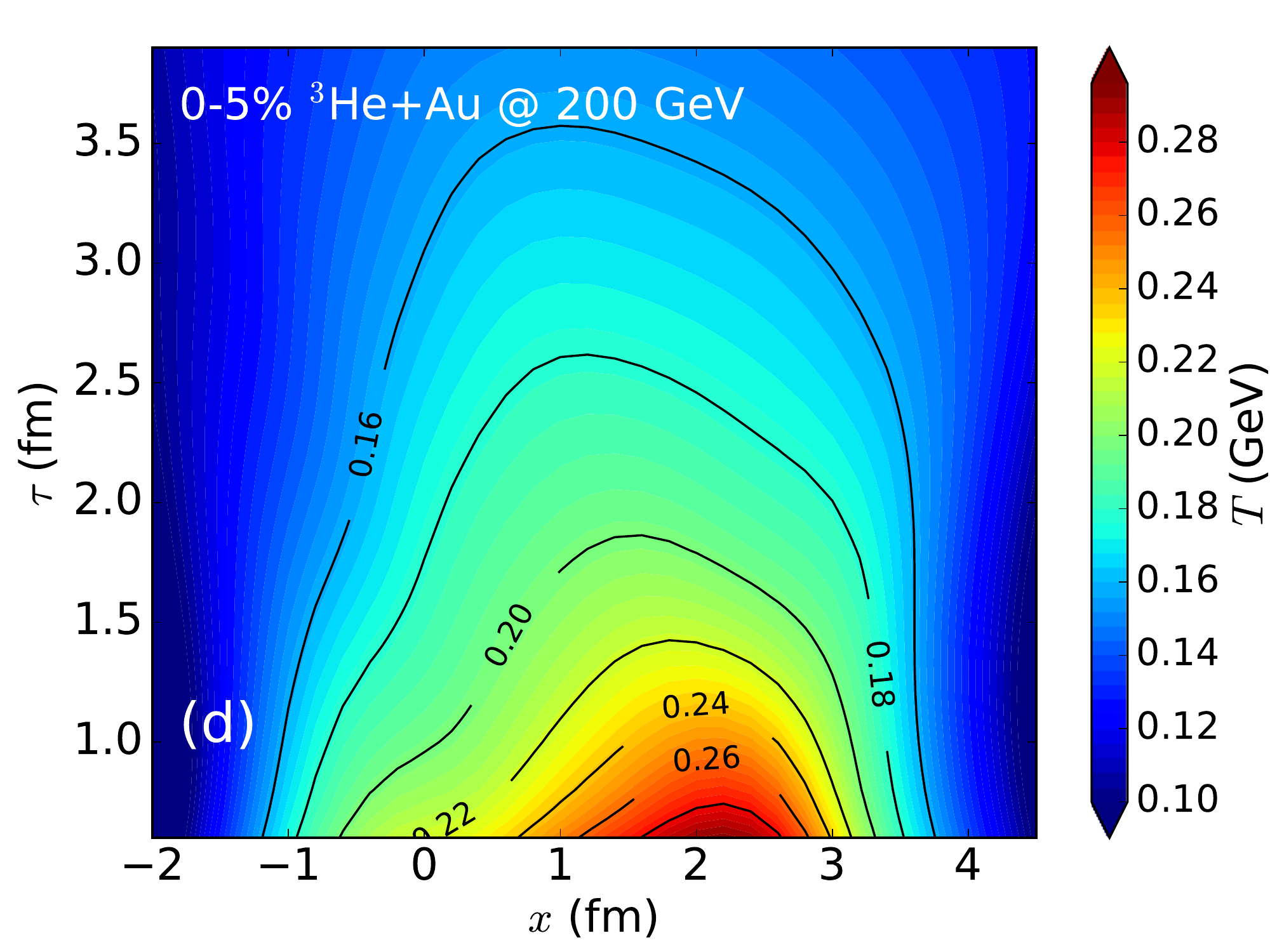} \\
  \includegraphics[width=0.42\linewidth]{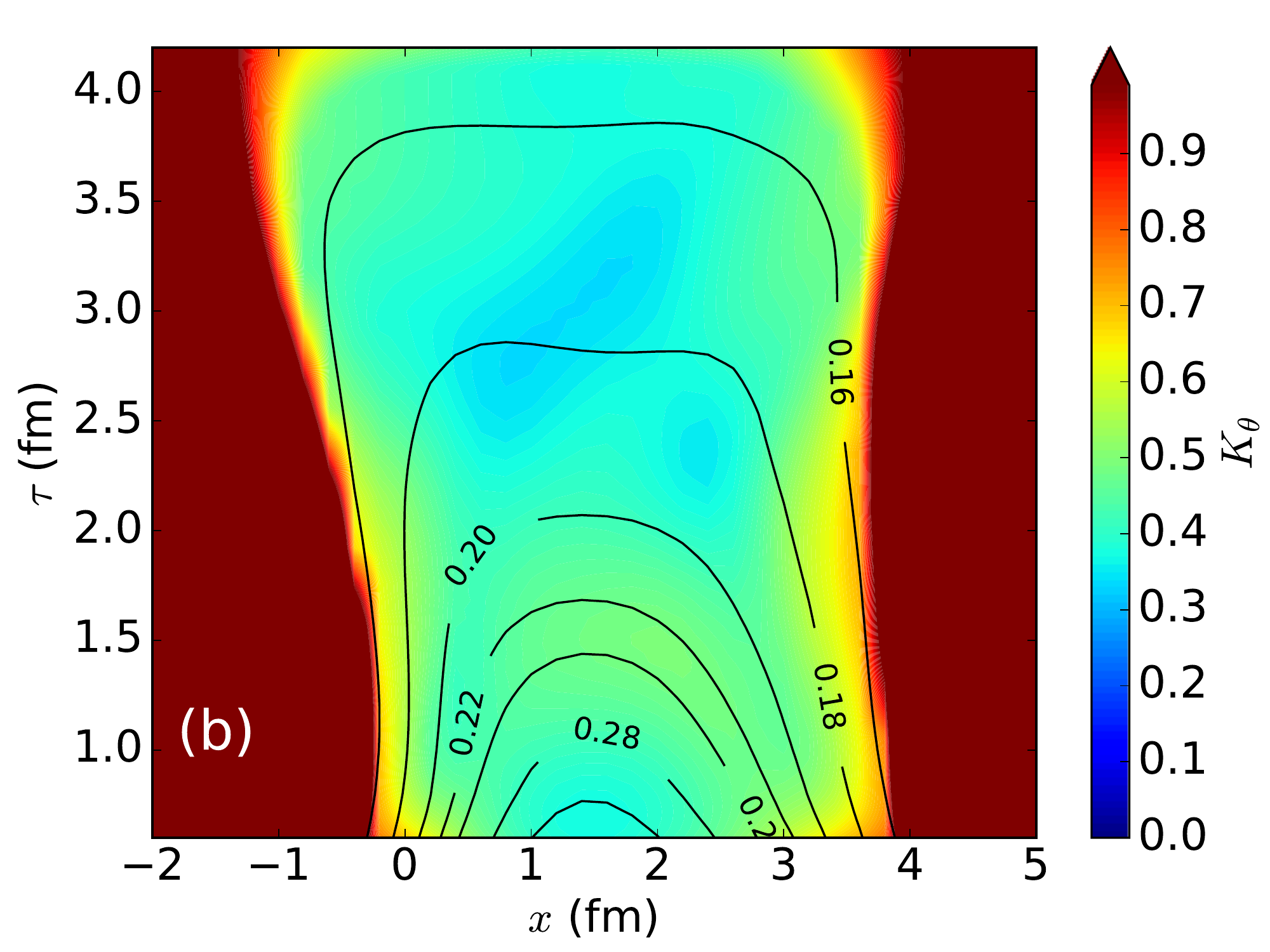} &
  \includegraphics[width=0.42\linewidth]{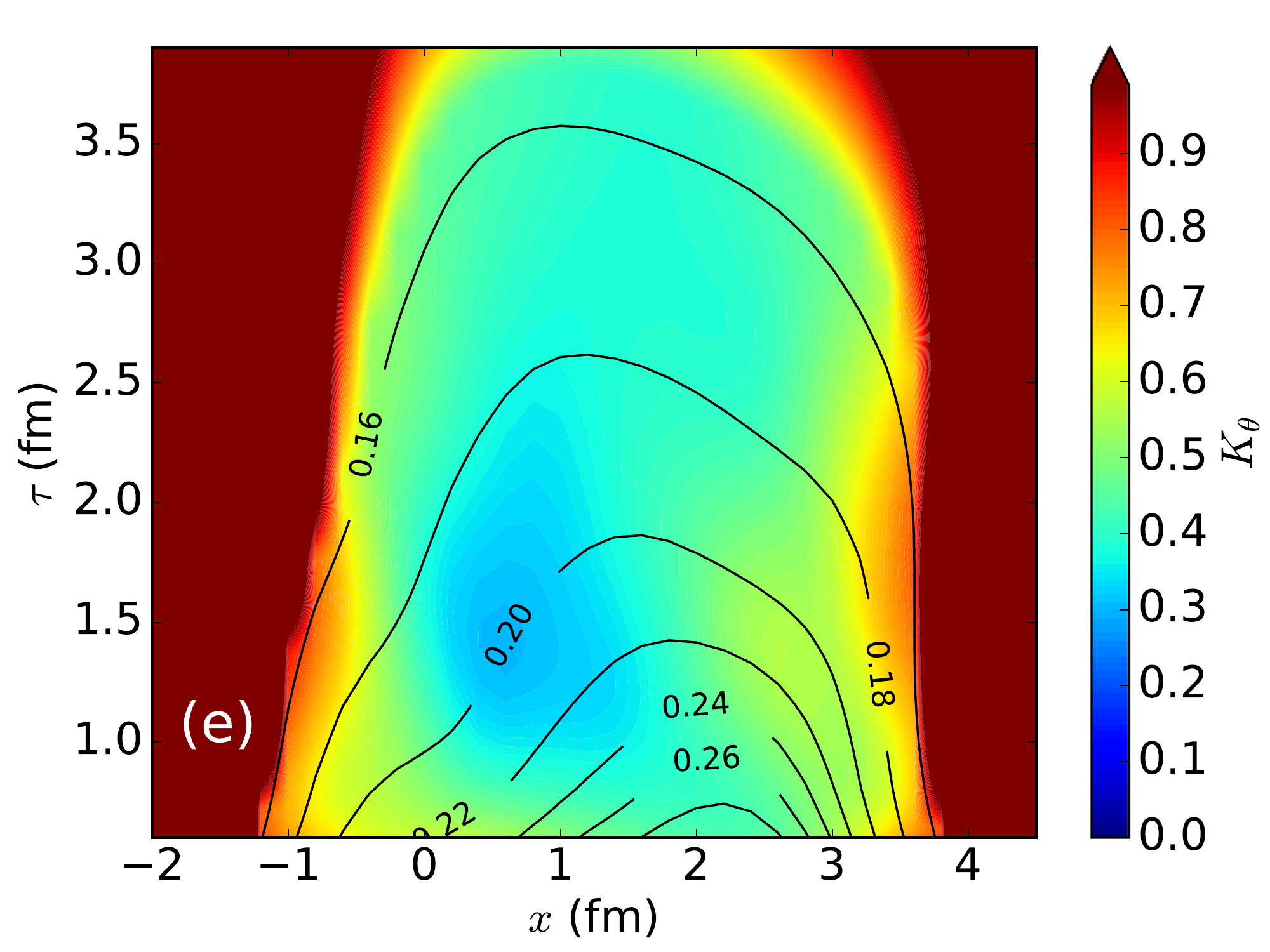} \\
  \includegraphics[width=0.42\linewidth]{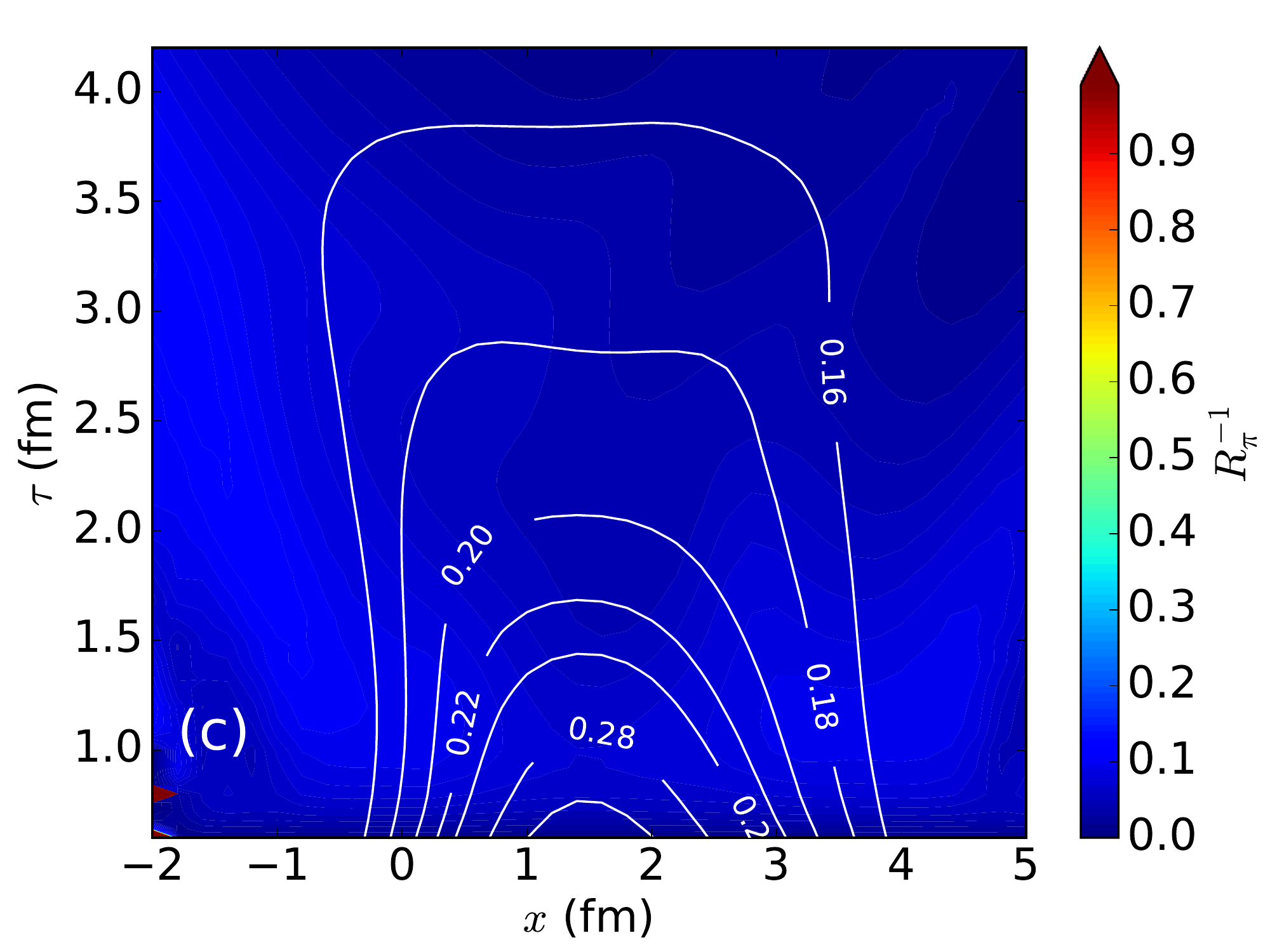} &
  \includegraphics[width=0.42\linewidth]{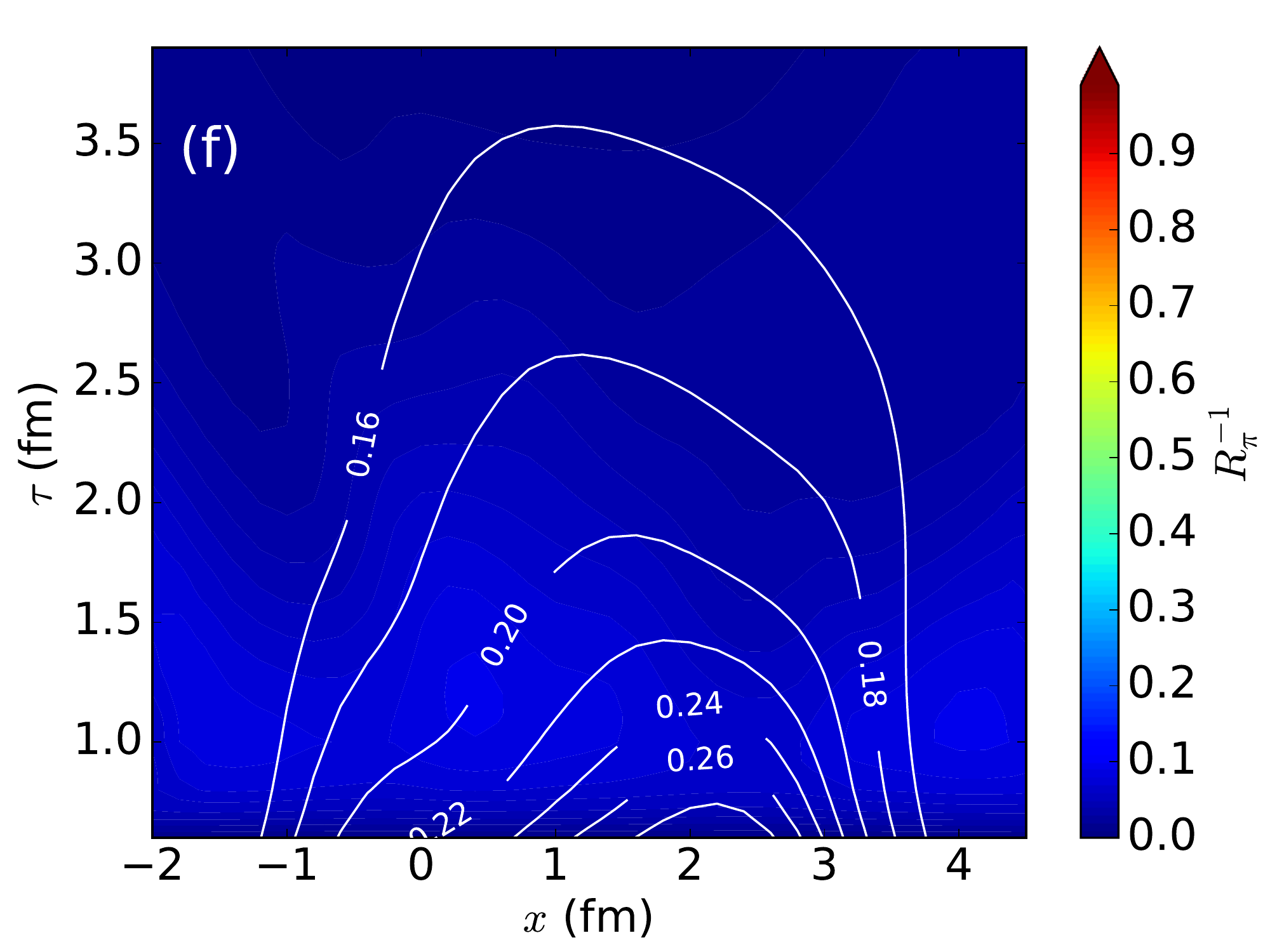}
  \end{tabular}
  \caption{(Colour online){\it Panels (a-c):} Contour plots for evolution of temperature, the Knudsen number, and the inverse Reynolds number in one fluctuating 0-1\% p+Pb collisions at 5.02 TeV. Black contour lines  are isotherms. {\it Panels (d-f):} Similar plots for one fluctuating event in 0-5\% $^3$He+Au collision at 200 GeV. }
  \label{fig12}
\end{figure*}
%
\begin{figure*}[ht!]
  \centering
  \centering
  \begin{tabular}{cc}
  \includegraphics[width=0.42\linewidth]{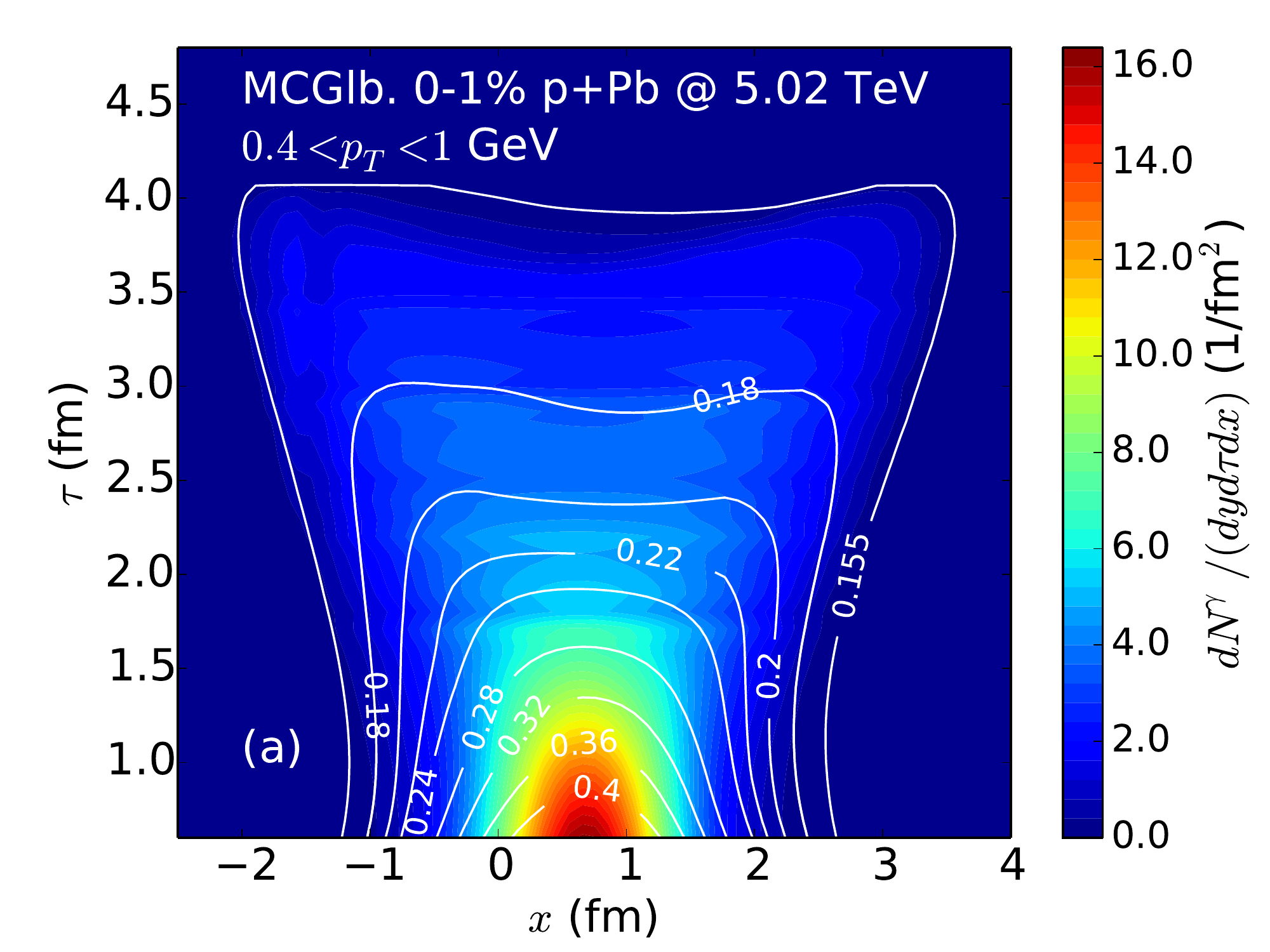} &
  \includegraphics[width=0.42\linewidth]{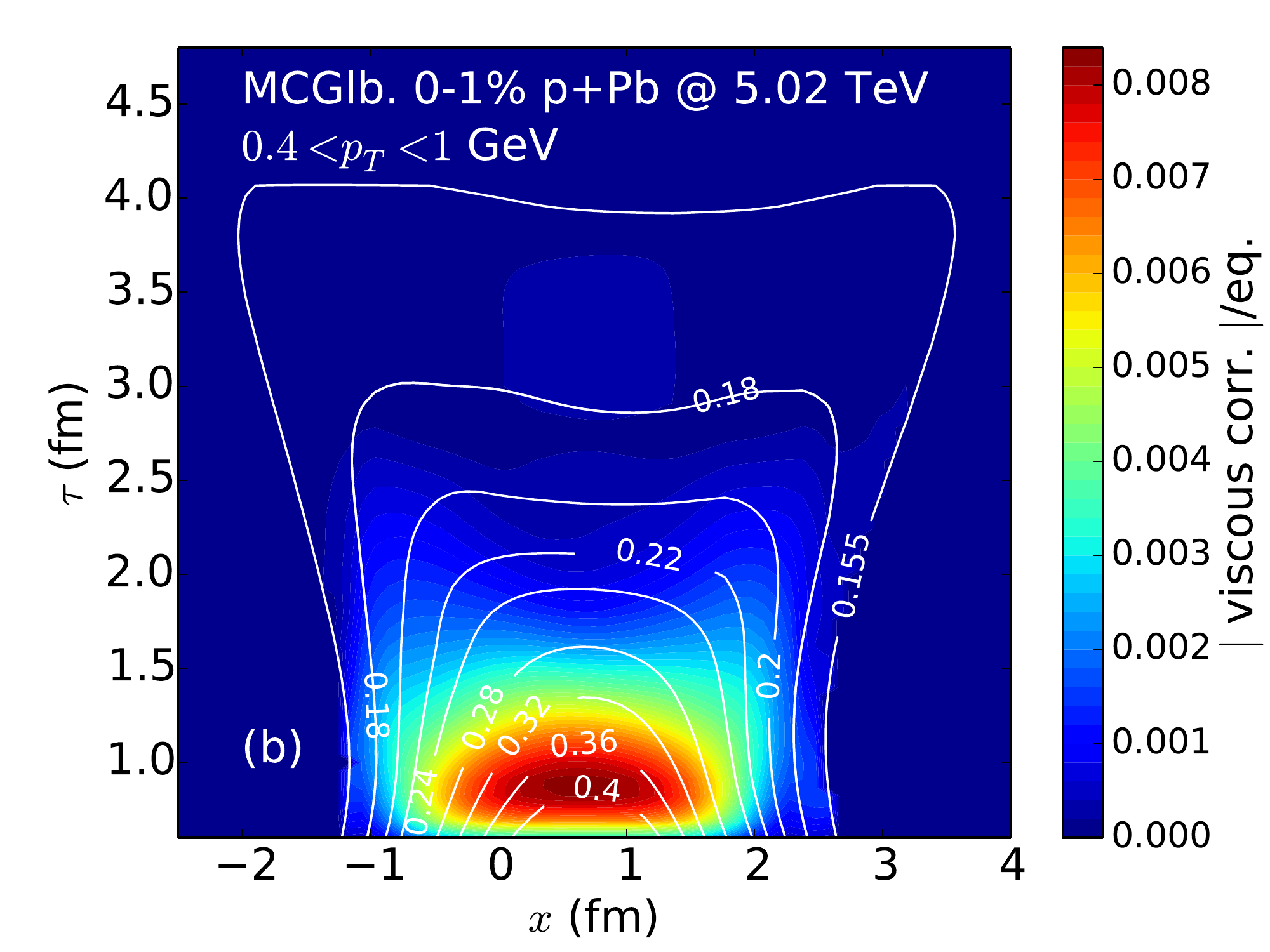} \\
  \includegraphics[width=0.42\linewidth]{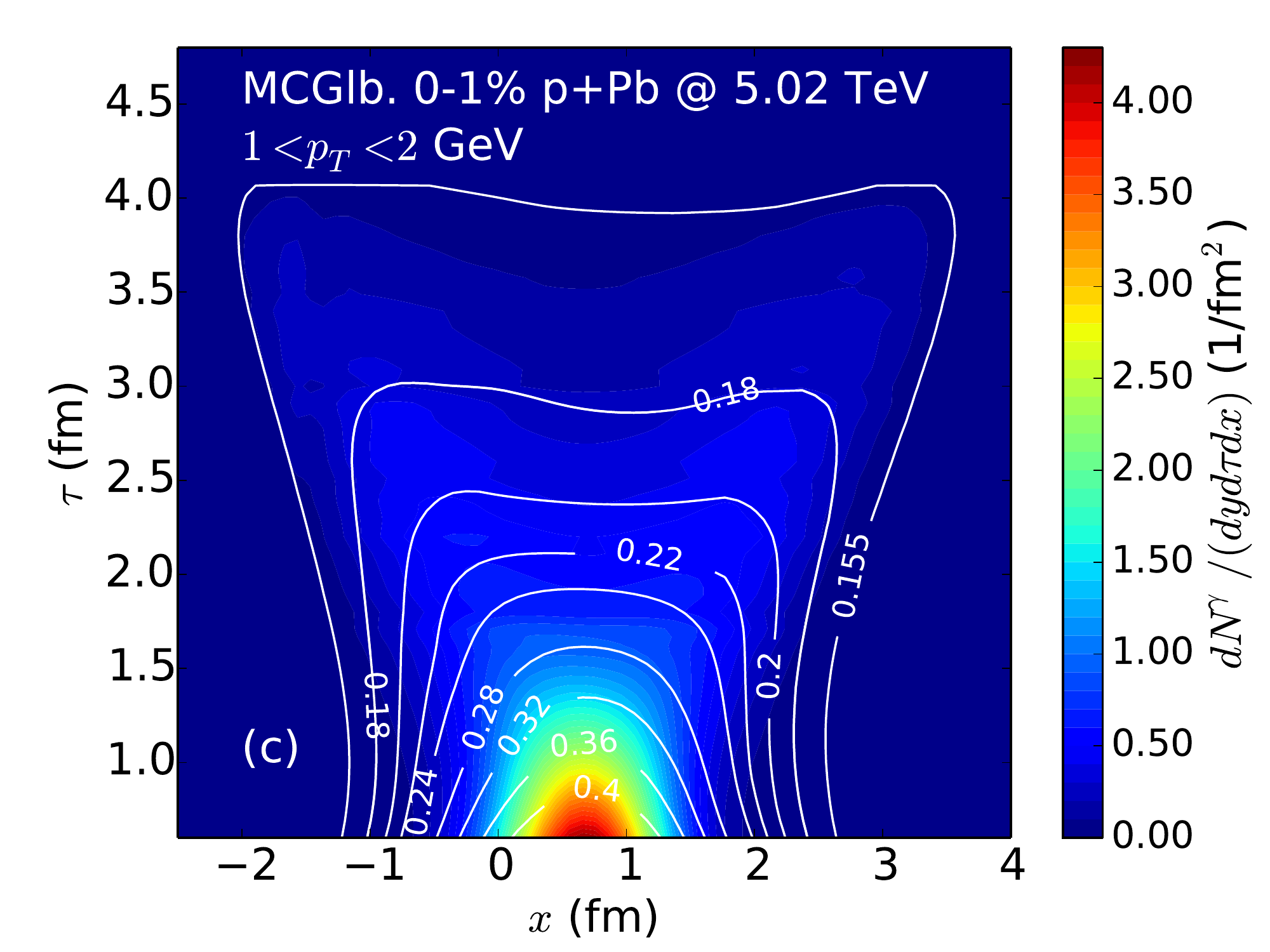} &
  \includegraphics[width=0.42\linewidth]{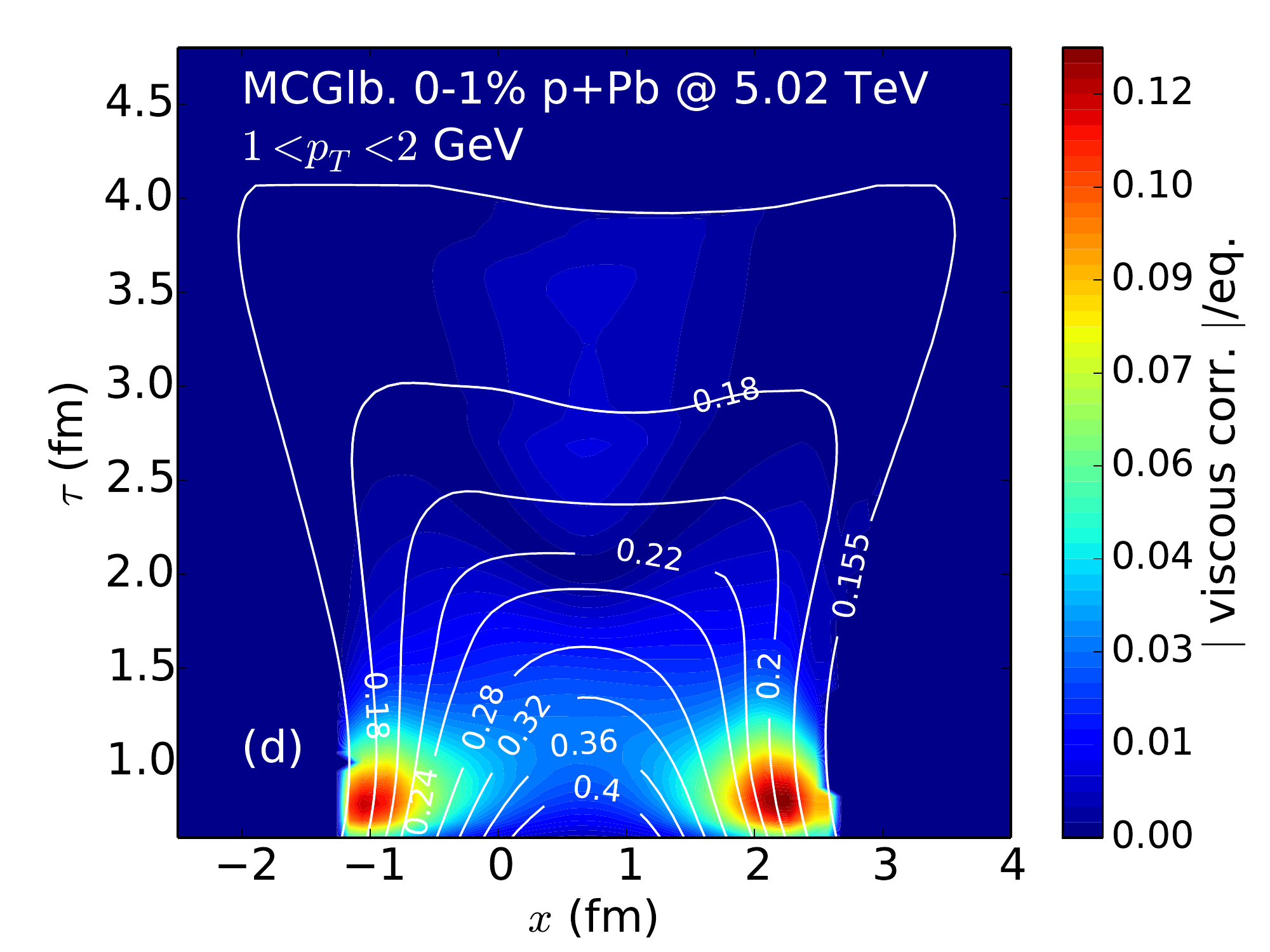} \\
  \includegraphics[width=0.42\linewidth]{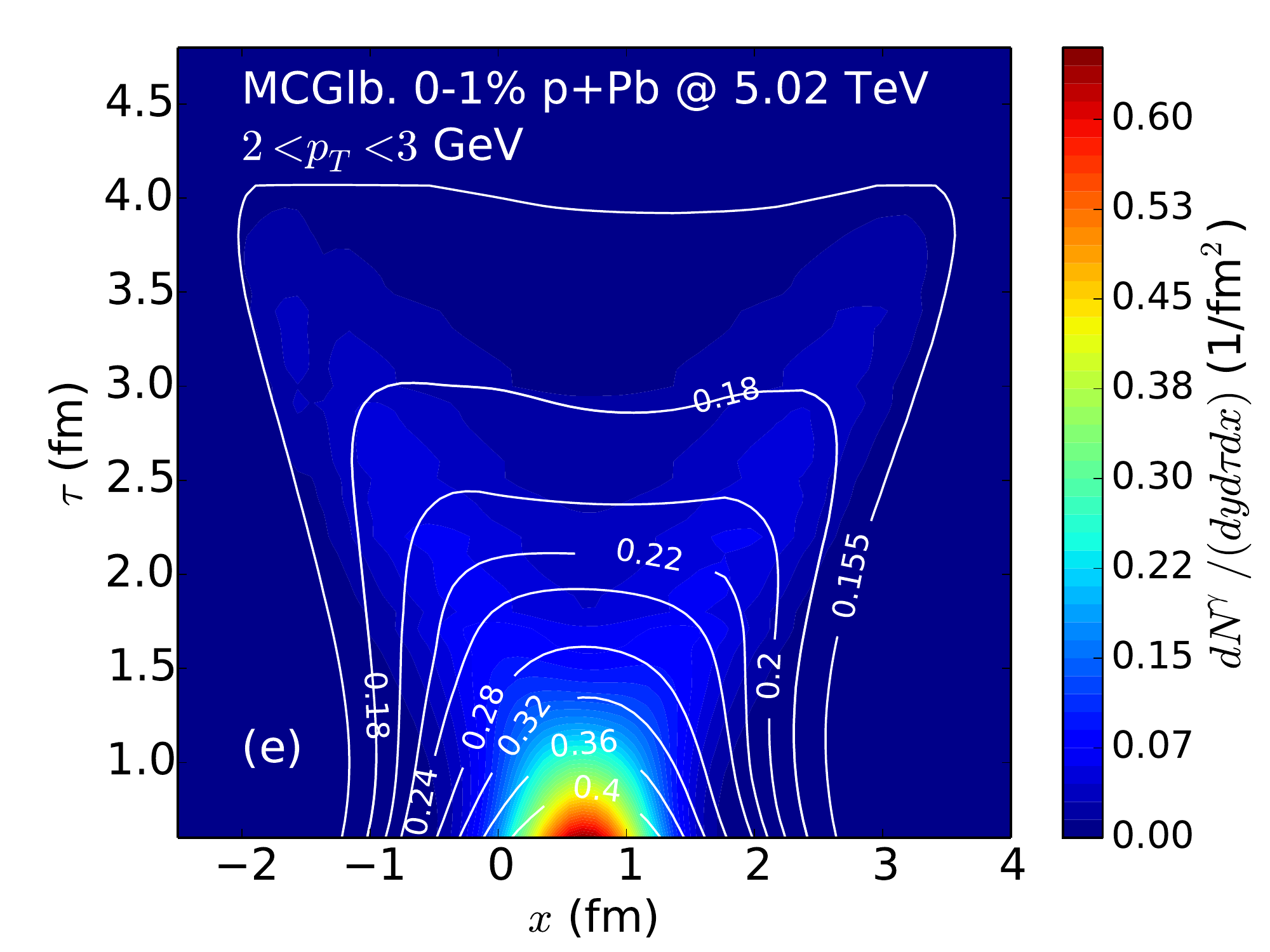} &
  \includegraphics[width=0.42\linewidth]{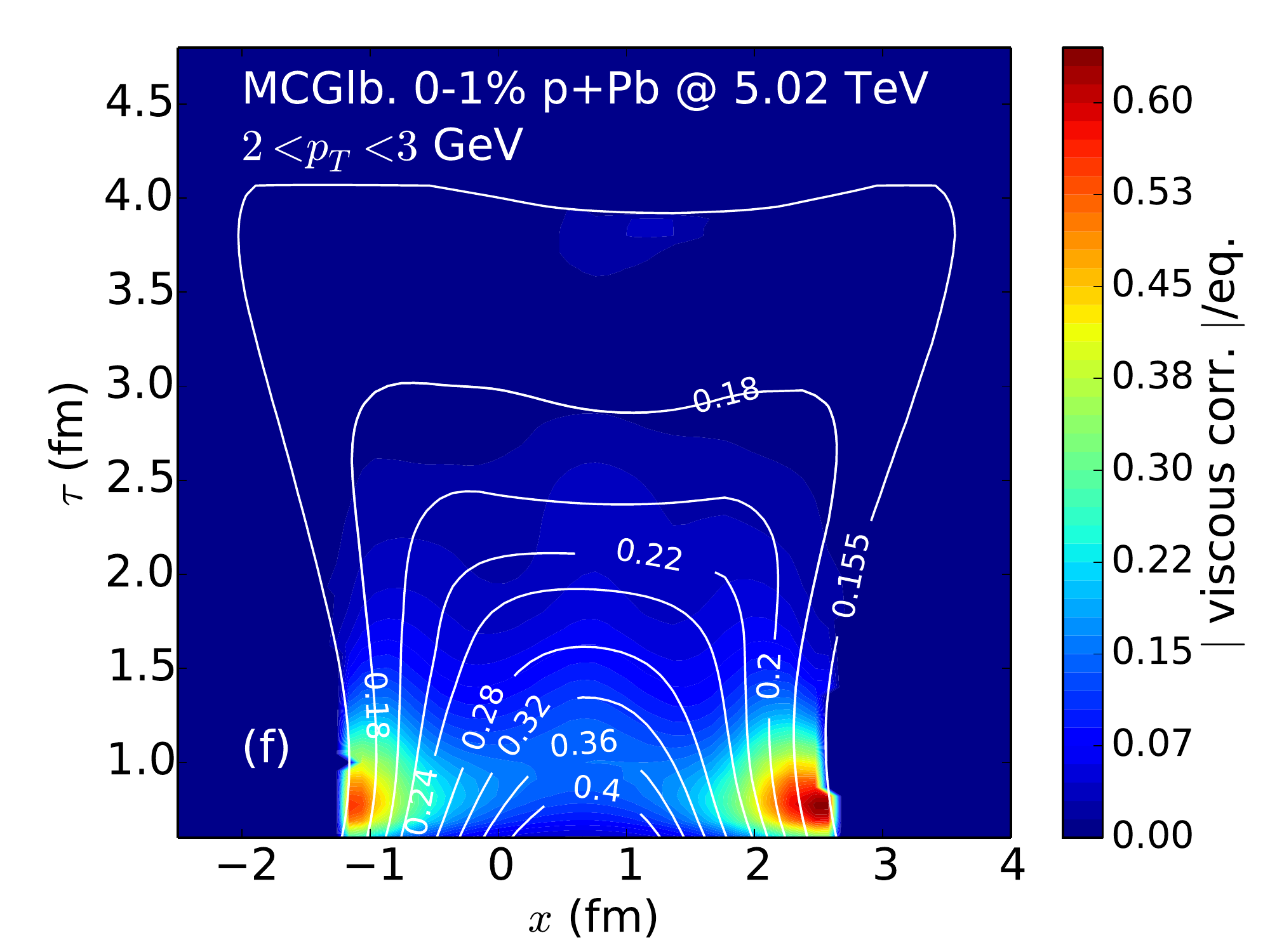}
  \end{tabular}
  \caption{(Color online) {\it Panel (a, c, e):} Contour plots for the space-time structure of thermal photon emission in one fluctuating 0-1\% p+Pb collision at 5.02 TeV. White contour lines are isotherms. {\it Panel (b, d, f):} The space-time structure of the relative size of shear viscous correction to thermal photon production compared to its corresponding equilibrium emission rate.}
  \label{fig13}
\end{figure*}
%

Semi-quantitative estimators of the validity of fluid dynamics are Knudsen and Reynolds numbers in numerical simulations. The Knudsen number is estimated according to Ref.~\cite{Niemi:2014wta},
\begin{equation}
{\rm Kn} = K_\theta = \tau_\pi \theta = \frac{5 \eta}{e + \mathcal{P}} \theta = \frac{5\eta}{sT} \theta,
\label{eqA1}
\end{equation}
where $\tau_\pi$ is the shear relaxation time and $\theta = \partial_\mu u^\mu$ is system's expansion rate. In order to have a realistic estimation of $K_\theta$ in the dilute hadronic phase, we adopt a temperature dependent $\eta/s(T)$ for $T < T_\mathrm{sw} = 155$ MeV \cite{Niemi:2011ix},
\begin{eqnarray}
\frac{\eta}{s}(T) &=& \left(\frac{\eta}{s}\right)_\mathrm{min} + 0.0594 \left(1 - \frac{T}{T_\mathrm{sw}} \right) \notag \\ 
 && + 0.544  \left(1 - \frac{T}{T_\mathrm{sw}} \right)^2.
\label{eqA2}
\end{eqnarray}
The smallness of the Knudsen number reflects how fast the system can evolve towards local thermal equilibrium. Complementarily, the distribution of the inverse Reynold's number \cite{Niemi:2014wta},
\begin{equation}
R_\pi^{-1} = \frac{\sqrt{\pi^{\mu\nu} \pi_{\mu\nu}}}{e + \mathcal{P}},
\label{eqA3}
\end{equation}
describes how far the system is out-of-equilibrium at every space-time point. Small values of Kn and $R_\pi^{-1}$ are needed for viscous effects to remain perturbative.

Figs.~\ref{fig12} shows the space-time distribution of the temperature profiles, together with the evolution of the Knudsen and of the inverse Reynold's number in two typical individual fluctuating events in central p+Pb and $^3$He+Au collisions. In spite of the small system size, the Knudsen numbers above the switching temperature, $T_\mathrm{sw} = 155$ MeV, remain $\sim 0.5$ in $^3$He+Au collisions at the top RHIC energy and reaches up to $0.6\sim0.7$ in the p+Pb collisions at 5.02 TeV. The result that $K_\theta < 1$ in the high temperature QGP phase suggests that a hydrodynamic description of these small systems is within the validity of the theory. In Figs.~\ref{fig12}(c) and (f), the evolution of the inverse Reynold's number is shown for the two events. The values of $R_\pi^{-1}$ remain small during the entire evolutions which means that the relative size of the shear stress tensor is small compared to its corresponding ideal part.

In Figs.~\ref{fig13}, we study the space-time structure of the thermal photon emission in one p+Pb collisions at 5.02 TeV. In the left panels, the thermal photon emission is shown for three $p_T$ cuts. 
In the right panels, the relative size of shear viscous correction to the photon yield is shown for the same momentum cuts. The largest viscous corrections are commonly found at the early stage of the evolution for all three $p_T$ bins. For $0.4 < p_T < 1$\,GeV, the maximum size of viscous correction is less 1\% compared to its equilibrium part. The relative size of the $\delta \Gamma$ correction increases quadratically as a function of $p_T$. For $2 < p_T < 3$ GeV, the shear viscous correction can reach up to 60\% of its equilibrium part. 
However, the corresponding left-hand-side plot indicates that the region where viscous corrections are large, at early time and large radius, are also regions of low photon emissivity. Thus very few photons are produced from these regions where viscous correction are large,  and viscous corrections are under control in space-time regions important for the thermal photon spectrum. This is in line with the findings of Section~\ref{section:viscousEffects}.

\bibliography{references}

\end{document}